\documentclass[twocolumn,twocolappendix]{aastex631}

\usepackage{amsmath,amstext}
\usepackage[utf8]{inputenc}
\usepackage[T1]{fontenc}
\usepackage{footmisc}

\newcommand{\D}{$^\circ$}
\newcommand{\per}{$^{-1}$}
\newcommand{\pers}{$^{-2}$}
\newcommand{\kms}{\mbox{km\,s$^{-1}$}}

\newcommand{\mJybeam}{\mbox{mJy\,beam$^{-1}$}}

\newcommand{\msun}{M$_\odot$}

\newcommand{\HII} {\mbox{\rm H{\small II}}}
\newcommand{\hcn}{HCN~$4-3$}
\newcommand{\httcn}{H$^{13}$CN~$4-3$}
\newcommand{\cs}{CS~$7-6$}

\newcommand{\cooz}{CO~$1-0$}
\newcommand{\cott}{CO~$3-2$}
\newcommand{\hcop}{HCO$^+~4-3$}
\newcommand{\htwo}{H$_2$}

\newcommand{\tclean}{{\fontfamily{cmtt}\selectfont tclean}}

\newcommand{\emcee}{{\fontfamily{cmtt}\selectfont emcee}}

\newcommand{\xone}{$x_1$}
\newcommand{\xtwo}{$x_2$}

\newcommand{\change}{}

\shorttitle{Super Star Cluster Architecture in NGC\,253}
\shortauthors{Levy et al.}

\begin{document}

\title{The Morpho-Kinematic Architecture of Super Star Clusters in the Center of NGC\,253}

\correspondingauthor{Rebecca C. Levy}
\email{rebeccalevy@email.arizona.edu}

\author[0000-0003-2508-2586]{Rebecca C. Levy}
\altaffiliation{NSF Astronomy and Astrophysics Postdoctoral Fellow}
\affiliation{Steward Observatory, University of Arizona, Tucson, AZ 85721, USA}
\affiliation{Department of Astronomy, University of Maryland, College Park, MD 20742, USA}

\author[0000-0002-5480-5686]{Alberto D. Bolatto}
\affiliation{Department of Astronomy, University of Maryland, College Park, MD 20742, USA}
\affiliation{Joint Space-Science Institute, University of Maryland, College Park, MD 20742, USA}
\affiliation{Visiting Scholar at the Flatiron Institute, Center for Computational Astrophysics, NY 10010, USA}
\affiliation{Visiting Astronomer, National Radio Astronomy Observatory, VA 22903, USA}

\author[0000-0002-2545-1700]{Adam K. Leroy}
\affiliation{Department of Astronomy, The Ohio State University, Columbus, OH 43210, USA}

\author[0000-0001-6113-6241]{Mattia C. Sormani}
\affiliation{Universit\"at Heidelberg, Zentrum f\"ur Astronomie, Institut f\"ur theoretische Astrophysik, Albert-Ueberle-Str. 2, 69120 Heidelberg, Germany}

\author[0000-0001-6527-6954]{Kimberly L. Emig}
\altaffiliation{Jansky Fellow of the National Radio Astronomy Observatory} 
\affiliation{National Radio Astronomy Observatory, 520 Edgemont Road, Charlottesville, VA 22903-2475, USA}

\author[0000-0001-9300-354X]{Mark Gorski}
\affiliation{Department of Space, Earth and Environment, Chalmers University of Technology, Onsala Space Observatory, 439 92 Onsala, Sweden}

\author[0000-0003-4023-8657]{Laura Lenki\'{c}}
\affiliation{SOFIA Science Center, USRA, NASA Ames Research Center, M.S. N232-12, Moffett Field, CA 94035, USA}

\author[0000-0001-8782-1992]{Elisabeth A. C. Mills}
\affiliation{Department of Physics and Astronomy, University of Kansas, 1251 Wescoe Hall Dr., Lawrence, KS 66045, USA}

\author[0000-0003-1356-1096 ]{Elizabeth Tarantino}
\affiliation{Department of Astronomy, University of Maryland, College Park, MD 20742, USA}

\author[0000-0003-1774-3436]{Peter Teuben}
\affiliation{Department of Astronomy, University of Maryland, College Park, MD 20742, USA}

\author[0000-0002-3158-6820]{Sylvain Veilleux}
\affiliation{Department of Astronomy, University of Maryland, College Park, MD 20742, USA}

\author[0000-0003-4793-7880]{Fabian Walter}
\affiliation{Max-Planck-Institut f\"ur Astronomie, K\"onigstuhl 17, 69120 Heidelberg, Germany}



\begin{abstract}

The center of the nearby galaxy NGC\,253 hosts a population of more than a dozen super star clusters (SSCs) which are still in the process of forming. The majority of the star formation of the burst is concentrated in these SSCs, and the starburst is powering a multiphase outflow from the galaxy. In this work, we measure the 350~GHz dust continuum emission towards the center of NGC\,253 at 47~milliarcsecond (0.8 pc) resolution using data from the Atacama Large Millimeter/submillimeter Array (ALMA). We report the detection of 350~GHz (dust) continuum emission in the outflow for the first time, associated with the prominent South-West streamer. In this feature, the dust emission has a width of $\approx$~8~pc, is located at the outer edge of the CO emission, and corresponds to a molecular gas mass of $\sim~(8-17)\times10^6$~M$_\odot$. In the starburst nucleus, we measure the resolved radial profiles, sizes, and molecular gas masses of the SSCs. Compared to previous work at somewhat lower spatial resolution, the SSCs here break apart into smaller substructures with radii $0.4-0.7$~pc. In projection, the SSCs, dust, and dense molecular gas appear to be arranged as a thin, almost linear, structure roughly 155~pc in length. The morphology and kinematics of this structure can be well explained as gas following $x_2$ orbits at the center of a barred potential. We constrain the morpho-kinematic arrangement of the SSCs themselves, finding that an elliptical, angular momentum-conserving ring is a good description of the both morphology and kinematics of the SSCs.

\end{abstract}


\section{Introduction}
\label{sec:intro}

Nuclear starburst regions in galaxies are thought to be fueled by inflows of cold gas to their centers. These gas inflows can be driven by a strong bar, a merger, or tidal interaction. In the case of a barred system, the bar efficiently funnels gas toward to center, where it settles onto a nuclear ring/disk with typical radii in the range $100-1000$~pc \citep[e.g.,][]{contopoulos77,contopoulos89,binney91,athanassoula92a,athanassoula92b,buta96,knapen99,perez-ramirez00,regan03,kormendy04,sormani22}. These collections of gas undergo shocks, causing them to collapse and form stars more efficiently than elsewhere in the disk leading to a nuclear starburst.

Studies of our own Galaxy and others suggest that the detailed morphology and kinematics of the nuclear region may play an important role in the formation and evolution of massive star clusters. However relatively few studies have dissected the star forming central molecular zones of galaxies at the high ($\sim 1$~pc scales) resolution needed to distinguish the locations of cluster formation. In this paper we conduct such an analysis targeting NGC\,253---the nearest bar-fed nuclear starburst to the Milky Way---by measuring the sizes and masses of the forming massive star clusters and constraining their 3D distribution and kinematics in relation to the bar. We also measure the properties of dust detected in the large scale outflow, a result of a central starburst.

In the case of the nearby galaxy NGC\,253, a strong bar fuels the nuclear starburst \citep[e.g.,][]{sorai00,paglione04}. As a result of the inflowing gas along the bar, the central few hundred parsecs of the galaxy contains $\approx3.5\times10^8$~\msun\ of \htwo\ and is forming stars at a rate of $\sim 2$~\msun~yr\per, resulting in a molecular gas depletion time of $\approx300$~Myr \citep{leroy15a,krieger19}. The nuclear starburst is responsible for launching a massive, multiphase wind which is detected across the electromagnetic spectrum, in X-rays \citep[e.g.,][]{strickland00,strickland02}, ionized gas \citep[e.g.,][]{heckman00,sharp10,westmoquette11}, polycyclic aromatic hydrocarbons \citep[e.g.,][]{mccormick13}, molecular gas \citep[e.g.,][]{sugai03,sturm11,bolatto13a,walter17,zschaechner18,krieger19}, and radio continuum emission \citep[e.g.,][]{turner85,heesen11}. The molecular gas in the outflow is concentrated along the edges of the biconical wind, and the most prominent feature is the so-called South-West (SW) streamer \citep{bolatto13a,walter17,zschaechner18,krieger19}. The deprojected molecular mass outflow rate of the wind is $\sim14-39$~\msun~yr\per, with large uncertainties due to the geometry \citep{krieger19}. With a mass loading factor (defined as the ratio of the mass outflow rate to the SFR) of $\sim7-20$, the outflow plays a critical role in regulating the star formation activity in the nucleus.

The nuclear region of NGC\,253 is a chemically-rich environment \citep{krieger20a,martin21,haasler22} and hosts a number of massive, dense molecular clouds \citep[e.g.,][]{sakamoto11,leroy15a}, radio continuum sources \citep[likely \HII\ regions and supernova remnants;][]{turner85,watson96,ulvestad97,kornei09}, and masers \citep[e.g.,][]{gorski17,gorski19,humire22}. The overwhelming majority of the star formation in the nuclear starburst is concentrated in massive forming "super" star clusters (SSCs; \citealt{ando17,leroy18,rico-villas20,mills21,levy21}). These proto-SSCs are responsible for 3\% of the total infrared emission of the galaxy \citep{martin21}. The gas content and star formation activity in this region resemble a scaled-up version of that found in the Central Molecular Zone (CMZ) of the Milky Way (MW; \citealt[e.g.,][]{sakamoto11,martin21}). The SSCs in NGC\,253 have stellar masses $\approx10^{4.0-6.0}$~\msun\ and gas masses $\approx10^{3.6-5.7}$~\msun\ \citep{leroy18,mills21}. At the resolution of these studies ($\sim2-5$~pc), however, multiple SSCs are blended together, as revealed by very high (0.5~pc) resolution data of this region \citep{levy21}.

From high resolution images of the SSCs taken using ALMA, the SSCs are embedded in an extended background of dust and molecular gas, and this structure measures ${\rm \approx155~pc~\times~15~pc}$ in projected length and width \citep[e.g.,][]{ando17,leroy18,levy21,mills21}. The observed nearly linear arrangement of the SSCs is almost certainly a projection effect, as NGC\,253 has an inclination of $\approx 78^\circ$ \citep[e.g.,][]{pence80,westmoquette11,krieger19}. In 3D, it is possible that the SSCs trace a circumnuclear ring which may be connected to the bar. A promising hint in this direction is that the location of the inner inner Lindblad resonance (IILR) --- inside which gas is expected to concentrate --- is located at a radius of $\sim350$~pc from the center, though the uncertainty on this radius is likely substantial \citep{sorai00}\footnote{\citet{sorai00} assumed a galaxy distance of 2.5~Mpc, whereas a more recent and accurate determination of the distance is 3.5~Mpc \citep{rekola05}. We re-calculate all physical sizes from \citet{sorai00} using this updated galaxy distance.\label{foot:sorai}}. Qualitatively, the IILR is on the same scale as the SSC and dense gas structures. While \citet{paglione04} find weaker evidence of an inner Lindblad resonance (ILR) than \citet{sorai00}, they do find that the dense molecular gas in the center is consistent with the locations of \xtwo\ orbits (see e.g., their Figure 11). The \xtwo\ orbits are expected to lie between the outer ILR (OILR) and IILR and are oriented perpendicular to the bar major axis \citep[e.g.,][]{contopoulos89,buta96,das01}. 

Given the nearly edge-on inclination of NGC\,253, inferring a connection with the bar and constraining the geometry of the SSC structure from the 2D morphology alone would be nearly impossible. In this study, we use new images of the dust continuum and dense gas emission in the center of NGC\,253, which combine multiple ALMA configurations to cover a wide range of spatial scales. These data allow us to simultaneously resolve the compact SSCs and the more extended molecular gas and dust emission. We combine the dust continuum images with the systemic velocities of the clusters measured from very high resolution spectral line data by \citet{levy21}. This velocity information adds a third dimension to the data, allowing us to better constrain the morpho-kinematic architecture of the SSCs and their connection to the larger scale gas flows in this galaxy.

This article is organized as follows. We describe the observations, data processing steps, and imaging in Section \ref{sec:obs4}. We report the detection of dust emission associated with the SW streamer of the molecular outflow in Section \ref{ssec:swstreamer}. The methods used to measure the cluster sizes, fluxes, and gas masses are described in Section \ref{ssec:clustersizes}. In Section \ref{sec:morphokin}, we quantitatively compare the arrangement and kinematic structure of the clusters to an elliptical, angular momentum-conserving ring. We summarize our findings in Section \ref{sec:summary4}.

The precise center of NGC\,253 is not known and the location of its central supermassive black hole has not been definitively identified. Throughout this paper, we refer to the galaxy center at $(\alpha,~\delta)_{\rm J2000}$ = (${0^{\rm h}47^{\rm m}33.06^{\rm s},}\ {-25^\circ17^\prime18.3^{\prime\prime}}$) measured using ionized gas kinematics traced by H92$\alpha$ at $\approx1.8$\arcsec\ resolution \citep{anantharamaiah96}. The 1-$\sigma$ uncertainty on this center position is $\sim0.3$\arcsec.

\section{Observations and Data Processing}
\label{sec:obs4}

Data for this project were taken with ALMA as part of projects 2015.1.00274.S and 2017.1.00433.S (P.I. A. Bolatto). We observed the central 16.64\arcsec\ (280~pc) of NGC\,253 at Band 7 ($\nu\sim350$~GHz, $\lambda\sim0.85$~mm) using the main 12-m array in the C43-9, C43-6, and C43-4 configurations and the 7-m (ACA) array. These configurations cover baselines of ${\rm 113~m - 13.9~km}$, ${\rm 15.1~m - 1.8~km}$, ${\rm 15.1~m - 783.5~m}$, and ${\rm 8.9~m - 49.0~m}$, respectively \citep{krieger19,levy21}. Including the ACA data, the maximum recoverable scale is 24.8\arcsec\ (421~pc); excluding the ACA data, the the maximum recoverable scale is 3.9\arcsec\ (66~pc). The spectral setup spans frequency ranges of $342.08-345.78$~GHz in the lower sideband and $353.95-357.66$~GHz in the upper sideband. The visibilities were pipeline calibrated (L. Davis et al. in prep.) using the Common Astronomy Software Application (\textsc{casa}; \citealt{casa}). More information on these observations has been published previously \citep[see][]{leroy18,krieger19,levy21}.

To extract the 350~GHz continuum data, we flagged channels that may contain strong lines in the band, assuming a recessional velocity of 243~\kms\ \citep{koribalski04}. Lines included in the flagging are $^{12}$\cott, \hcn, \httcn, \cs, \hcop, and $^{29}$SiO $8-7$, and channels within $\pm$200~\kms\ of the rest frequencies of these lines were flagged.

\subsection{Imaging the Multi-Configuration Data Sets}
\label{ssec:multiconfigimaging}

In this work, we make two different combinations of the multi-configuration datasets. First, we combine the ACA data and three 12-m configurations together to make what we will refer to as the "12-m+ACA map" or the "2.55~pc resolution map." The objective of this map is to recover the most extended dust continuum emission in the nuclear region. We also make a second multi-configuration data set using only the three 12-m configurations, which we will refer to as the "12-m map" or the "0.81~pc resolution map." The objective of this data set is to recover the dust emission associated with the clusters. The calibrated and line-flagged visibilities were combined for imaging using the \texttt{concat} task in \textsc{casa}. We spectrally averaged the combined measurement set to have 10 channels per sideband, so that each channel covers $\sim$40~MHz. 

Since the 12-m+ACA and 12-m maps have different objectives, we used different deconvolution strategies to produce the final images, which we describe below. All of the visibilities were imaged using the \textsc{casa} version 5.4.1 \tclean\ task. 

\subsubsection{Imaging the 12-m+ACA Data}
\label{sssec:12mACAim}

\begin{figure*}
    \centering
    \includegraphics[width=0.49\textwidth]{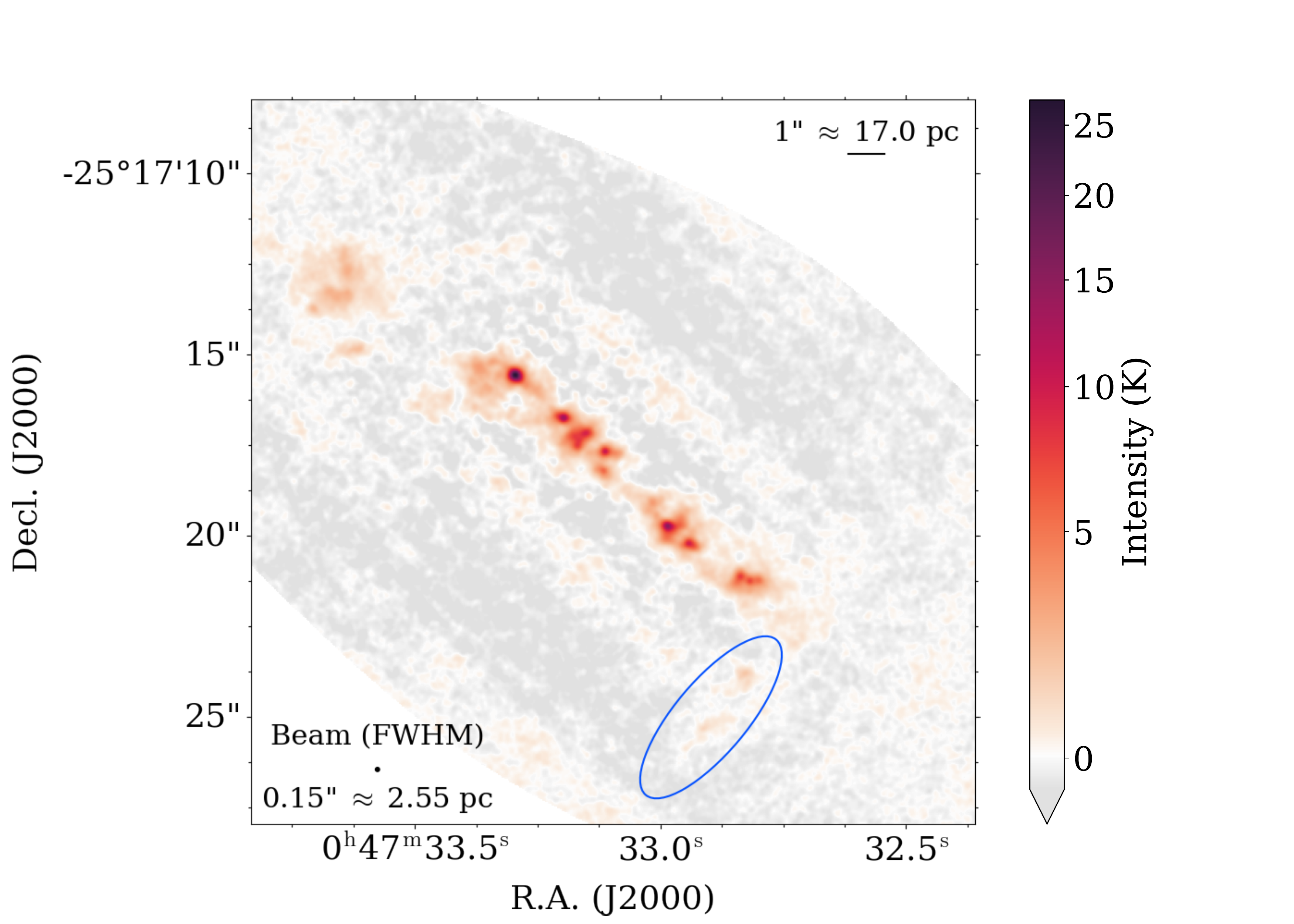}
    \includegraphics[width=0.49\textwidth]{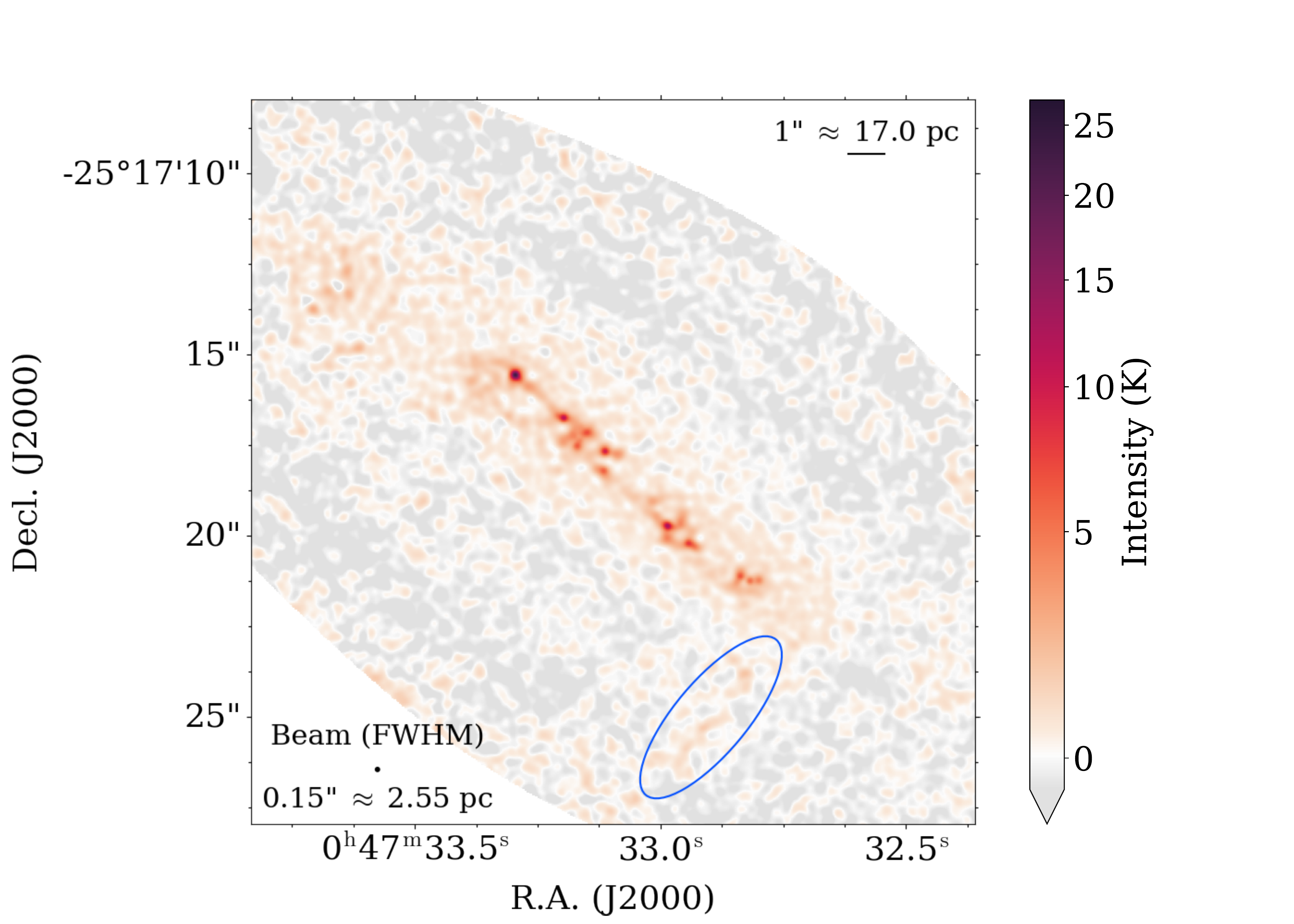}
    \includegraphics[width=0.49\textwidth]{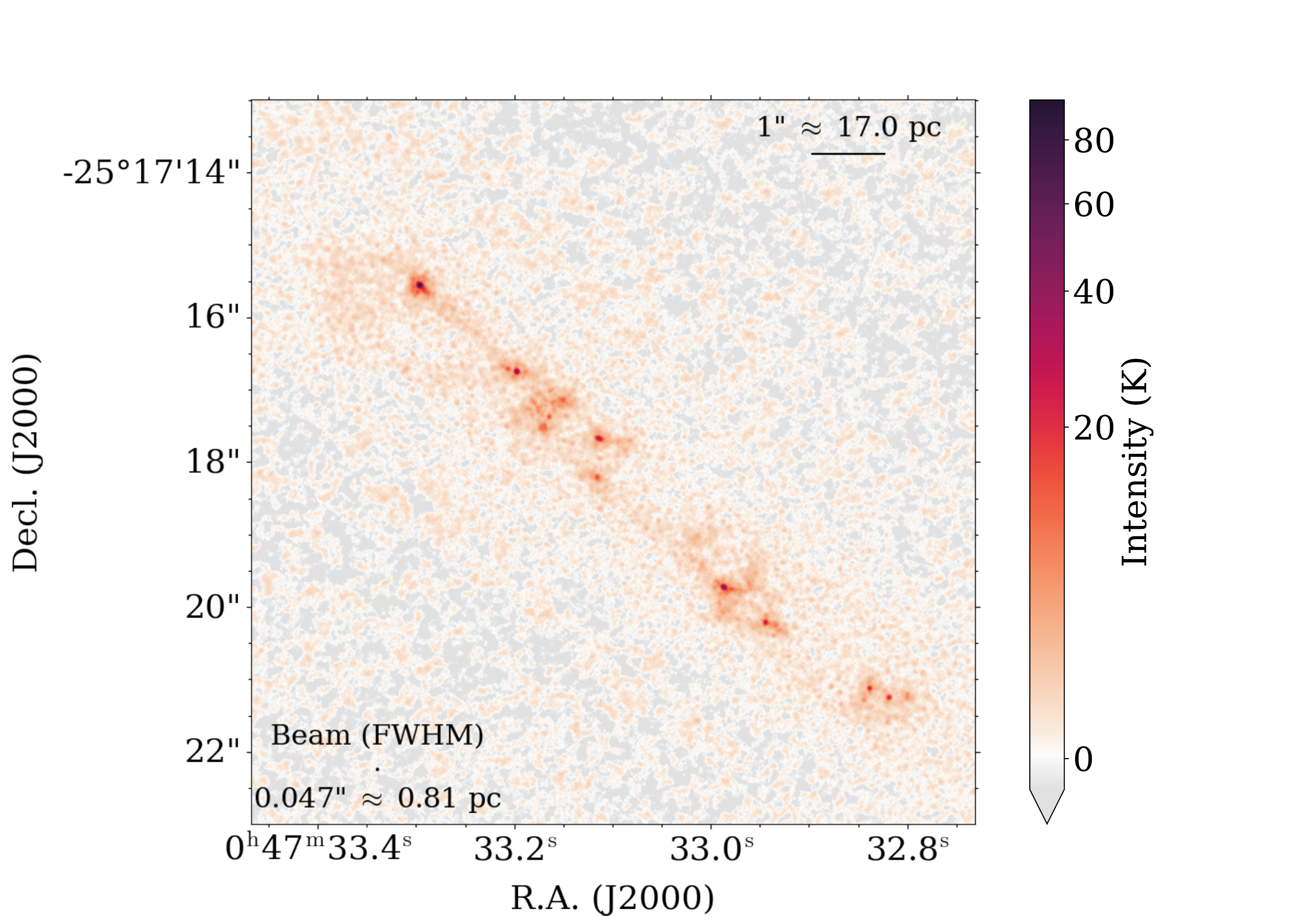}
    \includegraphics[width=0.49\textwidth]{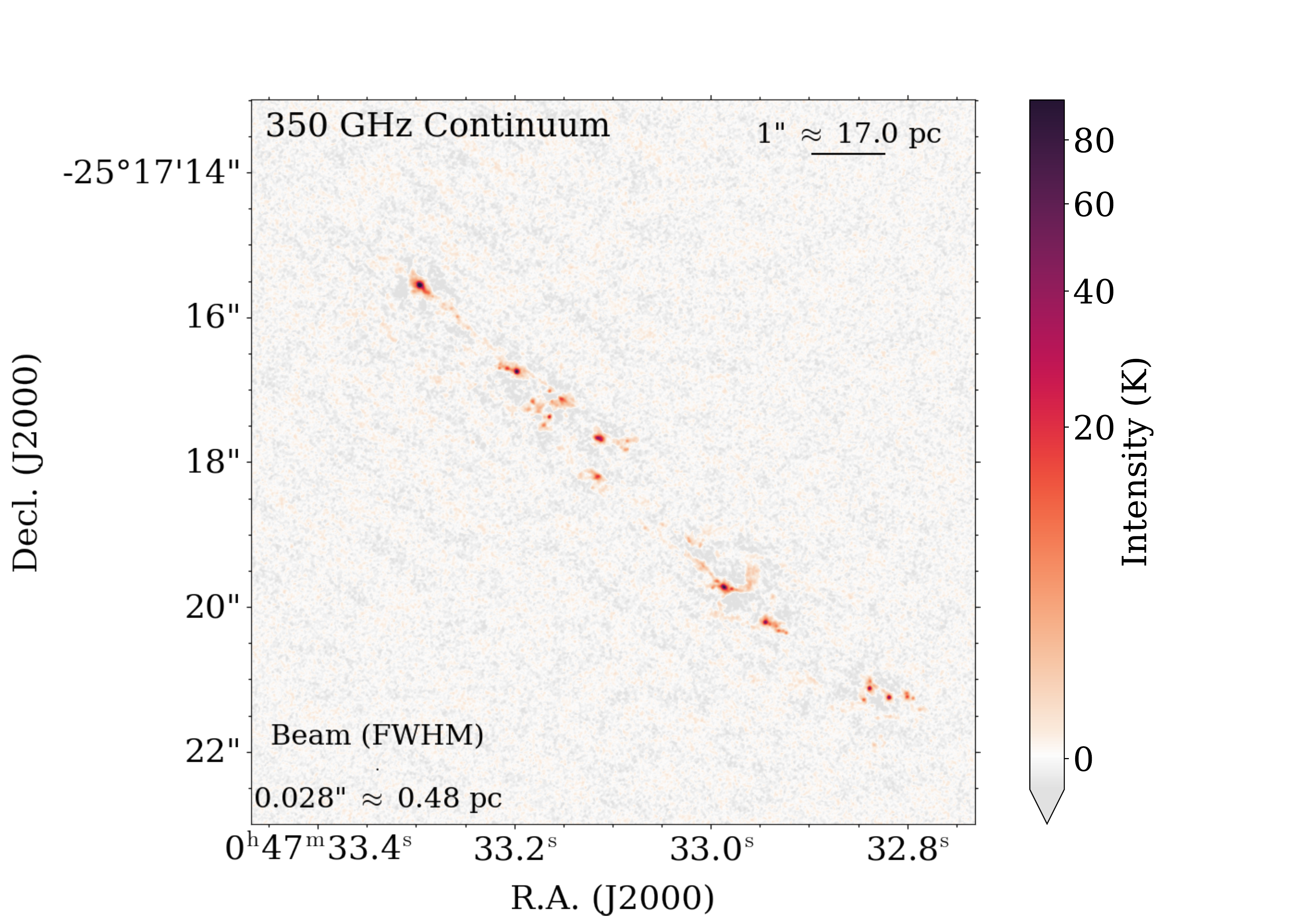}

    \caption{{\em Top row:} The 350~GHz 12-m+ACA dust continuum emission in the central 20\arcsec\ (340~pc) of NGC\,253, made by combining three 12-m configurations and one 7-m (ACA) configuration. The top left image shows the dirty map and the top right shows the cleaned map both convolved to a circular 0.15\arcsec\ beam as described in Section \ref{sssec:12mACAim}. The blue ellipses highlight the SW streamer seen in the dust continuum emission (see Section \ref{ssec:swstreamer}). {\em Bottom left:} The 350~GHz continuum image in the central 10\arcsec\ made by combining the three 12-m configurations, which has a final resolution of 0.047\arcsec. The cleaning has been optimized for the cluster-scales, as described in Section \ref{sssec:12mim}. {\em Bottom right:} The 350~GHz continuum image in the central 10\arcsec\ of NGC\,253 using only the highest resolution (0.028\arcsec) data from \citet{levy21}.}
    \label{fig:ngc253_cont}
\end{figure*}

We imaged the central 48\arcsec$\times$48\arcsec\ of the line-flagged, channel averaged, combined 12-m+ACA visibilities interactively using \tclean. Since we are interested in the more extended dust continuum emission, we choose a coarser cell (pixel) size of 0.04\arcsec\ than \citet{levy21} used to image the high resolution continuum data. In all iterations, we used \texttt{specmode=`mfs'}, \texttt{deconvolver=`multiscale'}, Briggs weighting with \texttt{robust=0.5}, and apply the primary beam correction. The clean components were modeled using a linear spectral slope (\texttt{nterms=2}) to account for any continuum slope over the bandpass. The "dirty" image (\texttt{niter=0}) is shown in Figure \ref{fig:ngc253_cont} (top left) for the inner 20\arcsec$\times$20\arcsec. The dirty map has a FWHM Gaussian beam size of 0.110\arcsec$\times$0.095\arcsec. This image is convolved to a circular 0.15\arcsec\ beam, to match the resolution of the cleaned, tapered image described below.

Before cleaning the extended emission, it was necessary to carefully clean the point source-like clusters, otherwise the algorithm had a tendency to over-subtract these regions leaving deep negative bowls. We cleaned the emission from the clusters using \texttt{scales=[0]} and interactively controlled the threshold and number of iterations to avoid over-cleaning. We cleaned the point sources until they were no longer point-like in the residual map and so that the extended residual emission near the point sources was similar to the larger scale emission in the map.

Due to the range of spatial scales covered by these combined data sets, the algorithm tends to favor small scales, making cleaning the extended emission time consuming. Since, for the 12-m+ACA map, we are interested in the larger scale more diffuse emission, we used a $uv$-taper of 0.2\arcsec, \texttt{scales=[0,8,16]}, \texttt{smallscalebias=0} which gives equal weight to all scales to more efficiently clean the map. We used a circular 0.15\arcsec\ restoring beam. To avoid over-cleaning, we reduced the gain of each major cycle to 0.05 and interactively lowered the threshold. We interactively cleaned the map until the residuals stopped changing. The final cleaned 12-m+ACA map is shown in Figure \ref{fig:ngc253_cont} (top right), which has an rms of 1.1~\mJybeam\ (0.5~K) in regions away from emission.

\subsubsection{Imaging the 12-m Data}
\label{sssec:12mim}

We imaged the central 48\arcsec$\times$48\arcsec\ of the line-flagged, channel averaged, combined 12-m visibilities interactively using \tclean. Since we are interested in the dust continuum emission associated with the clusters, we use a different imaging strategy from the one described above. We use a cell (pixel) size of 0.0046\arcsec, the same as was used for the high resolution continuum map \citep{levy21} which is shown for comparison in Figure \ref{fig:ngc253_cont} (bottom right). In all iterations, we used \texttt{specmode=`mfs'}, \texttt{deconvolver=`multiscale'}, Briggs weighting with \texttt{robust=0.5}, and apply the primary beam correction. The baseline was fit with a linear function (\texttt{nterms=2}) to account for any continuum slope over the bandpass.

As with the 12-m+ACA map, before cleaning the extended emission, it was necessary to carefully clean the point source-like clusters, otherwise the algorithm had a tendency to over-subtract these regions leaving deep negative bowls. We cleaned the emission from the clusters using \texttt{scales=[0]} and interactively controlled the threshold and number of iterations to avoid over-cleaning. We cleaned the point sources until they were no longer point-like and so that the emission in those regions was similar to the larger scale emission in the map.

\begin{figure*}
    \centering
    \includegraphics[width=\textwidth]{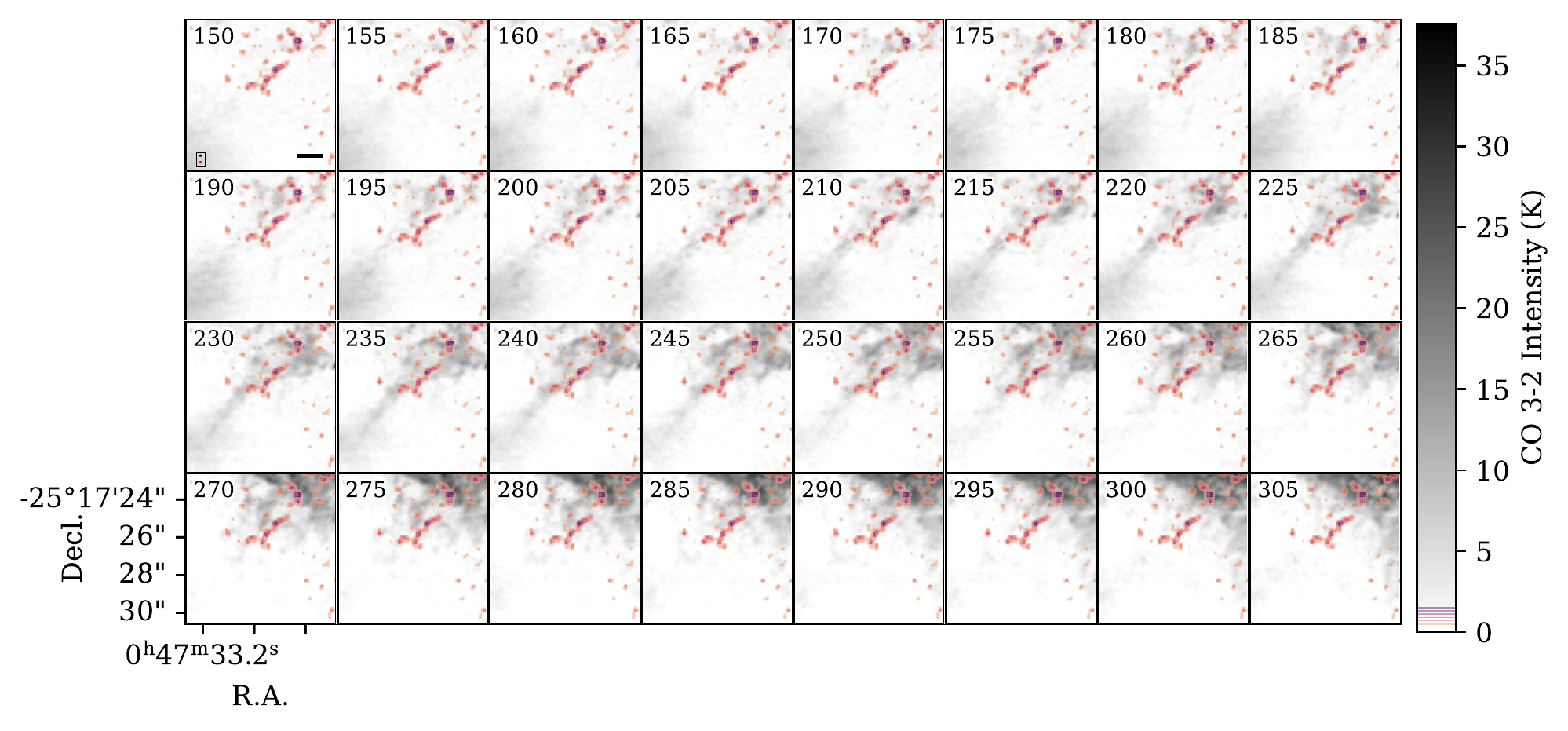}
    \caption{Channel maps of the CO $3-2$ emission showing the SW streamer \citep[grayscale;][]{krieger19}. The LSRK velocity of each channel is shown in the upper left corner. The red/orange filled contours show the 350~GHz dust continuum emission from the 12-m+ACA data (top right panel of Figure \ref{fig:ngc253_cont}). The contours span from $0.5-1.5$~K in steps of 0.2~K. We have removed data near the map edges where the signal is increased due to the primary beam correction. The FWHM beam sizes are shown in the box in the lower left corner of the first panel. Each panel is 135~pc (8\arcsec) on a side. The scale bar in the lower right corner of the first panel shows 20~pc. The 350~GHz continuum emission tends to trace the outer (southwest) edge of the SW streamer seen in CO.}
    \label{fig:SWstreamer_chanmaps}
\end{figure*}

Once the point sources were "cleaned", we carefully cleaned the more extended emission, starting from large scales and moving to smaller ones. For each iteration, we used \texttt{smallscalebias=0} and no $uv$-taper. We first started with \texttt{scales=[16,32,64]} (corresponding to  $\approx$0.07\arcsec, 0.15\arcsec, and 0.30\arcsec). We interactively cleaned these scales until the maximum residual and cleaned flux no longer changed significantly. We then added \texttt{scales=[8]} ($\approx$0.04\arcsec) to the existing scales and continued to clean interactively as before. After this scale was cleaned, the overall residuals resembled noise. The map had a FWHM Gaussian beam size of 0.045\arcsec$\times$0.041\arcsec. We convolved the image to a circular 0.0475\arcsec\ beam. The final cleaned 12-m map is shown in Figure \ref{fig:ngc253_cont} (bottom left), which has an rms of 0.2~\mJybeam\ (0.8~K) in regions away from emission.

\section{South-West Streamer Detected in Dust Continuum Emission}
\label{ssec:swstreamer}

In Figure \ref{fig:ngc253_cont} (top), we detect the SW streamer in 350~GHz dust continuum emission for the first time. The SW streamer is the brightest feature of the molecular outflow component and corresponds to the SW edge of the outflow cone. It was first detected in \cooz\ by \citet{bolatto13a} and has subsequently been detected in other CO transitions and dense molecular gas tracers \citep[e.g.,][]{walter17,zschaechner18,krieger19}. High line ratios of HCN/CO in the SW streamer indicate that this component of the outflow originates from the central starburst \citep{walter17}. The SW streamer has an estimated age of $\sim1$~Myr \citep{walter17}, in good agreement with the approximate ages of the massive star-forming regions in the starburst nucleus (see Section \ref{ssec:agegrad}). 

We compare the location of the dust component of the SW streamer to the \cott\ from \citet{krieger19} in Figure \ref{fig:SWstreamer_chanmaps}. The \cott\ data have a similar beam size (0.17\arcsec~$ = 2.88$~pc) as the dust map (0.15\arcsec~$ = 2.55$~pc; Figure \ref{fig:ngc253_cont}, top right). The dust continuum emission has a similar morphology as the \cott, but is offset to the southwest. In the context of the larger-scale outflow, this means that the dust is found towards the outer edge of the outflow cone.

That the dust emission is primarily at the edge of the outflow cone may be an optical depth effect. We would expect the dust to be distributed as a hollow cone, like the outflowing molecular gas, which confines the hot outflowing material \citep[e.g.,][]{leroy15b,meier15}. In projection, the dust will have the highest optical depth along the line of sight in a streamer-like structure at the very edge of the projected outflow. This effect is illustrated in Figure 8 of \citet[][see especially the purple regions of this figure]{bolatto21}. It is still unclear, however, precisely why there is such a large offset between the dust and \cott\ in the SW streamer in NGC\,253. While both CO and dust show temperature effects, the dust is more likely to remain a simple optically thin column density tracer and hence may better trace the true "spine" of the cone. 

\subsection{Inferred \htwo\ Column Density and Mass in the SW Streamer}
We estimate the flux of dust emission in the SW streamer using the 12-m+ACA maps within the blue ellipse shown in Figure \ref{fig:ngc253_cont} (top right). The flux density in this region is $\approx 540\pm180$~mJy. We convert this flux density to an estimated average \htwo\ column density and mass, assuming a dust-to-gas ratio of 1/100, a dust mass absorption coefficient of 1.9~cm$^2$~g\per\ following \citet{leroy18}. We assume a dust temperature of 34~K \citep{gao04,weiss08,mangum13}, but we note that the dust temperature is not well constrained in the outflow itself. We adopt a minimum dust temperature of 11~K \citep{zschaechner18} and a maximum dust temperature of 50~K \citep{walter17}, which we propagate into our uncertainty on the column density and mass. This calculation yields N$_{\rm H_{2}}\sim(4^{+6}_{-2})\times10^{23}$~cm~\per\ and
M$_{\rm H_{2}}\sim(1.7^{+2.6}_{-0.8})\times10^7$~\msun. The uncertainties are dominated by uncertainties on the dust temperature in the outflow. Since the outflow is expected to be warm \citep[e.g.,][]{leroy15b,walter17,zschaechner18}, the true column density and molecular gas mass are more likely to be N$_{\rm H_{2}}\sim(2-4)\times10^{23}$~cm~\per\ and M$_{\rm H_{2}}\sim(8-17)\times10^6$~\msun

\citet{walter17} estimated the \htwo\ column density in the SW streamer in two ways: from the \cooz\ and from the H$\alpha$/Pa$\beta$ line ratio. From the CO, they found N$_{\rm H_{2}}\sim4\times10^{21}$~cm\pers, which is consistent with their extinction-based estimate of N$_{\rm H_{2}}\sim5\times10^{21}$~cm\pers. They note, however, that the detection of bright emission from HCN and other molecules in the SW streamer implies a larger density of N$_{\rm H_{2}}\sim5\times10^{22}$~cm\pers, in better agreement with our estimate. Our dust-based estimate of the \htwo\ mass in the SW streamer is consistent with the minimum mass of the SW streamer of $\sim10^6$~\msun\ found by \citet{walter17} and other CO-based measurements  \citep[i.e.,][]{bolatto13a,zschaechner18,krieger19}.

\begin{figure}
    \centering
    \includegraphics[width=\columnwidth]{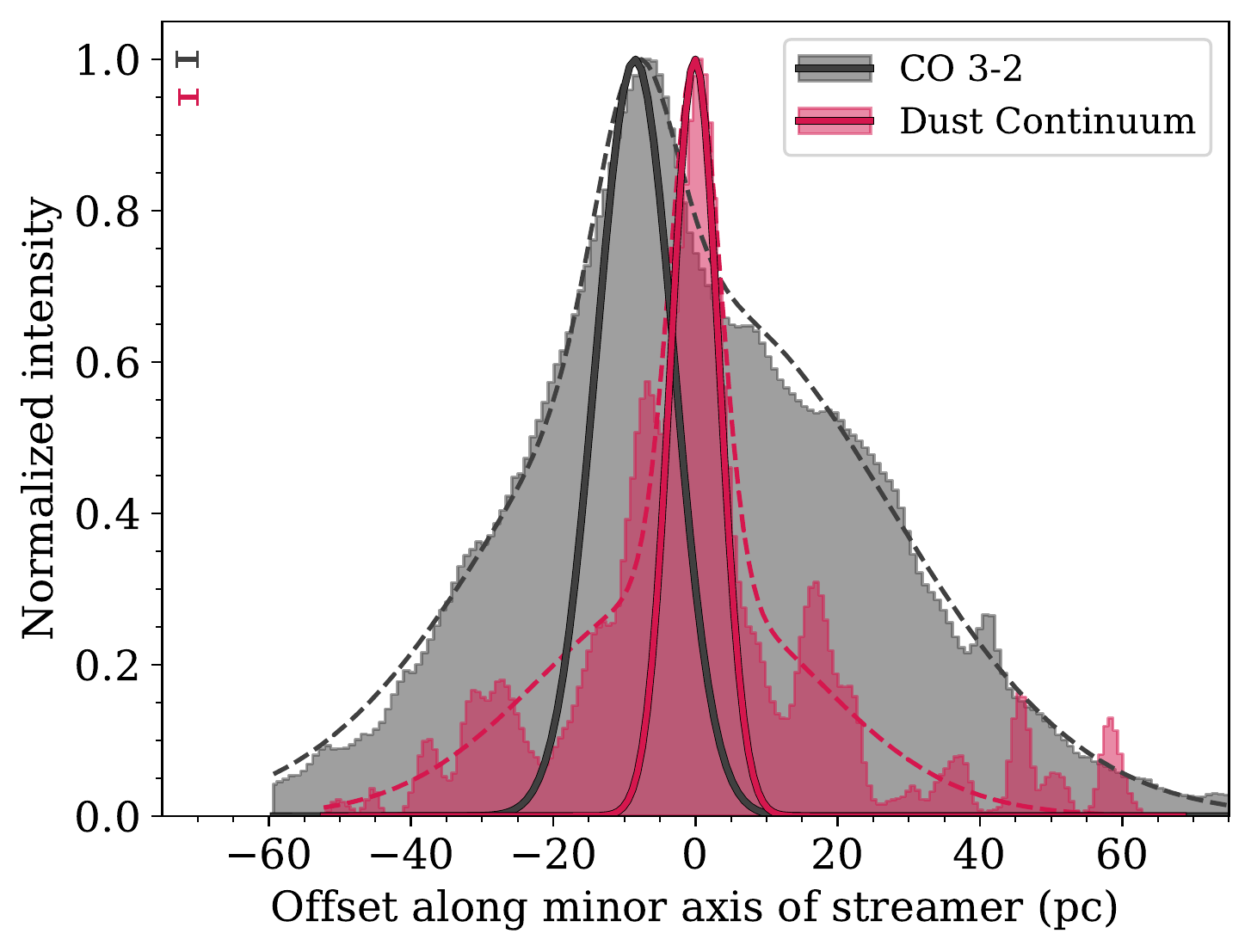}
    \caption{The normalized width profiles of the SW streamer, where the streamer has been summed along its major axis. The shaded histograms show the width profiles for the dust continuum (red) and \cott\ (gray). For the CO, this profile is the integrated intensity over the velocity channels shown in Figure \ref{fig:SWstreamer_CO}. These profiles are fit with a two-component Gaussian, shown as the dashed curves. The narrow components (corresponding to the outflow) of the fits are shown as the solid curves, where the respective beam sizes have been deconvolved. The FWHM beam sizes are shown in the upper left corner. The x-axis is the offset along the minor axis of the streamers relative to the centroid of the narrow dust component. The dust continuum component of SW streamer is intrinsically narrower than and offset from the \cott.}
    \label{fig:SWstreamer_width}
\end{figure}

\subsection{Width of the SW Streamer in Dust and \cott}
We estimate the width of the dust emission in the SW streamer, summing the emission along the major axis of the streamer (PA~$\approx 140$\D). This yields a profile of the summed intensity along the minor axis of the SW streamer (red histogram in Figure \ref{fig:SWstreamer_width}). We fit a two component Gaussian to this width profile, with one narrow component for the streamer and a broad component for any disk emission. We remove the beam in quadrature from the width of the narrow component and show this beam-deconvolved narrow component corresponding to the outflow in Figure \ref{fig:SWstreamer_width}. The dust streamer has a beam-deconvolved FWHM of 8~pc.

\begin{figure*}
    \centering
    \includegraphics[width=\textwidth]{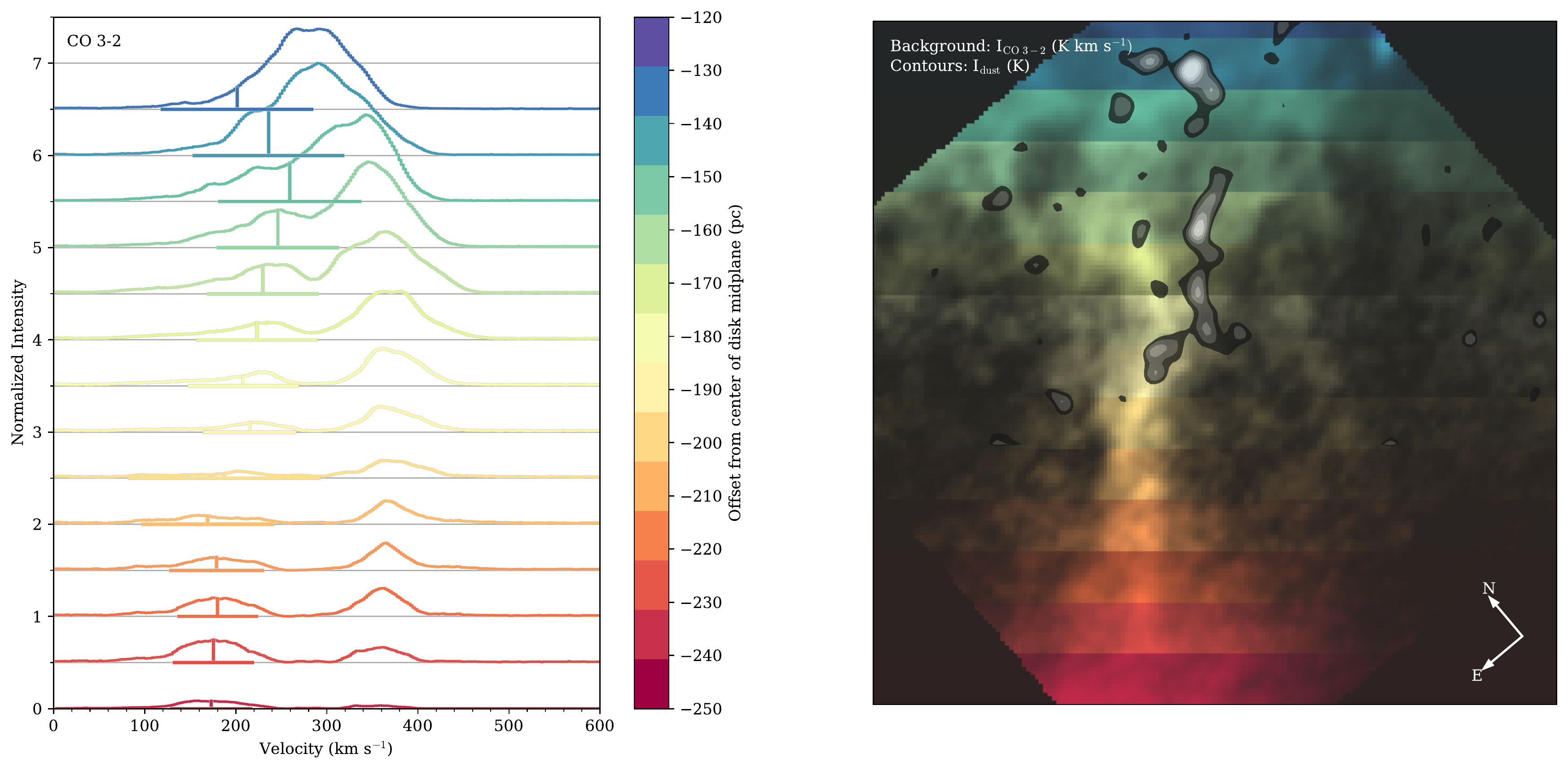}
    \caption{{\em Left:} \cott\ spectra in slices along the SW streamer. The \cott\ data are binned in 10~pc bins along the major axis of the outflow, as indicated by the colorbar. The data are summed in each bin and normalized based on the maximum intensity. The spectra are offset artificially along the y-axis. The velocity component corresponding to the outflow is fit with a Gaussian, where the mean and FWHM velocity are indicated by the vertical and horizontal line segments. {\em Right:} The SW streamer seen in \cott\ (colored background) and the 350~GHz dust emission (grayscale contours), rotated so the major axis is vertical and the top of the image is closest to the midplane. For the \cott, data in each 10~pc bin is color coded based on its offset from the midplane according to the colorbar in the left panel. We find the integrated intensity over the channels within one FWHM from the central velocity (i.e., over velocities indicated by the horizontal line segments in the left panel). The dust continuum contours are the same as in Figure \ref{fig:SWstreamer_chanmaps}.}
    \label{fig:SWstreamer_CO}
\end{figure*}

We compare the width of the dust streamer to that of the \cott. First, we kinematically identify the components of the \cott\ emission associated with the SW streamer, using a method similar to that of \citet{walter17} for the \cooz\ (see their Section 3.2). We take position-velocity slices through the CO data cube over the field-of-view (FOV) shown in Figure \ref{fig:SWstreamer_chanmaps}. Each slice is 10~pc wide along the major axis of the SW streamer. Pixels within each slice are summed to produce the spectra shown in Figure \ref{fig:SWstreamer_CO} (left). Away from the midplane, the \cott\ has two velocity components, where the blueshifted component traces the outflow and the redshifted component primarily traces emission from the central starburst and disk. We fit the outflow component with a Gaussian at each offset, where the mean velocity and FWHM are indicated by the vertical and horizontal lines for each spectrum in Figure \ref{fig:SWstreamer_CO} (left). Using these fits, we calculate the integrated \cott\ intensity over the FWHM velocity range of the outflow component for each slice. This integrated intensity map is shown in Figure \ref{fig:SWstreamer_CO} (right) where the color coding of the image indicates the offset from the midplane as in the left panel. We overplot contours of the dust continuum emission to again highlight the offset between the dust and CO in the SW streamer. We measure the width of the \cott\ SW streamer using this integrated intensity map following the same procedure as for the dust described in the previous paragraph. We note that the beam sizes of the 350 GHz continuum and \cott\ are very similar (2.55~pc and 2.88~pc respectively). The width profile and Gaussian fits for the \cott\ are shown in Figure \ref{fig:SWstreamer_width} (gray). The beam-deconvolved FWHM of the \cott\ associated with the SW streamer is 13~pc.  The SW streamer seen in the dust continuum is narrower than in the CO by a factor of $\sim1.6$. The peak of the dust continuum is offset to the southwest of the peak of the \cott\ by 8~pc ($\sim3\times$ the FWHM beam size).

\section{Cluster Size, Flux, and Mass Measurements}
\label{ssec:clustersizes}

Accurate sizes and masses for the SSCs are crucial to understand their physical properties and compare to numerical simulations \citep[e.g.,][]{grudic21}. Previous studies which have measured these parameters used data which marginally resolved the SSCs \citep[e.g.,][]{leroy18,rico-villas20,mills21}. However, using 0.5~pc resolution data, \citet{levy21} showed that the SSCs seen in the 350~GHz dust continuum emission break apart into multiple components once they are spatially resolved. Those data, however, used only the most extended ALMA configurations and hence lacked the short spacings which are sensitive to extended emission. This short spacing information is important to accurately measure the flux of the clusters as well as the background of extended emission in which they are embedded. Accurate measurements of the SSC sizes and masses, therefore, require both high spatial resolution and complete sampling of the Fourier plane.

\begin{figure*}
    \centering
    \includegraphics[width=\textwidth]{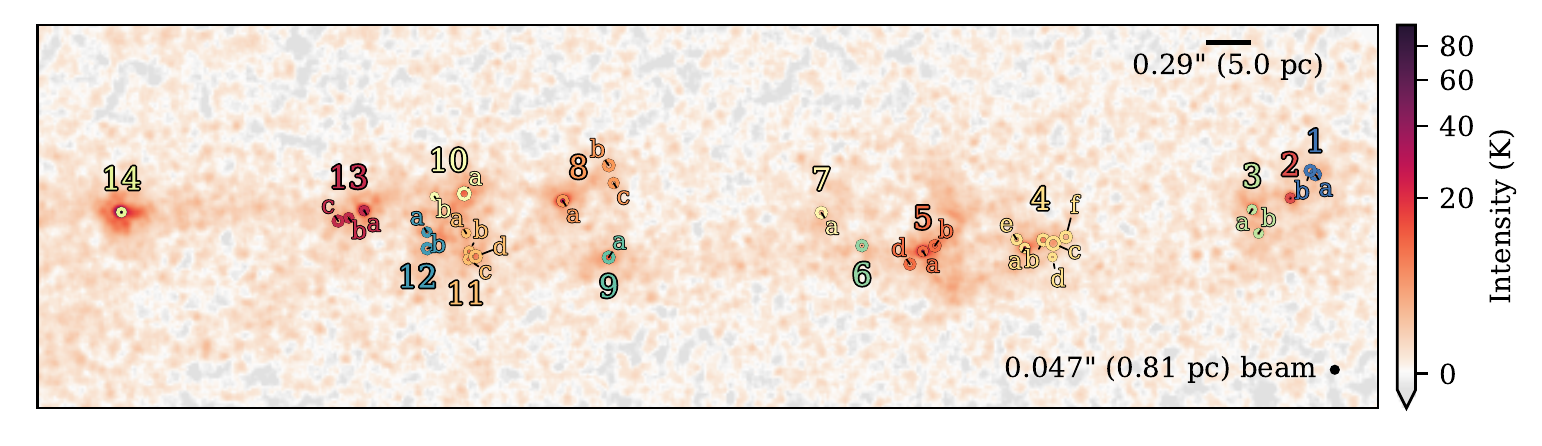}
    \caption{The continuum sizes of the SSCs over the high resolution combined 350 GHz continuum map. The map has been rotated counterclockwise by 42\D\ so that the SSC structure is horizontal. The colored circles show the deconvolved radii from the Gaussian fits to the radial profiles (Table \ref{tab:sscparams_radialprofile_deconv}). The cluster groups are labeled and colored following the nomenclature of \citet{leroy18}. Clusters that break apart in the higher resolution images are denoted with letters in order of decreasing brightness.}
    \label{fig:sscsizes}
\end{figure*}

As described in Section \ref{sec:obs4}, in this work we combine three configurations of 12-m data from ALMA. These data have a resolution of 0.8~pc, which allows us to spatially resolve the SSCs, and a maximum recoverable scale of 66~pc, allowing us to measure the flux of the SSCs and background emission. We, therefore, re-measure the cluster positions and sizes using the 0.8~pc resolution map (Figure \ref{fig:ngc253_cont} bottom left) as described below. 

\subsection{Identifying the Continuum Sources}
\label{ssec:findsources}

\begin{deluxetable*}{ccccccc}\tablecaption{Deconvolved SSC Parameters \label{tab:sscparams_radialprofile_deconv}}\tablehead{SSC \# & R.A. & Decl. & $I_{\mathrm{peak}}$ & $r_{\mathrm{deconv}}$ & Flux & log$_{10}$M$_{\mathrm{gas}}$ \\ & (J2000) & (J2000) & (K)  & (pc) & (mJy) & (log$_{10}$M$_\odot$)}
\startdata
1a & $0^\mathrm{h}47^\mathrm{m}32.801^\mathrm{s}$ & $-25^\circ17{}^\prime21.242{}^{\prime\prime}$ & $6.4\pm0.4$ & $0.58\pm0.04$ & $2.1\pm0.1$ & $4.3\pm0.1$ \\
1b & $0^\mathrm{h}47^\mathrm{m}32.801^\mathrm{s}$ & $-25^\circ17{}^\prime21.197{}^{\prime\prime}$ & $5.0\pm2.6$ & $0.60\pm0.18$ & $1.8\pm0.9$ & $4.2\pm0.3$ \\
2 & $0^\mathrm{h}47^\mathrm{m}32.819^\mathrm{s}$ & $-25^\circ17{}^\prime21.248{}^{\prime\prime}$ & $23.3\pm1.6$ & $0.44\pm0.02$ & $4.4\pm0.3$ & $4.6\pm0.1$ \\
3a & $0^\mathrm{h}47^\mathrm{m}32.839^\mathrm{s}$ & $-25^\circ17{}^\prime21.122{}^{\prime\prime}$ & $22.5\pm1.9$ & $0.43\pm0.03$ & $4.0\pm0.3$ & $4.6\pm0.1$ \\
3b & $0^\mathrm{h}47^\mathrm{m}32.845^\mathrm{s}$ & $-25^\circ17{}^\prime21.285{}^{\prime\prime}$ & $7.1\pm0.9$ & $0.43\pm0.04$ & $1.3\pm0.2$ & $4.1\pm0.1$ \\
4a & $0^\mathrm{h}47^\mathrm{m}32.945^\mathrm{s}$ & $-25^\circ17{}^\prime20.212{}^{\prime\prime}$ & $27.2\pm1.9$ & $0.48\pm0.03$ & $6.0\pm0.4$ & $4.8\pm0.1$ \\
4b & $0^\mathrm{h}47^\mathrm{m}32.934^\mathrm{s}$ & $-25^\circ17{}^\prime20.259{}^{\prime\prime}$ & $7.5\pm0.4$ & $0.63\pm0.05$ & $2.9\pm0.2$ & $4.4\pm0.1$ \\
4c & $0^\mathrm{h}47^\mathrm{m}32.932^\mathrm{s}$ & $-25^\circ17{}^\prime20.327{}^{\prime\prime}$ & $5.1\pm0.4$ & $0.74\pm0.10$ & $2.7\pm0.2$ & $4.4\pm0.1$ \\
4d & $0^\mathrm{h}47^\mathrm{m}32.937^\mathrm{s}$ & $-25^\circ17{}^\prime20.398{}^{\prime\prime}$ & $4.1\pm4.3$ & $0.38\pm0.16$ & $0.6\pm0.6$ & $3.7\pm0.5$ \\
4e & $0^\mathrm{h}47^\mathrm{m}32.945^\mathrm{s}$ & $-25^\circ17{}^\prime20.126{}^{\prime\prime}$ & $4.7\pm0.6$ & $0.54\pm0.09$ & $1.3\pm0.2$ & $4.1\pm0.1$ \\
4f & $0^\mathrm{h}47^\mathrm{m}32.924^\mathrm{s}$ & $-25^\circ17{}^\prime20.354{}^{\prime\prime}$ & $3.7\pm0.3$ & $0.62\pm0.09$ & $1.4\pm0.1$ & $4.1\pm0.1$ \\
5a & $0^\mathrm{h}47^\mathrm{m}32.987^\mathrm{s}$ & $-25^\circ17{}^\prime19.727{}^{\prime\prime}$ & $49.4\pm3.5$ & $0.47\pm0.03$ & $10.5\pm0.7$ & $5.0\pm0.1$ \\
5b & $0^\mathrm{h}47^\mathrm{m}32.980^\mathrm{s}$ & $-25^\circ17{}^\prime19.756{}^{\prime\prime}$ & $10.5\pm8.5$ & $0.59\pm0.47$ & $3.6\pm3.7$ & $4.5\pm0.6$ \\
5d & $0^\mathrm{h}47^\mathrm{m}32.997^\mathrm{s}$ & $-25^\circ17{}^\prime19.734{}^{\prime\prime}$ & $5.1\pm0.7$ & $0.57\pm0.10$ & $1.6\pm0.2$ & $4.1\pm0.1$ \\
6 & $0^\mathrm{h}47^\mathrm{m}33.010^\mathrm{s}$ & $-25^\circ17{}^\prime19.395{}^{\prime\prime}$ & $3.4\pm0.5$ & $0.58\pm0.10$ & $1.1\pm0.2$ & $3.6\pm0.1$ \\
7a & $0^\mathrm{h}47^\mathrm{m}33.014^\mathrm{s}$ & $-25^\circ17{}^\prime19.015{}^{\prime\prime}$ & $3.0\pm0.2$ & $0.59\pm0.06$ & $1.0\pm0.1$ & $4.0\pm0.1$ \\
8a & $0^\mathrm{h}47^\mathrm{m}33.114^\mathrm{s}$ & $-25^\circ17{}^\prime17.675{}^{\prime\prime}$ & $27.7\pm1.8$ & $0.57\pm0.04$ & $8.7\pm0.6$ & $4.9\pm0.1$ \\
8b & $0^\mathrm{h}47^\mathrm{m}33.083^\mathrm{s}$ & $-25^\circ17{}^\prime17.707{}^{\prime\prime}$ & $3.9\pm0.3$ & $0.61\pm0.07$ & $1.4\pm0.1$ & $4.1\pm0.1$ \\
8c & $0^\mathrm{h}47^\mathrm{m}33.087^\mathrm{s}$ & $-25^\circ17{}^\prime17.828{}^{\prime\prime}$ & $2.8\pm0.4$ & $0.51\pm0.09$ & $0.7\pm0.1$ & $3.8\pm0.1$ \\
9a & $0^\mathrm{h}47^\mathrm{m}33.116^\mathrm{s}$ & $-25^\circ17{}^\prime18.211{}^{\prime\prime}$ & $13.0\pm0.5$ & $0.61\pm0.03$ & $4.7\pm0.2$ & $4.5\pm0.1$ \\
10a & $0^\mathrm{h}47^\mathrm{m}33.151^\mathrm{s}$ & $-25^\circ17{}^\prime17.149{}^{\prime\prime}$ & $9.1\pm0.8$ & $0.71\pm0.07$ & $4.4\pm0.4$ & $4.6\pm0.1$ \\
10b & $0^\mathrm{h}47^\mathrm{m}33.164^\mathrm{s}$ & $-25^\circ17{}^\prime17.018{}^{\prime\prime}$ & $5.3\pm2.4$ & $0.35\pm0.10$ & $0.6\pm0.3$ & $3.7\pm0.3$ \\
11a & $0^\mathrm{h}47^\mathrm{m}33.165^\mathrm{s}$ & $-25^\circ17{}^\prime17.376{}^{\prime\prime}$ & $13.1\pm1.7$ & $0.39\pm0.03$ & $1.9\pm0.2$ & $4.0\pm0.1$ \\
11b & $0^\mathrm{h}47^\mathrm{m}33.170^\mathrm{s}$ & $-25^\circ17{}^\prime17.491{}^{\prime\prime}$ & $9.0\pm1.2$ & $0.50\pm0.06$ & $2.2\pm0.3$ & $4.1\pm0.1$ \\
11c & $0^\mathrm{h}47^\mathrm{m}33.174^\mathrm{s}$ & $-25^\circ17{}^\prime17.530{}^{\prime\prime}$ & $8.5\pm4.0$ & $0.48\pm0.22$ & $1.9\pm1.0$ & $4.0\pm0.3$ \\
11d & $0^\mathrm{h}47^\mathrm{m}33.170^\mathrm{s}$ & $-25^\circ17{}^\prime17.550{}^{\prime\prime}$ & $7.9\pm0.6$ & $0.68\pm0.06$ & $3.5\pm0.3$ & $4.3\pm0.1$ \\
12a & $0^\mathrm{h}47^\mathrm{m}33.180^\mathrm{s}$ & $-25^\circ17{}^\prime17.177{}^{\prime\prime}$ & $5.4\pm0.5$ & $0.48\pm0.05$ & $1.2\pm0.1$ & $3.5\pm0.1$ \\
12b & $0^\mathrm{h}47^\mathrm{m}33.186^\mathrm{s}$ & $-25^\circ17{}^\prime17.268{}^{\prime\prime}$ & $3.4\pm0.5$ & $0.59\pm0.14$ & $1.1\pm0.2$ & $3.5\pm0.2$ \\
13a & $0^\mathrm{h}47^\mathrm{m}33.198^\mathrm{s}$ & $-25^\circ17{}^\prime16.750{}^{\prime\prime}$ & $52.0\pm3.6$ & $0.44\pm0.03$ & $9.5\pm0.7$ & $5.0\pm0.1$ \\
13b & $0^\mathrm{h}47^\mathrm{m}33.207^\mathrm{s}$ & $-25^\circ17{}^\prime16.712{}^{\prime\prime}$ & $11.8\pm0.8$ & $0.47\pm0.04$ & $2.5\pm0.2$ & $4.4\pm0.1$ \\
13c & $0^\mathrm{h}47^\mathrm{m}33.212^\mathrm{s}$ & $-25^\circ17{}^\prime16.678{}^{\prime\prime}$ & $4.8\pm1.4$ & $0.57\pm0.17$ & $1.5\pm0.5$ & $4.2\pm0.2$ \\
14 & $0^\mathrm{h}47^\mathrm{m}33.297^\mathrm{s}$ & $-25^\circ17{}^\prime15.560{}^{\prime\prime}$ & $90.3\pm6.7$ & $0.45\pm0.03$ & $17.6\pm1.3$ & $5.3\pm0.1$ \\
\enddata
\tablecomments{See Section \ref{ssec:clustersizes} for details.}
\end{deluxetable*}

From the 0.48~pc resolution continuum data, many of the candidate SSCs identified by \citet{leroy18} at 2~pc resolution break apart into smaller structures (Figure \ref{fig:ngc253_cont} bottom right; \citealt{levy21}). We find more than three dozen dust clumps by-eye in the 0.48~pc resolution dust image. The SSCs identified by \citet{leroy18} remain the largest and brightest structures. We, therefore, follow the SSC nomenclature of \citet{leroy18}, but add letters to sources that break apart in order of decreasing brightness, as described by \citet{levy21}. From there, we match the locations of the dust clumps from the 0.48~pc resolution image to the 0.81~pc resolution map (Figure \ref{fig:ngc253_cont} bottom left). The 0.81~pc resolution map combines three configurations of ALMA data and, therefore, better recovers the extended emission than the 0.48~pc resolution map. From the 0.81~pc map, we identify 33 clumps of dust emission, which are listed in Table \ref{tab:sscparams_radialprofile_deconv} and are shown in Figure \ref{fig:sscsizes}. Some of the very small sources previously identified in the 0.48~pc resolution image are no longer visible in the 0.81~pc resolution image due to the slightly lower resolution and extended emission (i.e., SSCs 1c, 3c, 5c, 7b, 9b). For SSCs 7a and 9a, we retain the "a" lettering to indicate that these clusters do break apart in the 0.48~pc resolution image, though these smaller clusters are not visible in the 0.i81~pc resolution map. Clusters without letters (SSCs 2, 6, 14) do not break apart even in the 0.48~pc resolution dust continuum map.

\subsection{Cluster Positions}
\label{sssec:gaussianfits}

To measure the precise centers of the SSCs, we fit the continuum intensity map with a 2D rotated elliptical Gaussian function. We include a constant background component since the clusters are embedded within more extended emission. Before fitting, we mask out other sources in the images. This is especially important for clusters in crowded fields. We automatically mask out primary clusters based on their half-flux radii ($r_{\rm half-flux}$) measured from the high resolution data \citep{levy21}, removing pixels within 2$\times r_{\rm half-flux}$ from the cluster centers. We remove contaminating subclusters by-eye and remove pixels within 1.5$\times$ the beam half-width-half-maximum (HWHM) from the cluster centers. 

For some of the weaker clusters, the 2D elliptical Gaussian fit does not converge. In these cases, we determine the center position based on the brightest pixel in the dust continuum near the center of the SSC. The best-fitting SSC centers are listed in Table \ref{tab:sscparams_radialprofile_deconv} and shown in Figure \ref{fig:sscsizes}. We estimate that the positional accuracy of this image is $\approx2.5$~milliarcseconds (0.04~pc)\footnote{See Section 10.5.2 of the the most recent version of the ALMA Technical Handbook: \url{https://almascience.nrao.edu/documents-and-tools/cycle9/alma-technical-handbook}. Since the SSCs in the image have SNR $\gtrsim20$, the positional accuracy is $\approx 5$\% of the synthesized beam. We note that the actual positional accuracy may be a factor of $\approx2$ poorer than this value due to degradation of atmospheric phase stability in the most extended configurations.}.

\subsection{Radial Profiles}
\label{sssec:radialprofiles}

\begin{figure}
    \centering
      \includegraphics[width=\columnwidth]{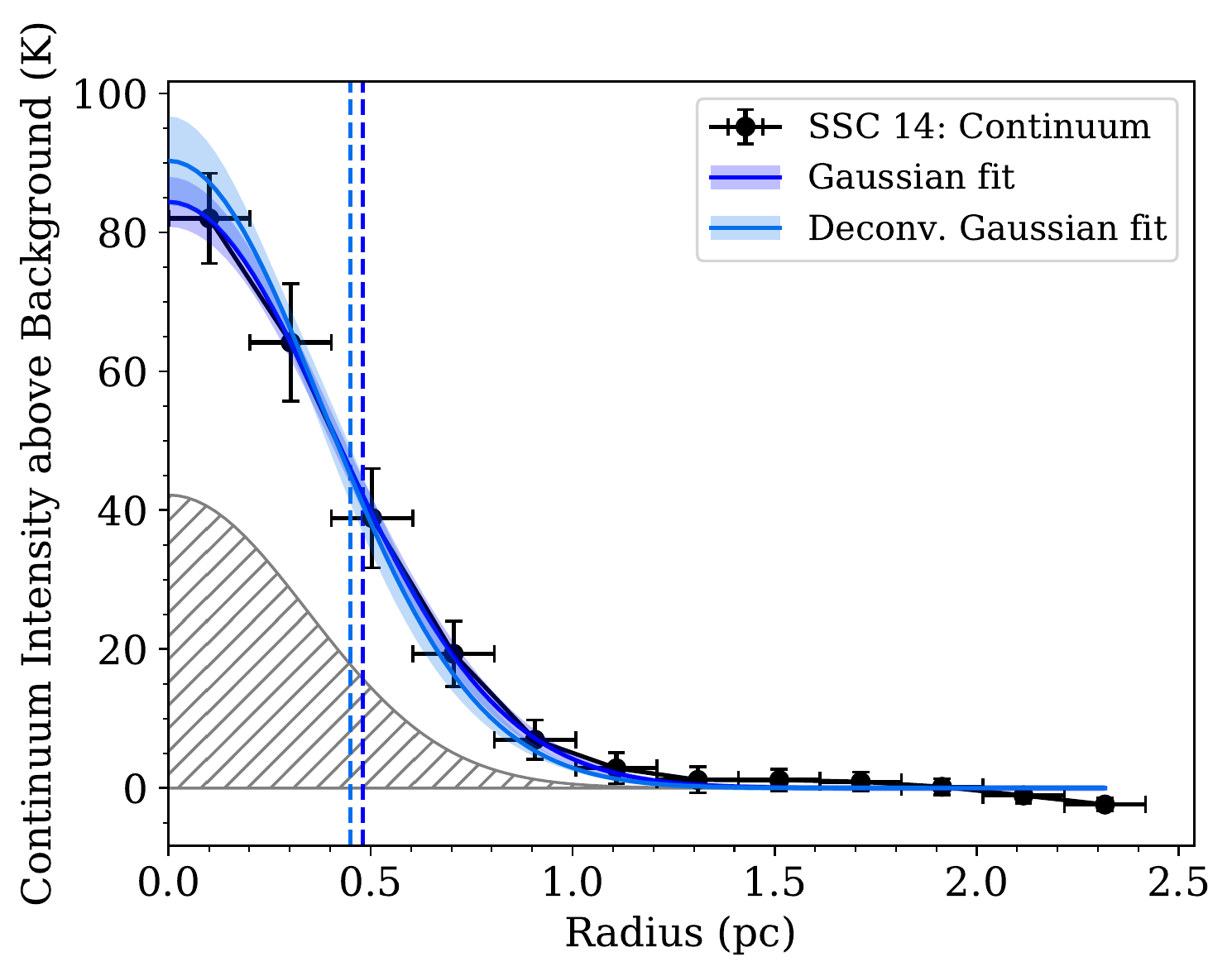}
    \caption{The extracted radial profile of SSC~14 (black data points). The gray hatched region shows the Gaussian beam with arbitrary vertical scaling. The dark blue curve shows the Gaussian fit to the radial profile. The dark blue shaded region around this curve reflects the standard deviation of 500 Monte Carlo realizations varying the data points within the errorbars. The vertical dark blue dashed line shows the HWHM radius. The light blue curve shows the beam-deconvolved (i.e., intrinsic) Gaussian radial profile, where the shaded region shows the propagated uncertainties on the fit and the vertical light blue dashed line shows the deconvolved HWHM radius. }
    \label{fig:radialprofiles}
\end{figure}

We construct radial profiles for each cluster. Before extracting the radial profiles, we mask the images in the same way as for the 2D Gaussian fitting. We use concentric circular\footnote{From the 2D Gaussian fitting, the median axis ratio of the rotated elliptical Gaussian fits is 0.9, so clusters only deviate from circular by 10\%. This means that extracting the radial profiles in circular annuli (rather than in ellipses) will not introduce major systematic errors.} annuli centered on the R.A. and Decl. from the 2D Gaussian fitting (Table \ref{tab:sscparams_radialprofile_deconv}). The width of the annuli is the beam HWHM (0.024\arcsec\ = 0.40~pc) and the last ring has a radius of 3$\times$ the beam FWHM ($3\times0.0475$\arcsec\ = 2.4~pc). We measure the median intensity in each annulus, which is shown in Figure \ref{fig:radialprofiles} for SSC~14; the uncertainty is the standard error in each annulus. 

We model the cluster radial profiles using a Gaussian of the form
\begin{equation}
    \label{eq:gauss}
    {\rm SB}(r) = ae^{-\frac{r^2}{2\sigma^2}}+c
\end{equation}
to model the radial surface brightness (SB) profile of the clusters. We also include a constant background component ($c$) since the clusters sit in an extended background of dust emission. An example of this model fit to the cluster radial profile is shown for SSC~14 in Figure \ref{fig:radialprofiles}, where the fitted background level ($c$) has been removed. We determine the uncertainty on the Gaussian fit using a Monte Carlo simulation where we randomly vary the data points within the errorbars. The dark blue shaded region in Figure \ref{fig:radialprofiles} reflects the standard deviation of 500 trials. 

In addition to a Gaussian function, we also modeled the radial profiles using \cite{king62} and \citet{plummer1911} profiles. These provide equally good fits to the cluster radial profiles. We proceed using the Gaussian fits to the radial profiles.

\subsection{Deconvolved Sizes, Fluxes, and Gas Masses}
\label{ssec:gasmass}

\begin{figure}
    \centering
    \includegraphics[width=\columnwidth]{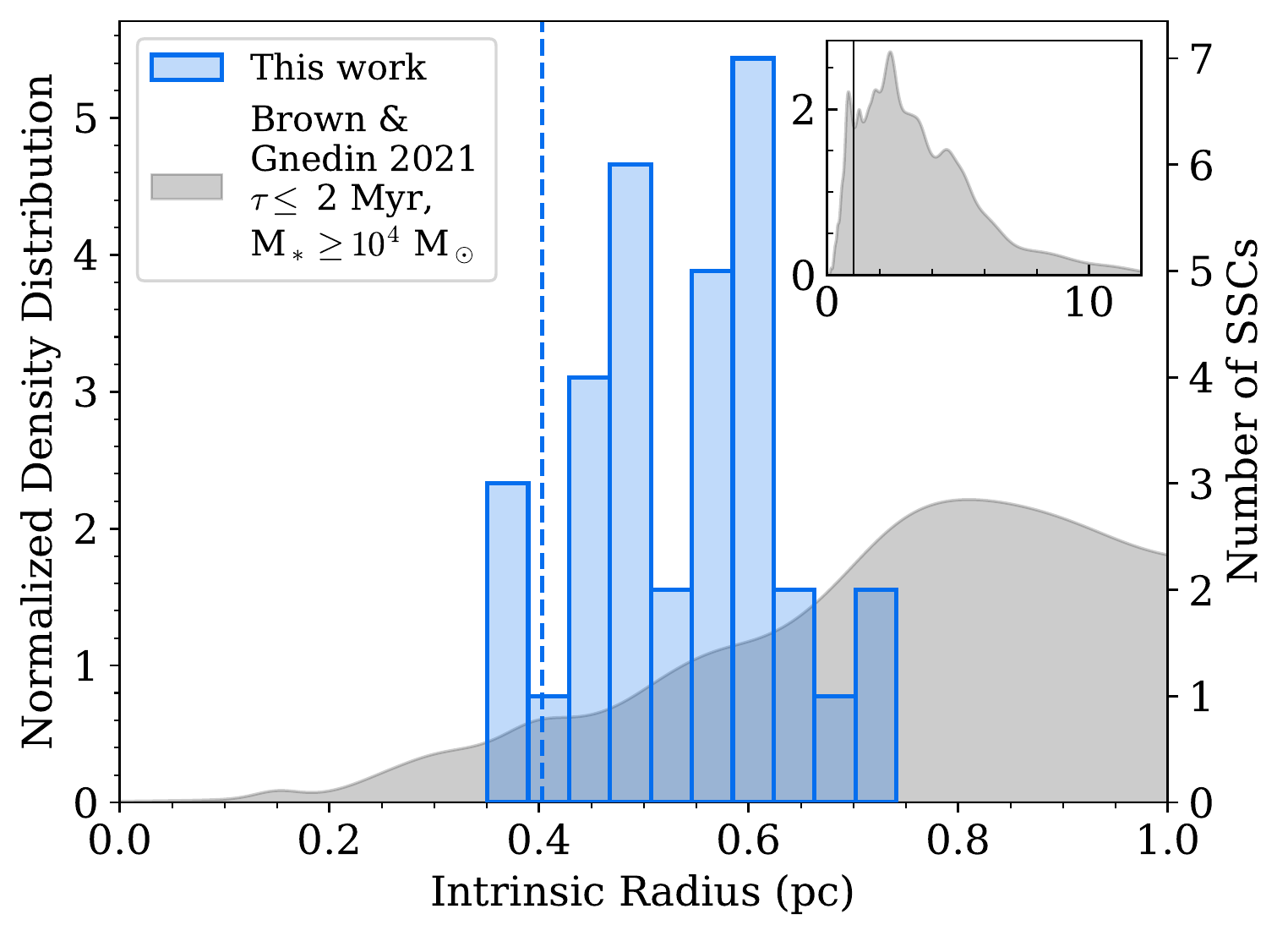}
    \caption{The light blue histogram shows the distribution of the intrinsic SSC radii, measured from the beam-deconvolved Gaussian fit to the radial profile. The left y-axis shows the histogram normalized to unit area, whereas the right y-axis shows the number of SSCs in each bin. The vertical blue dashed line shows the beam HWHM. The gray KDE shows the distribution of star cluster intrinsic effective radii measured in LEGUS galaxies normalized to unit area over the plotted radius range \citep{brown21}. We select clusters with ages $<2$~Myr and stellar masses $>10^4$~\msun\ to be most comparable to our sample of very young, massive clusters. The inset in the upper right shows the full distribution of the LEGUS radii with the same selection criteria, which peaks around radii of $2-3$~pc. The vertical black line marks 1~pc, the radial extent shown in the main panel.}
    \label{fig:intrinsicradii}
\end{figure}

We deconvolve the beam size from the fitted Gaussian profile by removing the (Gaussian) beam HWHM in quadrature. We produce deconvolved Gaussian radial profiles using the deconvolved radii and conserving the flux. An example of the deconvolved Gaussian radial profile is shown for SSC~14 in Figure \ref{fig:radialprofiles}. We report the deconvolved cluster radii (r$_{\rm deconv}$), peak intensities ($I_{\rm peak}$), and fluxes in Table \ref{tab:sscparams_radialprofile_deconv}. For $I_{\rm peak}$, we subtract the background level so this value reflects the peak intensity of the cluster above the background of surrounding material. We show the distribution of intrinsic cluster radii in Figure \ref{fig:intrinsicradii}. Our intrinsic radii cover a narrow range of radii from $\approx0.25-0.70$~pc. We also show the deconvolved radii for each of the clusters in Figure \ref{fig:sscsizes} over the continuum image.

We compare our measured cluster radii to those measured by \citet{brown21} from the Legacy Extragalactic UV Survey (LEGUS) survey \citep{calzetti15}. These clusters are identified in 31 nearby galaxies in five bands from the near-UV to near-IR. \citet{brown21} measure the intrinsic stellar half-light (effective) radii of the young star clusters in these galaxies from the "white light" (i.e. combined 5-filter) images. From their cluster catalog\footnote{\url{https://www.gillenbrown.com/LEGUS-sizes}}, we select clusters with reliable radius and mass measurements, ages $\leq2$~Myr, and stellar masses $\geq10^4$~\msun; see \citet{brown21} for the definitions of these quantities. We show a kernel density estimator (KDE) of the LEGUS cluster radii in gray in Figure \ref{fig:intrinsicradii}, where the inset shows the KDE over their full radius range. The LEGUS clusters tend to be larger than the SSCs studied here. The peak in the LEGUS radius distribution for clusters with the above selection criteria is between $2-3$~pc.

It is perhaps not unexpected that the clusters identified by \citet{brown21} are larger. The radii we measure for the clusters in NGC\,253 correspond to the size of the dust (and molecular gas) envelopes, whereas the radii measured for the LEGUS clusters come from the stellar light. The clusters in NGC\,253 are still in the process of forming \citep{leroy18,rico-villas20,mills21} and are, therefore, still very compact. Since the LEGUS clusters are typically older than most of the SSCs in this work and are no longer (deeply) embedded in their natal molecular clouds, it is possible that the stellar light would extend to larger radii than the compact dust emission from the SSCs. Simulations of star cluster evolution show an increase in the radius with age due primarily to mass loss from stellar winds of young massive stars \citep[e.g.,][and references therein]{portegieszwart10}. 

We estimate the gas masses of the clusters based on their 350~GHz dust continuum emission, following \citet[][see their Section 4.3.3 for more details]{leroy18}. Assuming a fiducial dust temperature of $T_{\rm dust} = 130$~K, we convert the deconvolved peak intensity at 350~GHz to a dust optical depth via
\begin{equation}
    \label{eq:taudust}
    \tau_{\rm 350~GHz} = -\ln\left[1-\frac{I_{\rm 350~GHz}}{B_\nu(T_{\rm dust})}\right]
\end{equation}
where $I_{\rm 350~GHz}$ is the peak intensity in Table \ref{tab:sscparams_radialprofile_deconv} and $B_\nu(T_{\rm dust})$ is the Planck function evaluated at 350~GHz for our adopted value of $T_{\rm dust}$. Though we do not yet have measurements of the dust temperatures towards these SSCs, the peak intensities of $\gtrsim25$~K we measure towards some of these clusters (Table \ref{tab:sscparams_radialprofile_deconv}) supports a high value of $T_{\rm dust}$. We convert the dust optical depth to a gas surface mass density where
\begin{equation}
    \label{eq:tau2Sigma}
    \Sigma_{\rm gas} = \frac{\tau_{\rm 350~GHz}}{\rm DGR\ \kappa_{\rm 350~GHz}}
\end{equation}
where the dust-to-gas ratio (DGR) is assumed to be 1/100. {\change The central 300~pc of NGC\,253 is known to have a somewhat super-solar gas-phase metallicity ($Z=2.2Z_\odot$; \citealt{galliano08,davis13}).} $\kappa_{\rm 350~GHz}$ is the mass absorption coefficient; we assume a value of 1.9~cm$^2$~g\per, but this value is uncertain by a factor of $\sim$2 \citep{ossenkopf94,leroy18}. Finally, we convert the gas mass surface density to a gas mass by multiplying by the area of the cluster: 
\begin{equation}
    {\rm M}_{\rm gas} = \Sigma_{\rm gas}{\rm A}_{\rm deconv}
\end{equation}
where
\begin{equation}
    {\rm A}_{\rm deconv} \equiv \frac{4\pi {\rm r}_{\rm deconv}^{2}}{2\ln2}
\end{equation}
is the area of a 2D Gaussian whose HWHM is equal to the beam-deconvolved radius measurement ($r_{\rm deconv}$; Table see \ref{tab:sscparams_radialprofile_deconv}).

\subsection{Flux Density and Gas Mass Distributions}
\label{ssec:cmf}

\begin{figure*}
    \centering
    \includegraphics[width=0.49\textwidth]{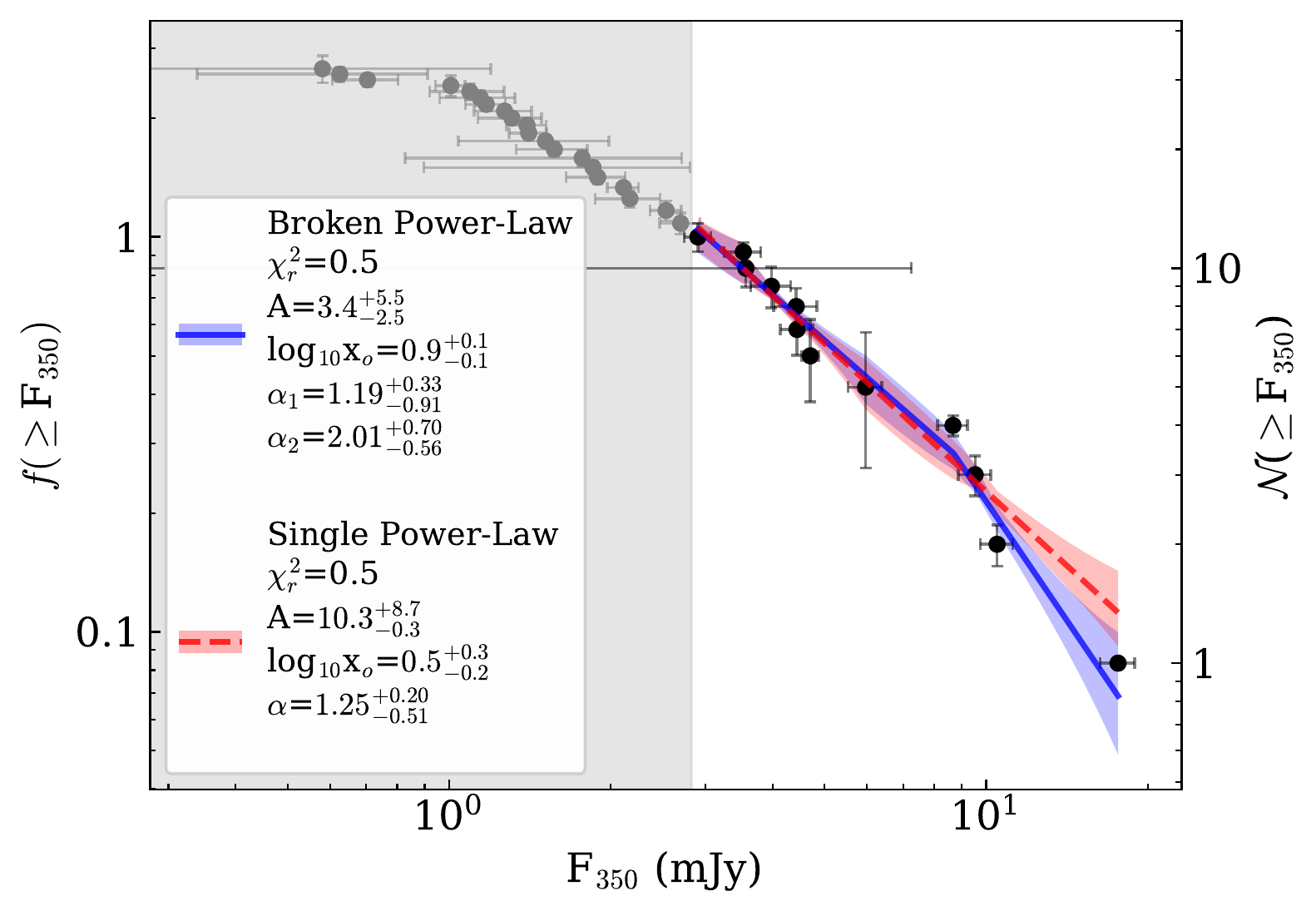}
    \includegraphics[width=0.49\textwidth]{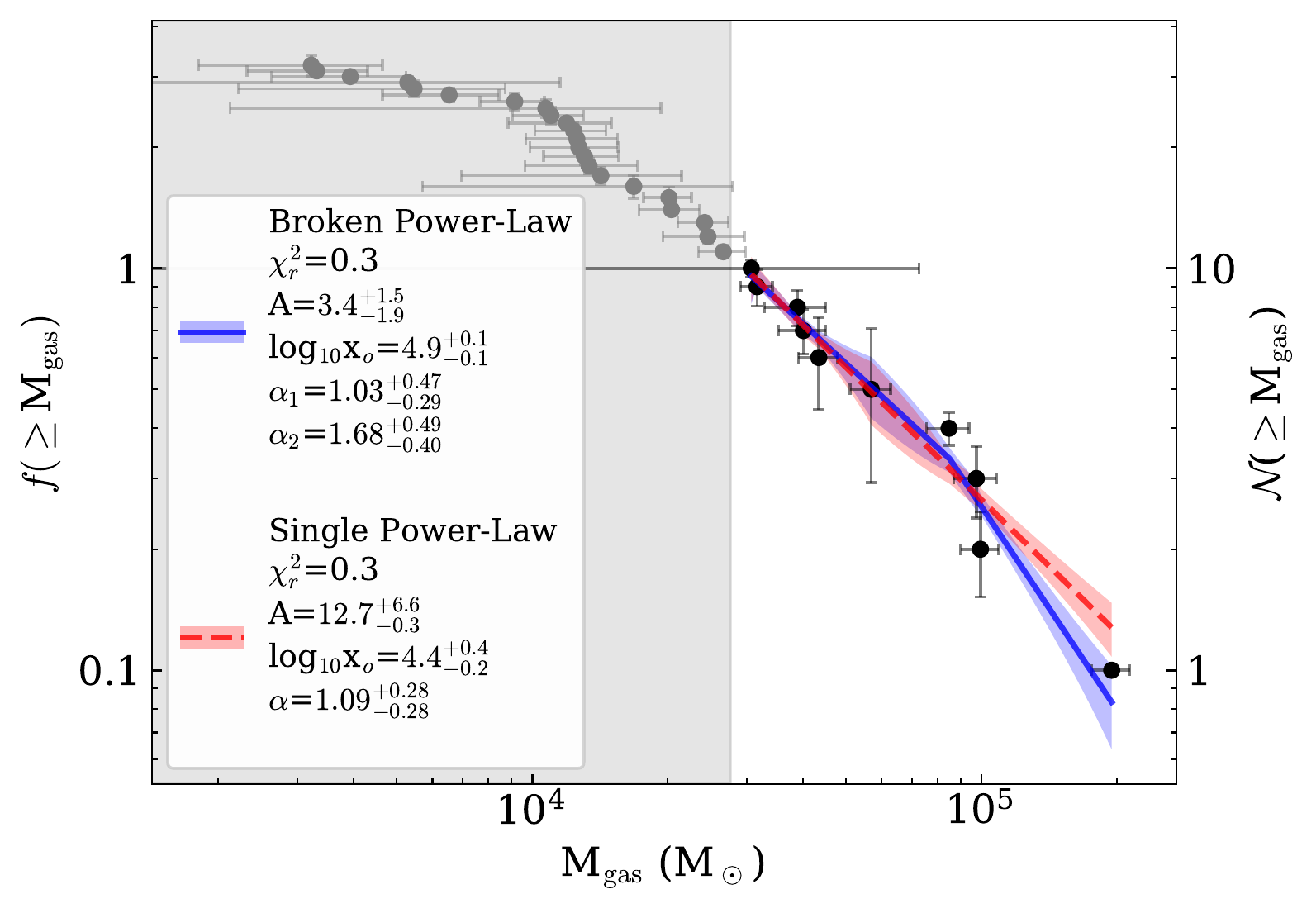}
    \caption{The cluster flux density ({\em left}) and molecular gas mass ({\em right}) functions. The right y-axis shows the number of clusters whereas the left y-axis shows the fraction of clusters with reliable measurements. The gray shaded regions and gray points show our estimate of confusion; measurements in this region are likely underestimated beyond the uncertainties. We fit the distributions with a broken power law (blue solid lines) and a single power law (red dashed lines).  See Section \ref{ssec:cmf} for details on the uncertainty calculations and the fitting.}
    \label{fig:CMF}
\end{figure*}

From the fluxes (Table \ref{tab:sscparams_radialprofile_deconv}), we construct the cluster flux density function, which is shown in Figure \ref{fig:CMF} (left). This is essentially a cumulative distribution function (CDF), where the ordinate counts the number of clusters with flux densities larger than the value on the abscissa. The horizontal error bars come from the uncertainties on the measured fluxes. To determine the vertical error bars, we use a Monte Carlo calculation, allowing the measured fluxes to vary uniformly within their uncertainties. This can change the ordering of the flux densities and hence the CDF. We perform 100 trials of the Monte Carlo, and report the standard deviation of the CDF for each point over those trials as the vertical error bars. We repeat this procedure with the gas masses which are shown in Figure \ref{fig:CMF} (right).

Both distributions appear to follow a broken power law, with a break at the very low flux/mass end; however, this break {\change is very likely} due to incompleteness at the low flux end of the distribution \citep[e.g.,][]{emig20}. The SSC identification is done by-eye based on the 0.5~pc resolution continuum image (Figure \ref{fig:ngc253_cont} bottom right). The radial profiles are measured from the 0.8~pc resolution image (Figure \ref{fig:ngc253_cont} bottom left) at the locations of the SSCs identified from the high resolution image, as long as they are still apparent in the lower resolution image. The major uncertainty matching the SSCs between these images are from the "speckles" seen in the 0.8~pc resolution continuum image (Figure \ref{fig:ngc253_cont} bottom left). These speckles arise in the imaging by modeling the extended emission as Gaussians matched to the beam size, and they can resemble small, compact clusters. We, therefore, use these speckles as our test particles to evaluate the completeness and confusion of our SSC flux and gas mass functions. We choose a speckle in the image and measure its radial profile, beam-deconvolved size, and peak intensity as described in Sections \ref{sssec:radialprofiles} and \ref{ssec:gasmass}. We find that a typical speckle has a flux of $\approx3.5$~mJy and hence an inferred gas mass of $\approx5\times10^{4.5}$~\msun. We show these values as the gray shaded regions in Figure \ref{fig:CMF}. Clusters with fluxes or masses near or below this limit are likely more uncertain than represented by the error bars, and we may be missing SSCs in this flux and mass regime.

For values above our completeness threshold, we fit the cluster flux density and gas mass functions using a broken power law of the form
\begin{equation}
    \label{eq:bpl}
    f(x) = \begin{cases}
        A\left(\frac{x}{x_0}\right)^{-\alpha_1} \text{; $x < x_0$,}
        \\
         A\left(\frac{x}{x_0}\right)^{-\alpha_2} \text{; $x > x_0$,}
        \end{cases}
\end{equation}
which is implemented using \texttt{BrokenPowerLaw1D} from Astropy. As a comparison, we also fit the cluster flux density function with a single power law of the form
\begin{equation}
    \label{eq:pl}
    f(x) = A\left(\frac{x}{x_0}\right)^{-\alpha}
\end{equation}
which is implemented using \texttt{PowerLaw1D} from Astropy.

We obtain the uncertainties on the fitted parameters using the same Monte Carlo approach described above. As the flux density or mass points are allowed to vary within their uncertainties, we re-fit Equation \ref{eq:bpl} at each iteration. The uncertainties listed in Figure \ref{fig:CMF} reflect the 16$^{\rm th}$ and 84$^{\rm th}$ percentiles (i.e., the inner 68\%) of the parameter distributions after the Monte Carlo. The same strategy is used to obtain the uncertainties on the model curves, shown as the colored shaded regions in Figure \ref{fig:CMF}.

When the completeness is accounted for, both the broken and single power law fits are able to reproduce the flux and gas mass distributions equally well ($\chi_r^2=0.3-0.5$; Figure \ref{fig:CMF}). We find a single power law slope of 1.25 (1.09) for the flux (gas mass) functions. When we fit a broken power law, we measure a slope of 2.01 (1.58) at the high flux (gas mass) end.

Previous literature studies typically investigate the cluster {\em stellar} mass function. Recently, \citet{mok19,mok20} studied the cluster stellar mass function of a sample of star-forming galaxies. The young clusters included in their studies are older on average and span a wider range of stellar masses than the SSCs in NGC\,253 \citep{leroy18,mills21}. \citet{mok19} found no evidence of a high-mass cut-off of the cluster stellar mass functions {\change though other studies have found evidence of a high-mass cut off \citep[e.g.,][]{gieles06,whitmore10,adamo15,adamo17,hollyhead16,johnson17,messa18a,messa18b}.}

Although our measurements are of the gas mass (not the stellar mass), we also do not see strong evidence for a high-mass cut-off up to $2\times10^5$~\msun, which would be indicated by a more apparent break in the gas mass distribution (Figure \ref{fig:CMF} right). {\change While our two highest mass clusters may hint at a break around $10^{4.9}$~\msun, statistically the single and broken power law fits are equally good (as indicated by $\chi_r^2$). This break (or truncation) mass is similar to what has been measured in other nearby galaxies \citep[see e.g., discussion and references in][]{messa18a}.} 

\citet{mok20} found that the mass functions are well fit by single power laws (Equation \ref{eq:pl}) with slopes $2.0\pm0.3$, in agreement with previous theoretical and observational studies \citep[e.g.,][]{zhang99,fall12,chandar17,krumholz19}. When the clusters are fit in age bins, the youngest clusters have somewhat shallower power law slopes, with an average of $1.7\pm0.5$, though this is still consistent with the full sample within the uncertainties \citep{mok19}\footnote{We note that the galaxy samples used by \citet{mok19,mok20} overlap but are not the same.}. {\change \citet{messa18a} studied the cluster mass function in M51. Although they preferred a Schechter function, they performed single power law fits as well. From their fits using a minimum mass of $10^4$~\msun\ (similar to our completeness limit), \citet{messa18a} found a single power law slope of $2.67\pm0.03$, steeper than found by \citet{mok20}. The power law slope is reduced to $2$ (within the uncertainties) if the power law is truncated and if the minimum mass is reduced (see their Table 8). }

{\change The average slopes for the youngest clusters found by both \citet{mok20} and \citet{messa18a} are steeper than, but close to, the slope we measure for the gas mass distribution in NGC\,253.  We caution, however, that we are measuring the gas mass whereas \citet{mok19,mok20} and \citet{messa18a} measured the stellar mass of the clusters.} Both \citet{leroy18} and \citet{mills21} found a median M$_{\rm gas}$/M$_*\approx1$ for the clusters, but with appreciable scatter (average $\approx 2$, standard deviation $\approx2.5$) using different tracers of the stellar and gas masses. This scatter from cluster to cluster means that the slope of the stellar function function in NGC\,253 may be different than measured for the gas mass.

\section{The Morpho-Kinematic Architecture of the SSCs}
\label{sec:morphokin}

The quasi-linear arrangement of the SSCs in the center of NGC\,253 is striking (e.g., Figures \ref{fig:ngc253_cont} and \ref{fig:sscsizes}). In projection, this structure measures $\sim155{\rm~pc}\times15{\rm~pc}$ in diameter with a major axis position angle (PA) of $\approx48^\circ$ east-of-north. This axis ratio of $\sim10$ may suggest that the structure is intrinsically very thin and/or that we are seeing this structure at a high inclination. The galactic disk of NGC\,253 is nearly edge-on, with an inclination of $\approx 78^\circ$  and a PA of $\approx50^\circ$ \citep[e.g.,][]{pence80,westmoquette11,krieger19}. On $\sim$~kpc scales, the bar also has an inclination of $\approx78^\circ$ but has a major-axis PA of $68^\circ$ \citep{scoville85,sorai00}. The quasi-linear arrangement of SSCs has approximately the same PA as the galaxy disk and is offset from the PA of the bar.

Using CO observations at 35~pc resolution, \citet{leroy15a} measure the geometry of the GMC structures in which these SSCs are embedded. They build 3D models of the GMC geometry as a disk, a linear bar-like arrangement, and a hybrid model. They find that the hybrid model provides the best fit to the data, where the inner $\sim100-150$~pc (diameter) is more disk-like and regions beyond this extending out to $\sim850-1400$~pc (diameter) have a more linear structure. They find that the maximum vertical thickness of the GMC structure is $<100$~pc for the molecular gas traced by CO and $<55$~pc for the denser molecular gas. Our measurement of the minor axis width of the SSC structure sets a maximum vertical extent of $<15$~pc, similar to the vertical extent of the MW CMZ \citep[e.g.,][]{molinari11,kruijssen15,shin17,henshaw22}.

\begin{figure*}
    \centering
    \gridline{\fig{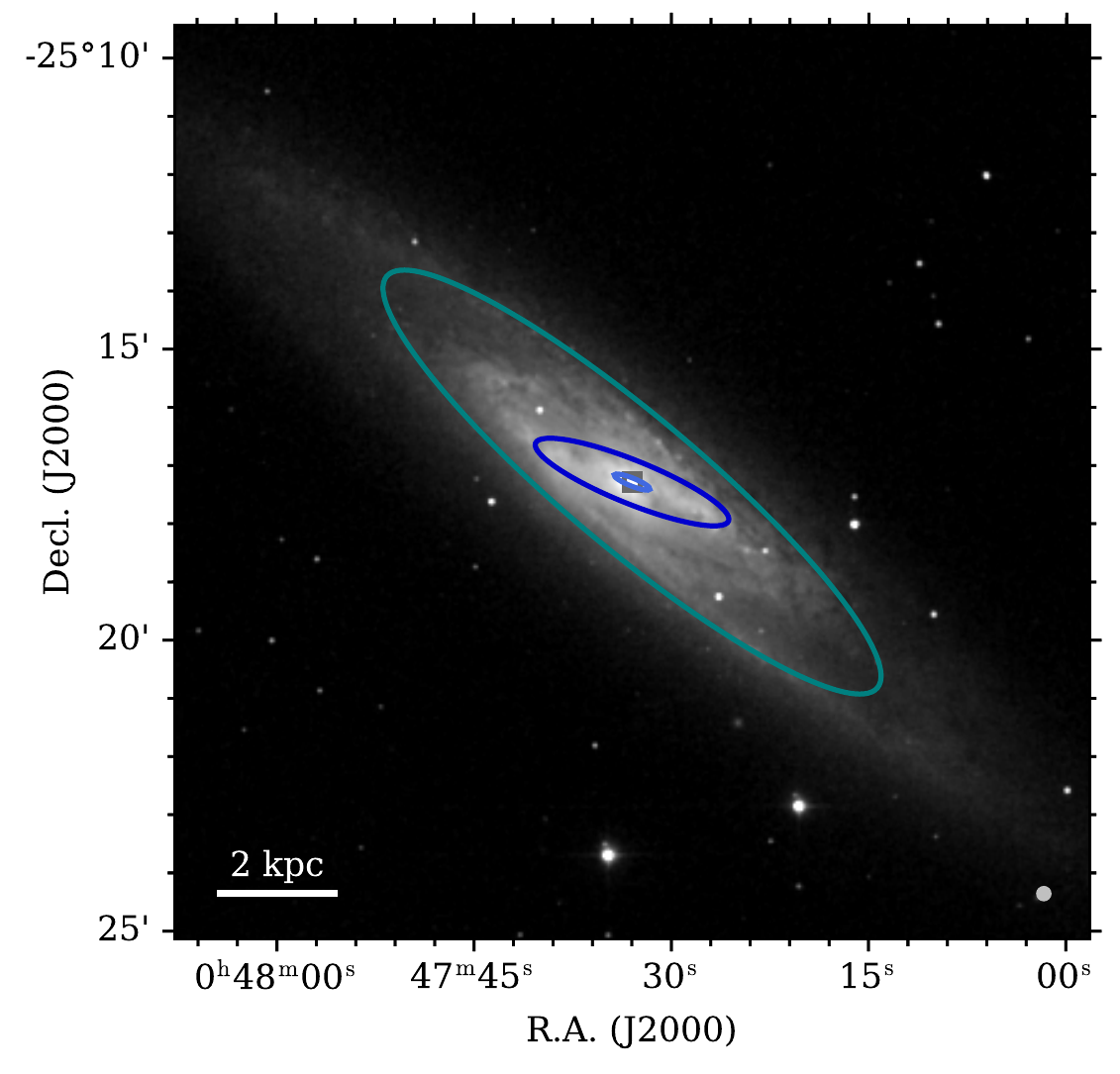}{0.49\textwidth}{}
    \fig{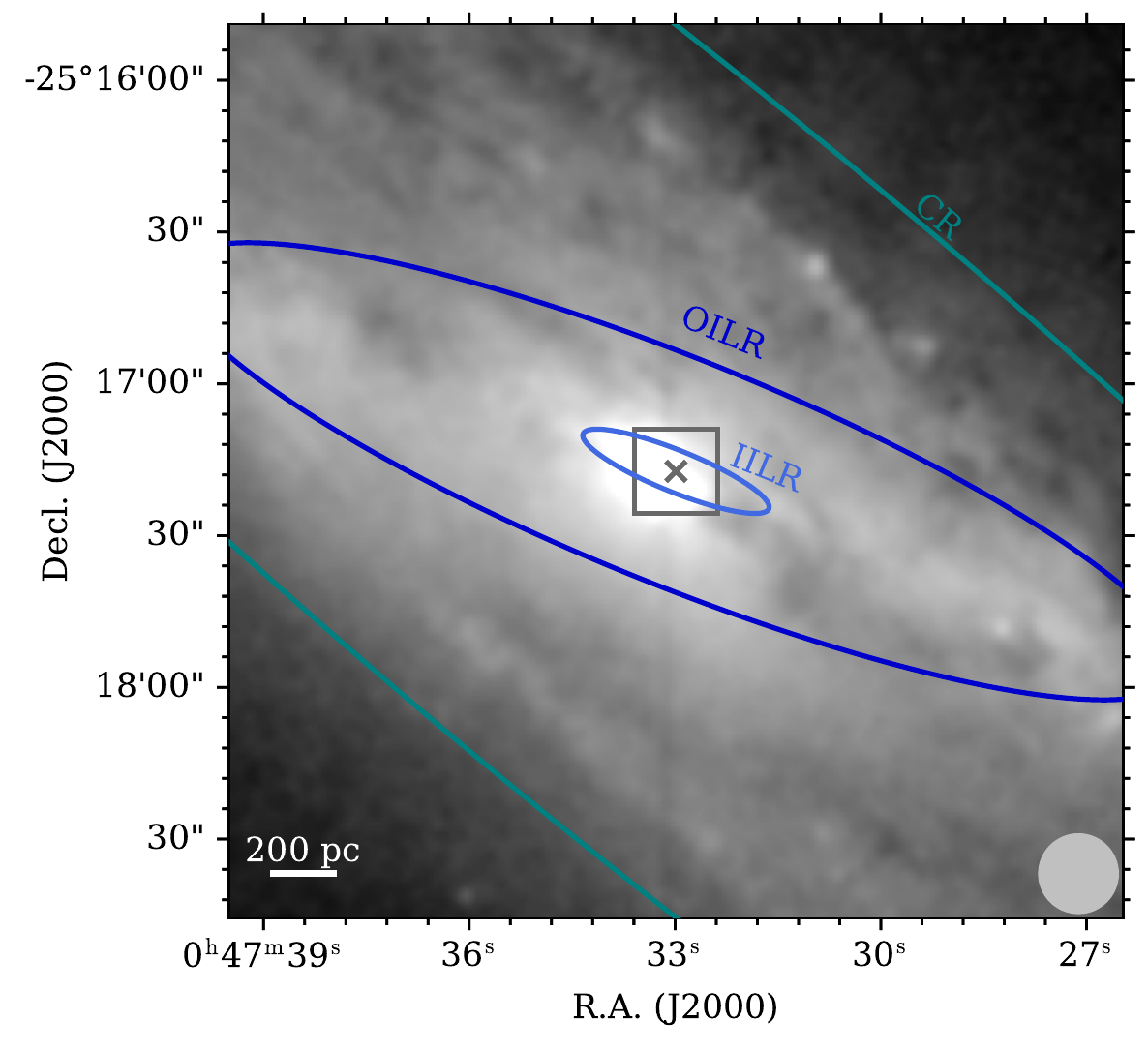}{0.49\textwidth}{}}
    \gridline{\fig{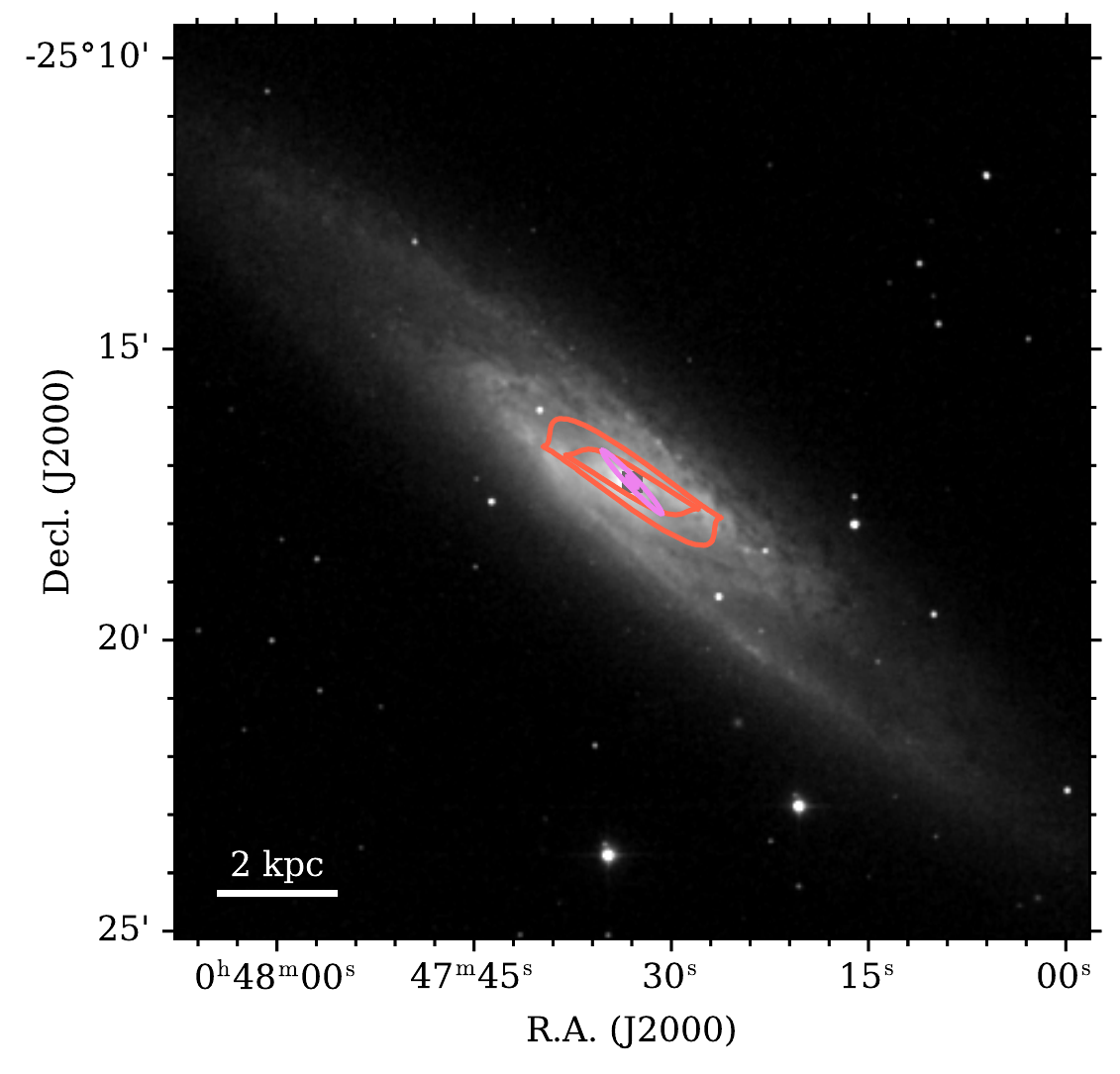}{0.49\textwidth}{}
    \fig{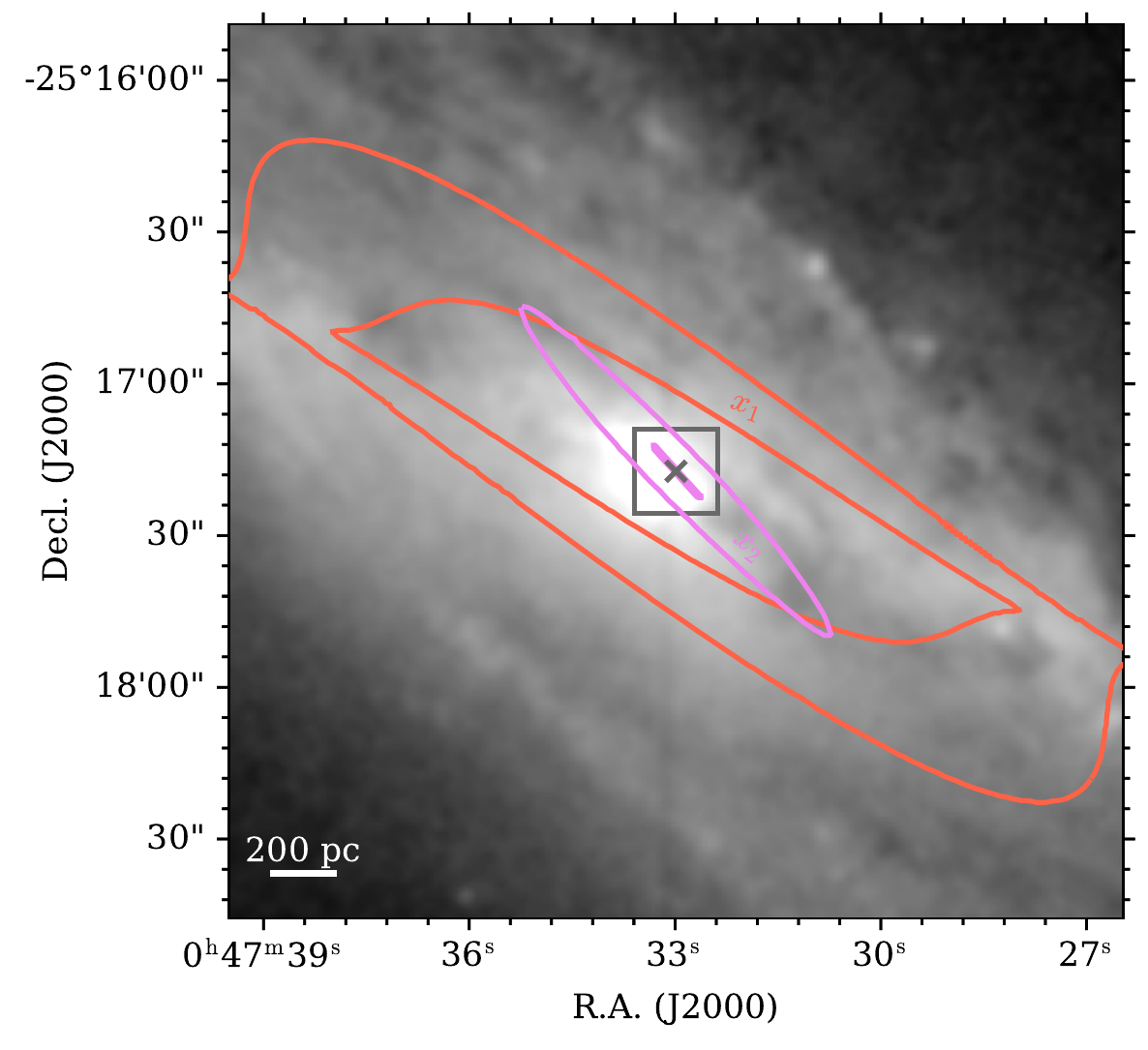}{0.49\textwidth}{}}
    \caption{NGC\,253 as seen in the $J$-band from 2MASS (grayscale). The images on the left are 16~kpc (15.7\arcmin) on a side; the images on the right are 3~kpc (2.9\arcmin) on a side. The gray squares mark the inner 280~pc, the FOV of the ALMA data. In the right column, the gray $\times$ shows the kinematic center determined by \citet{anantharamaiah96}. {\em Top row:} Overplotted are the locations of important resonances and features, determined by \citet{sorai00}. Working inward, the solid ellipses show the CR (green), the OILR (dark blue), and the IILR (light blue). We note that the location of the IILR is highly uncertain (see the discussion in Section \ref{ssec:resonances_orbits}). The gray circles in the lower right corners show the 16\arcsec\ ($\sim$270~pc) beam of the \cooz\ observations used by \citet{sorai00}. {\em Bottom row:} The outermost and innermost \xone\ orbits are shown in red, and the outermost and innermost \xtwo\ orbits are shown in pink \citep{das01}.}
    \label{fig:barresonances}
\end{figure*}

\subsection{Bar Resonances and Families of Orbits in NGC\,253}
\label{ssec:resonances_orbits}

\subsubsection{Bar Resonances}

Zooming out in scale, \citet{sorai00} used \cooz\ data with 16\arcsec\ ($\sim$270~pc) resolution from the Nobeyama 45-m telescope to study the bar resonances and kinematics in the center of NGC\,253. By fitting the CO rotation curve, they were able to estimate the locations of several key bar resonances. From their CO rotation curve, they measure the angular velocity, $\Omega(R)$, and the epicyclic frequency, $\kappa(R)$. We note that \citet{sorai00} used an older distance for NGC\,253 of 2.5~Mpc; here we have converted all of their measurements assuming a galaxy distance of 3.5~Mpc\footref{foot:sorai}. \citet{sorai00} used previous estimates of the pattern speed of the bar, $\Omega_p\approx35$~\kms~kpc\per. The radius at which $\Omega_p=\Omega$ is called the corotation radius (CR) and occurs at a radius of 5.6~kpc in NGC\,253 (green ellipse in Figure \ref{fig:barresonances}). Other resonances occur at harmonics of $\kappa$. For example, the outer and inner inner Lindblad resonances (OILR and IILR respectively) occur where $\Omega_p = \Omega-\kappa/2$.

These resonances are important because gas is expected to collect inside these resonance locations, with gas inside the IILR possibly collapsing to trigger the nuclear starburst \citep[e.g.,][]{goldreich79,wada92,sorai00,paglione04}. In NGC\,253, the OILR is located at a radius of 1.8~kpc, well matched to the extent of the bar where it intersects with the circumnuclear "2~kpc" ring (Figure \ref{fig:barresonances}). The IILR is located at a radius of 336~pc.

We caution, however, that the location of the IILR determined by \citet{sorai00} is especially uncertain because it is on the same scale as their spatial resolution. Moreover, \citet{sorai00} determined their CO rotation curve by finding the terminal velocities \citep[see e.g., Section 3.3 of][]{sofue01}. While this method is often used for highly inclined systems, rotation velocities within the central few resolution elements (especially where there are steep velocity gradients) are especially uncertain \citep[e.g.,][]{sofue01} Moreover, the central bar will produce strong non-circular motions that are not taken into account in the terminal velocities method. This can produce dramatic artifacts in the determination of the rotation curve \citep[see e.g., Section 5.3 of][]{binney91}. Uncertainties on the CO rotation curve will propagate into the determination of the resonance locations. Therefore, all of the resonance locations reported by \citet{sorai00} are likely uncertain, with the IILR being the most uncertain.

As an example of the uncertainty in the position of the IILR in NGC\,253, we compare the IILR location measured by \citet{sorai00} to that inferred by \citet{das01}. \citet{das01} model the central regions of NGC\,253 using a logarithmic bar potential (see their Equation 1). From this potential and the values in their Table 1, we calculate $\Omega(R)$ and $\kappa(R)$ to determine the CR, OILR, and IILR from this model. From this, we find that the CR is 5.8~kpc, the OILR has a radius of 1.8~kpc, and the IILR has a radius of 27~pc. While the locations of the CR and OILR are similar between \citet{sorai00} and \citet{das01}, the IILR differs by an order of magnitude. Therefore, because the position of the IILR is so uncertain, we should not attach particular meaning to its location relative to the SSCs and dense molecular gas in NGC\,253.

\subsubsection{Families of Orbits in a Barred Potential}

There are families of closed orbits in a bar potential. The \xone\ (bar) orbits are extended along the major axis of the bar whereas the \xtwo\ (antibar) orbits are perpendicular to the bar major axis \citep[e.g.,][]{contopoulos77,athanassoula92a,athanassoula92b,binney08}. The \xtwo\ orbits are closely related to the ILR (or an OILR and IILR; e.g., \citealt{vanalbada82,athanassoula92a}). At the intersections between these two orbital families, gas can collide and shock, lose angular momentum, and transition from the \xone\ to the \xtwo\ orbits, bringing it closer to the galactic center. \citet{das01} calculate the \xone\ and \xtwo\ orbit families in NGC\,253 assuming a logarithmic bar potential. In Figure \ref{fig:barresonances} (bottom) we show the outermost and innermost \xone\ orbits (red) and the outermost and innermost \xtwo\ orbits (pink), for a non-axisymmetry parameter of 0.8 (see also Figure 3 of \citealt{das01}). The \xtwo\ orbits intersect the \xone\ orbits and extend down to $\sim$100~pc scales. Therefore, these orbits can facilitate the transfer of gas from large to small scales, fueling the nuclear starburst.

\subsubsection{Connecting the Bar to the SSCs and Dense Molecular Gas}
From the discussion above, it is interesting to determine how the SSCs are arranged with respect to the bar orbits and resonances. Due to the nearly edge-on inclination, constraining the arrangement purely from the cluster locations is difficult. A similar challenge is faced in studying the MW CMZ, where the massive star forming regions are viewed nearly edge-on and thought to be arranged as spirals, streams with either open or closed orbits, or a ring \citep[e.g.,][]{sofue95,sawada04,molinari11,krumholz15,kruijssen15,henshaw16,ridley17,tress20,sormani22}. \citet{henshaw16} and \citet{henshaw22} provide excellent reviews and testing of these models in the CMZ (see especially Figures 18 and 19 of \citealt{henshaw16}). Without knowledge of how the SSC structure is dynamically linked to the bar (or not), it is impossible to tell whether the SSC structure is tilted in the same manner as the galaxy disk (i.e., where the near side of the disk is towards the northwest). We can, however, use the kinematic information from the cluster velocity measurements presented by \citet{levy21} together with the morphology to constrain the cluster arrangement. 

\subsection{SSC and Dense Molecular Gas Kinematics}
\label{ssec:ssckin}

\begin{figure*}
    \centering
    \begin{interactive}{animation}{ngc253_PPV_SSCs_CS.mp4}
    \includegraphics[width=0.6\textwidth]{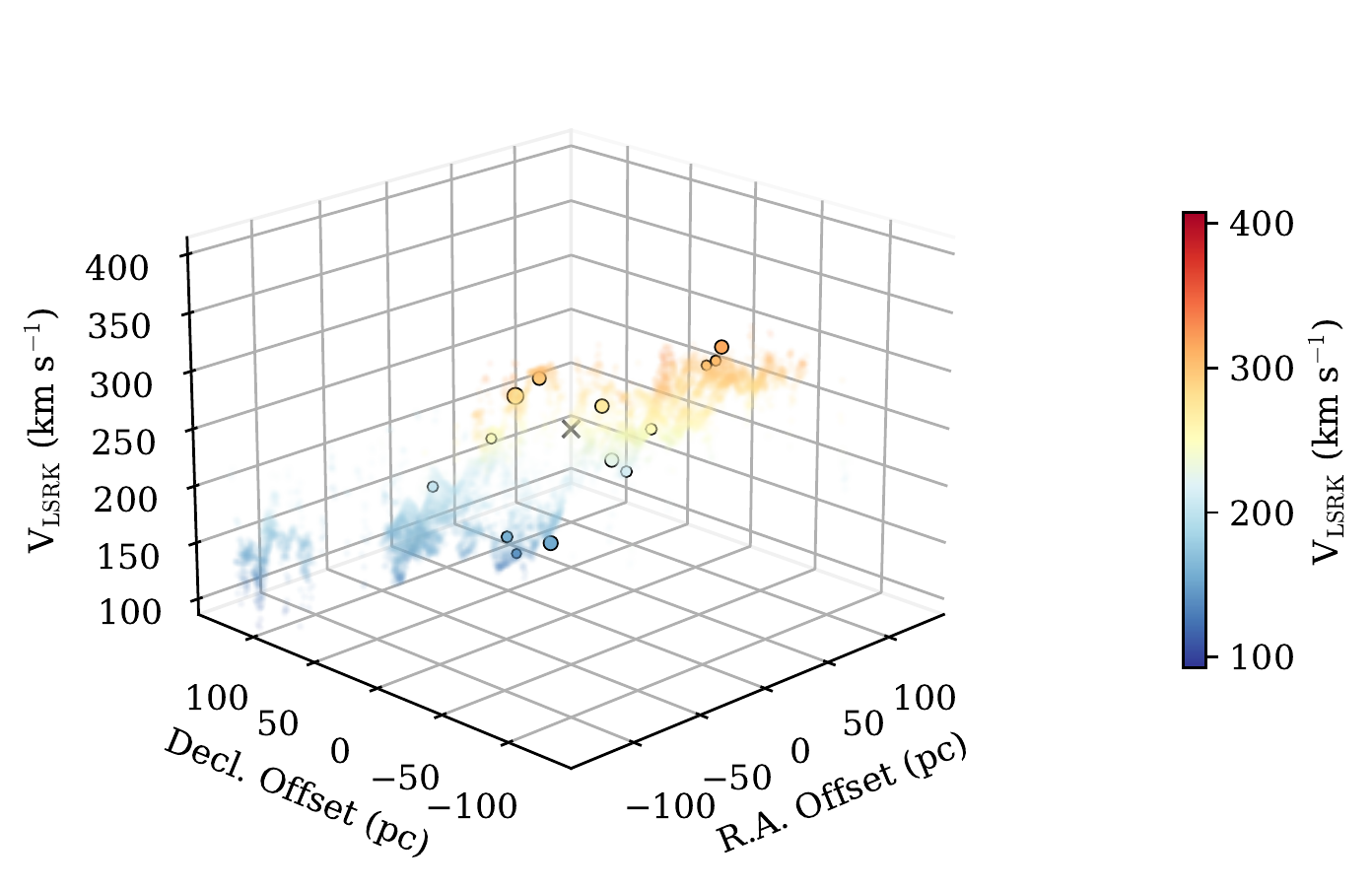}\\
    \end{interactive}
    \includegraphics[width=0.49\textwidth]{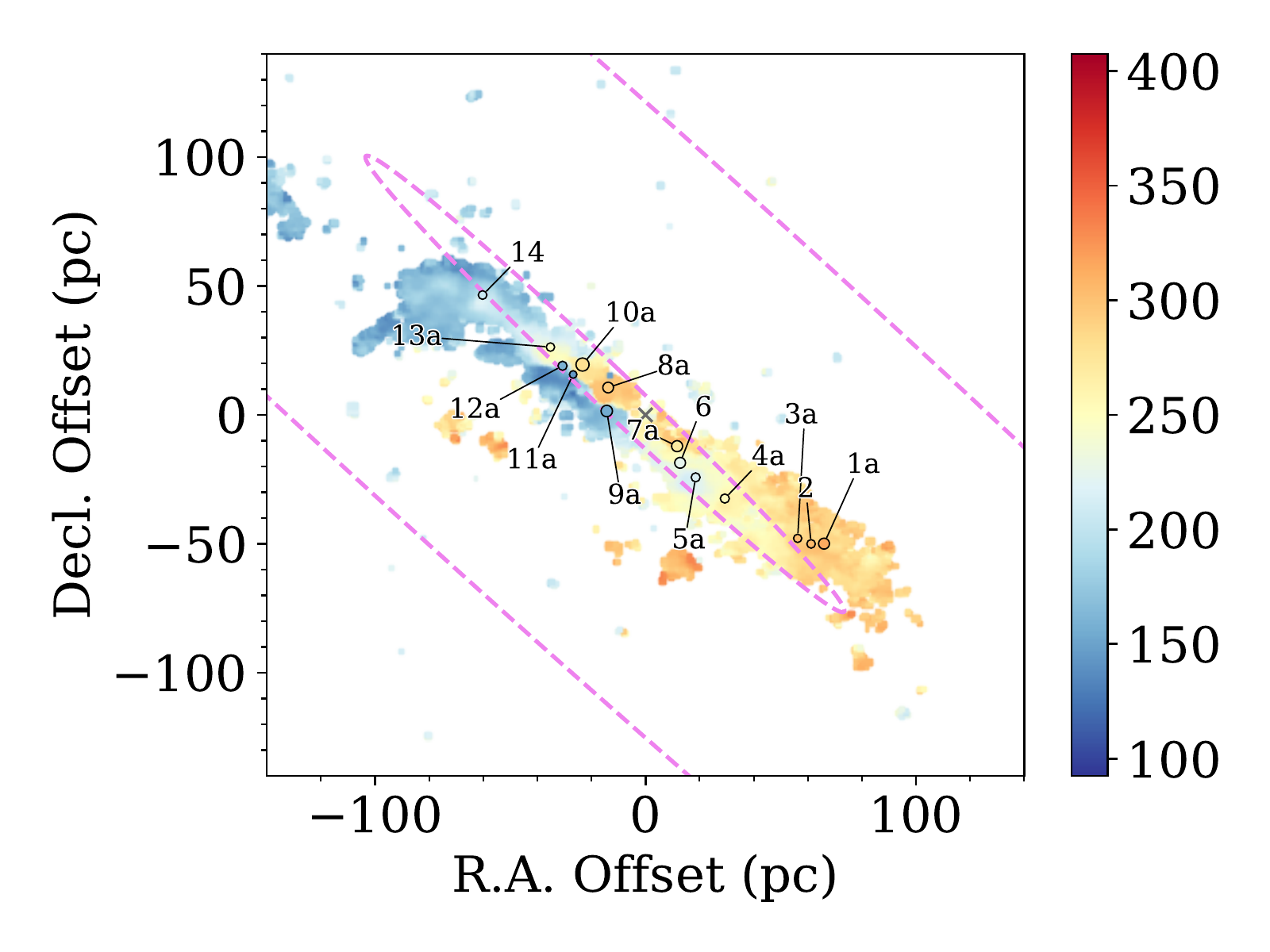} 
    \includegraphics[width=0.49\textwidth]{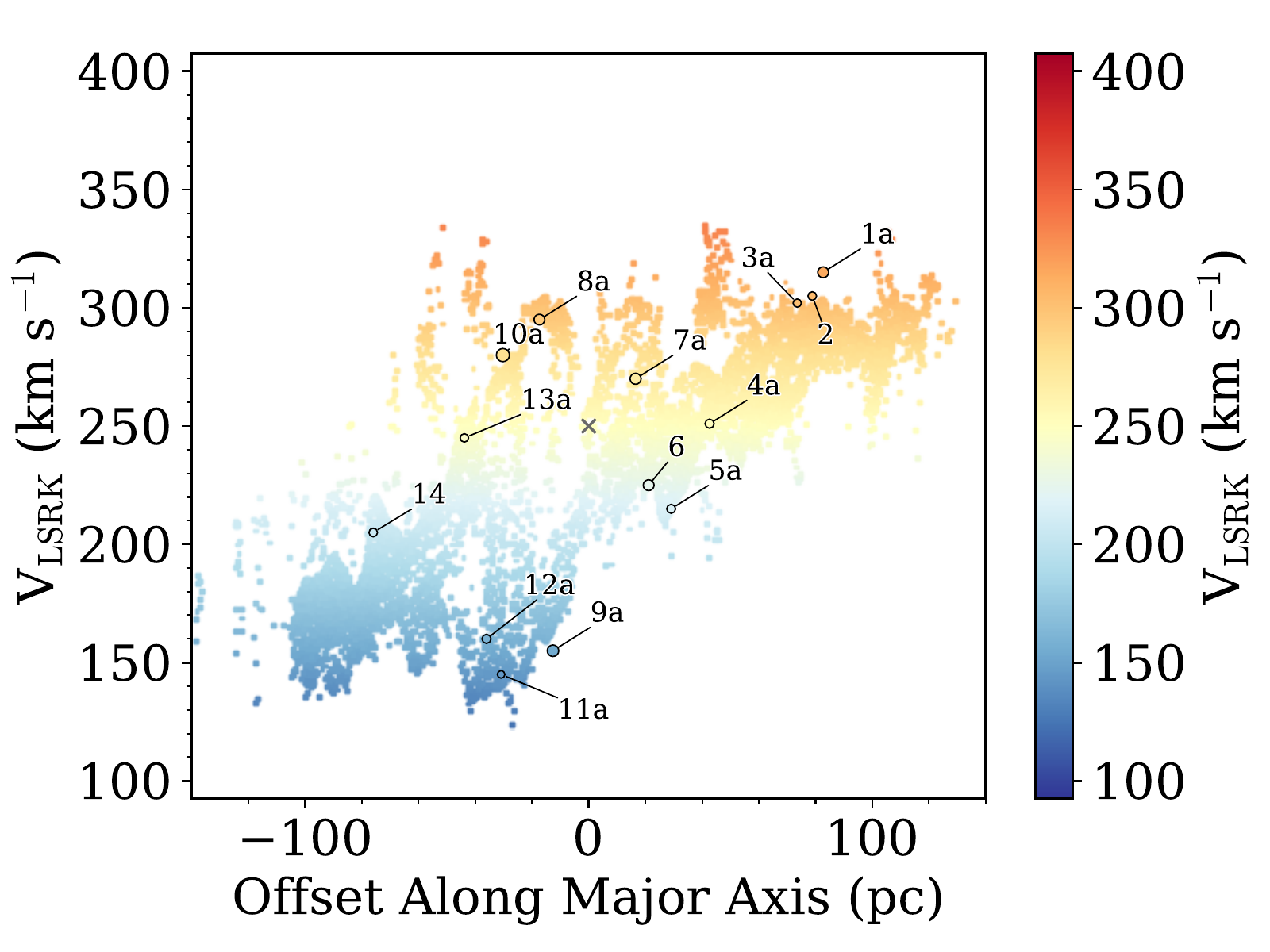}\\
    \caption{R.A.-Decl.-Velocity projections of the primary SSCs and \cs\ in the central 280~pc of NGC\,253 (the gray box in Figure \ref{fig:barresonances} and the FOV of the ALMA data). Pixels with SNR~<~5 in peak intensity are masked out in the \cs\ velocity map. The positional offsets are calculated based on the center determined by \citet[][shown as the gray $\times$ assuming this point has a velocity equal to the galaxy systemic velocity]{anantharamaiah96}. The color scale is the same in all panels and shows the LSRK velocity. The radii of the SSC markers are  $2\times r_{\rm deconv}$ (Table \ref{tab:sscparams_radialprofile_deconv}). {\em Top:} One orientation of the 3D R.A.-Decl.-Velocity view. (An animated version that rotates in azimuth is available in the online version.) {\em Bottom left:} The R.A.-Decl. projection. The pink dashed ellipses show the locations of the outermost and innermost \xtwo\ orbits, as in Figure \ref{fig:barresonances}, though we note that the calculations of these orbits is highly uncertain on these scales. {\em Bottom right:} A position-velocity projection, where the positional coordinate is along the major axis of the structure shown in the bottom-left panel (PA~$= 235$\D). The morphology, distribution, and kinematics of the SSCs and dense molecular gas traced by \cs\ agree very well.}
    \label{fig:sscs_PPV}
\end{figure*}

In Figure \ref{fig:sscs_PPV}, we show the locations and systemic velocities of the primary SSCs. The radius of each circle shows the deconvolved SSC size ($r_{\rm deconv}$, see Table \ref{tab:sscparams_radialprofile_deconv} and Section \ref{sssec:radialprofiles}). The systemic velocities were measured by \citet{levy21} using a multi-Gaussian fit to many spectral lines detected towards these clusters. The uncertainties on the SSC systemic velocities are better than $\pm$5~\kms. The velocity color scale is centered on the systemic velocity of the galaxy \citep[250~\kms;][]{muller-sanchez10,krieger19,krieger20b}.

We compare the distribution and kinematics of the primary SSCs to \cs\ observations from \citet{krieger19,krieger20a} in Figure \ref{fig:sscs_PPV}. Their cleaned cubes are a combination of 12-m, 7-m, and total power data and have a spectral resolution of 2.2~\kms\ and a spatial resolution of 0.17\arcsec~$\times$~0.13\arcsec\ (2.9~pc~$\times$~2.2~pc). We fit the \cs\ line with a Gaussian at each pixel. We mask the velocity map based on the peak intensity, where pixels with a signal-to-noise ratio (SNR) <~5 are removed. The top panel of Figure \ref{fig:sscs_PPV} shows a 3D position-position-velocity (R.A.-Decl.-Velocity) diagram of the \cs\ data and the primary SSCs. The bottom row of Figure \ref{fig:sscs_PPV} shows R.A.-Decl. and position-velocity projections. The panels in Figure \ref{fig:sscs_PPV} are cropped to the FOV of the ALMA data of the SSCs described in Section \ref{sec:obs4}, which cover the central 280~pc of the galaxy.

The positions and velocities of the SSCs agree well with the \cs\ emitting dense molecular gas. {\change Overplotted in the R.A.-Decl. panel of Figure \ref{fig:sscs_PPV} are the outermost and innermost \xtwo\ orbits (pink dashed ellipses; see also Figure \ref{fig:barresonances}).} The SSC structure and the dense molecular gas in which they are embedded are remarkably well aligned with the innermost \xtwo\ orbits, especially given the spatial resolution of data with which the \xtwo\ orbits were derived \citep[$\approx$300~pc;][]{das01}, though there appears to be a slight PA offset. {\change We note that the \cs\ data include short- and zero-spacing data and so recover emission on large spatial scales. This means that the concentration of \cs\ emission around the innermost \xtwo\ orbit is not a result of spatial filtering by the interferometer.} This is compelling evidence that the SSCs and the dense molecular gas from which they form are located at the nexus of the family of orbits which transfer gas from the bar on large scales to small scales where the gas can become dense enough to form massive young stellar clusters.

\subsection{3D Structure of the SSCs and Dense Molecular Gas}
\label{ssec:3dstructure}

\begin{figure*}
    \centering
    \begin{interactive}{animation}{ngc253_SSCs_AngMomConsRing_XYZV.mp4}
    \includegraphics[width=0.55\textwidth]{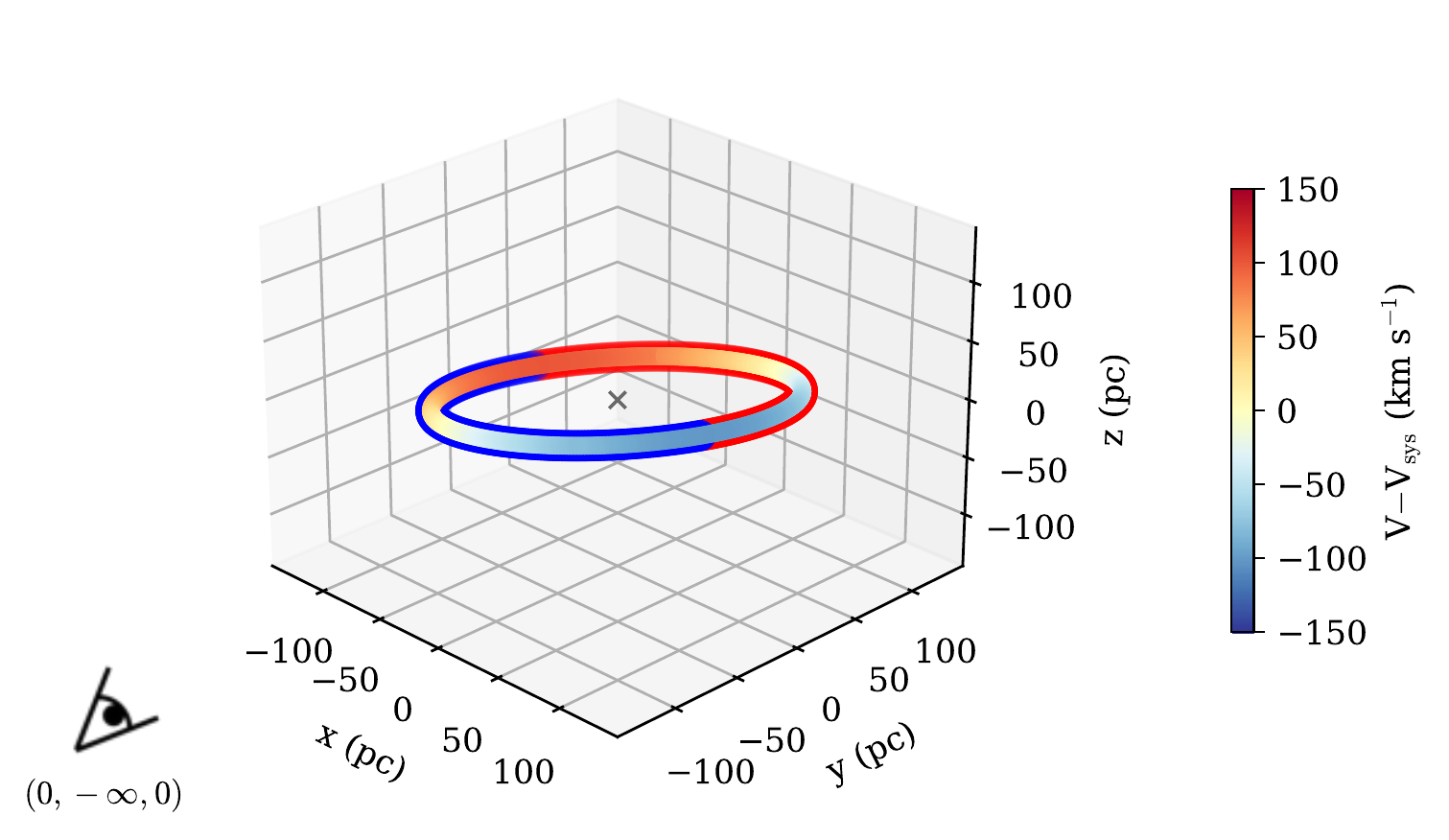}\\
    \end{interactive}
    \includegraphics[width=0.45\textwidth]{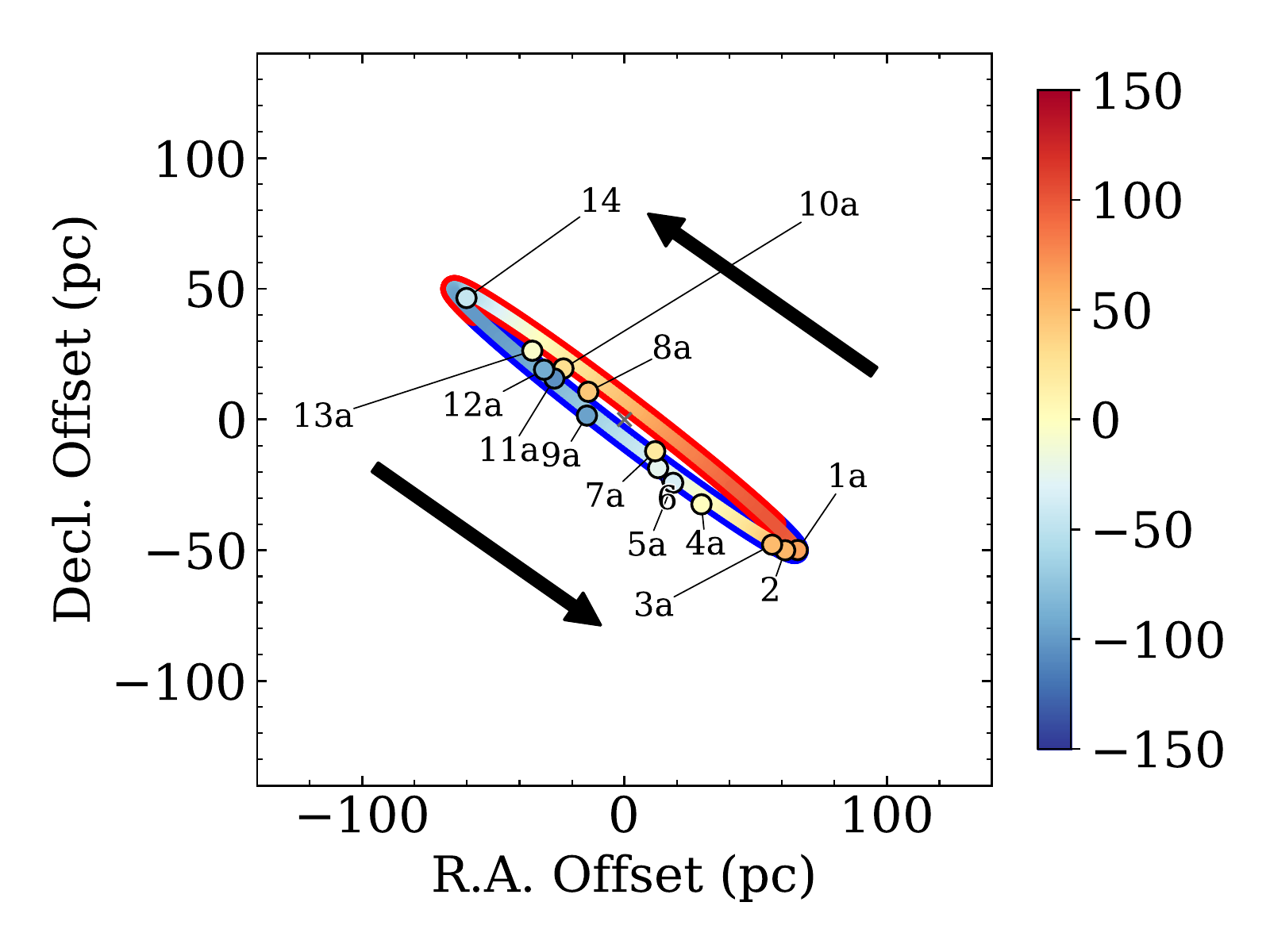}
    \includegraphics[width=0.45\textwidth]{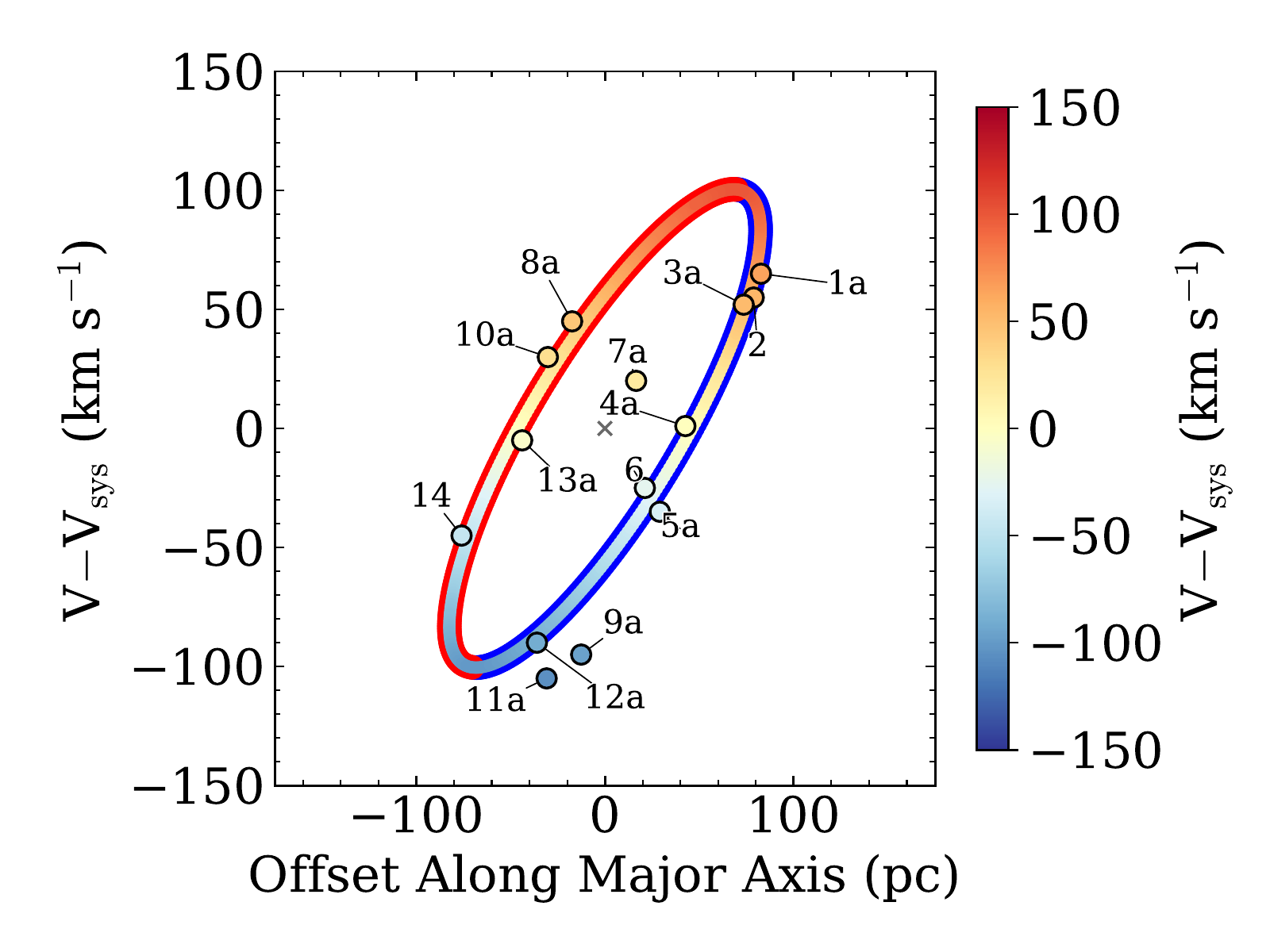}
    \includegraphics[width=0.45\textwidth]{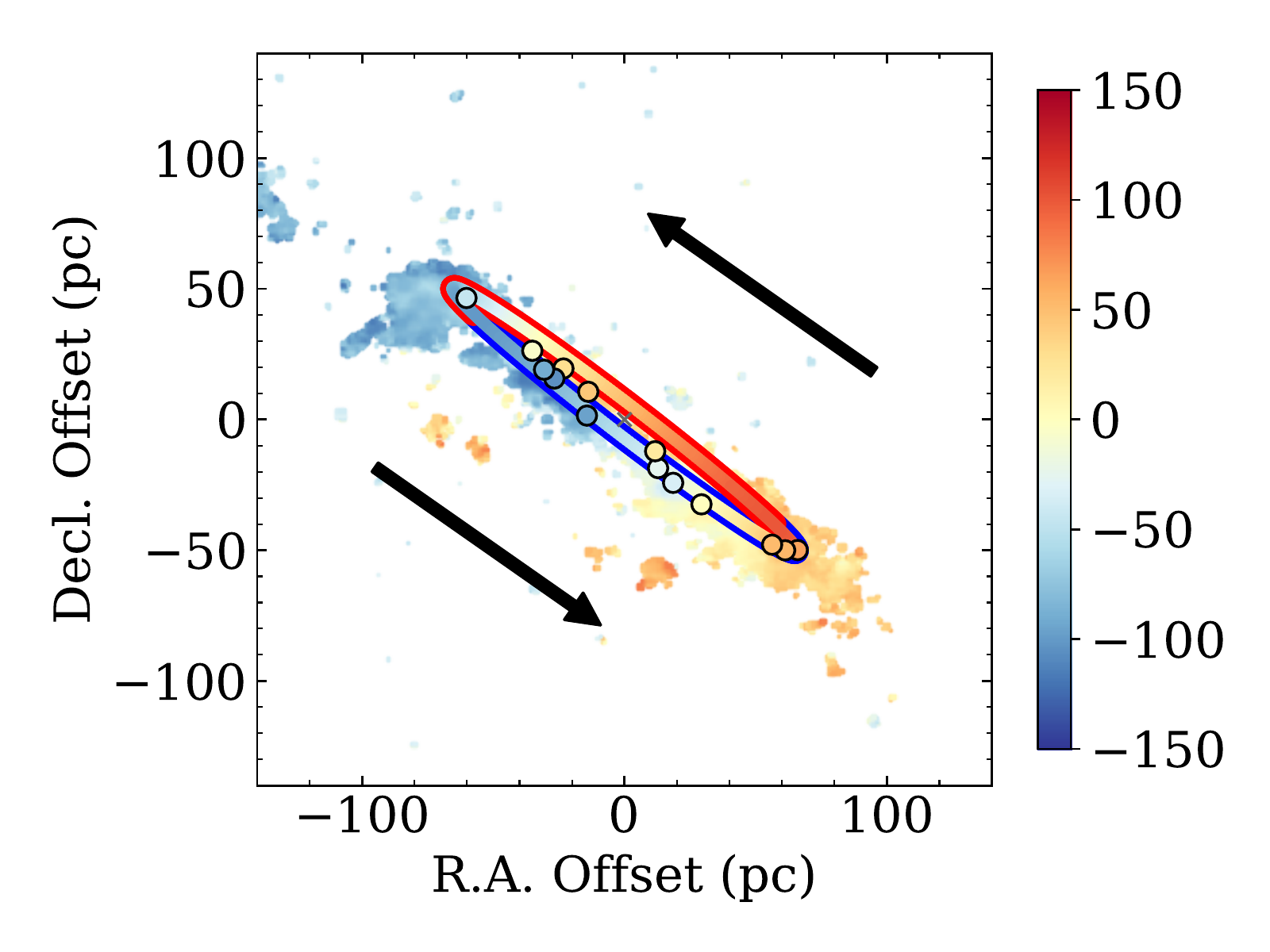}
    \includegraphics[width=0.45\textwidth]{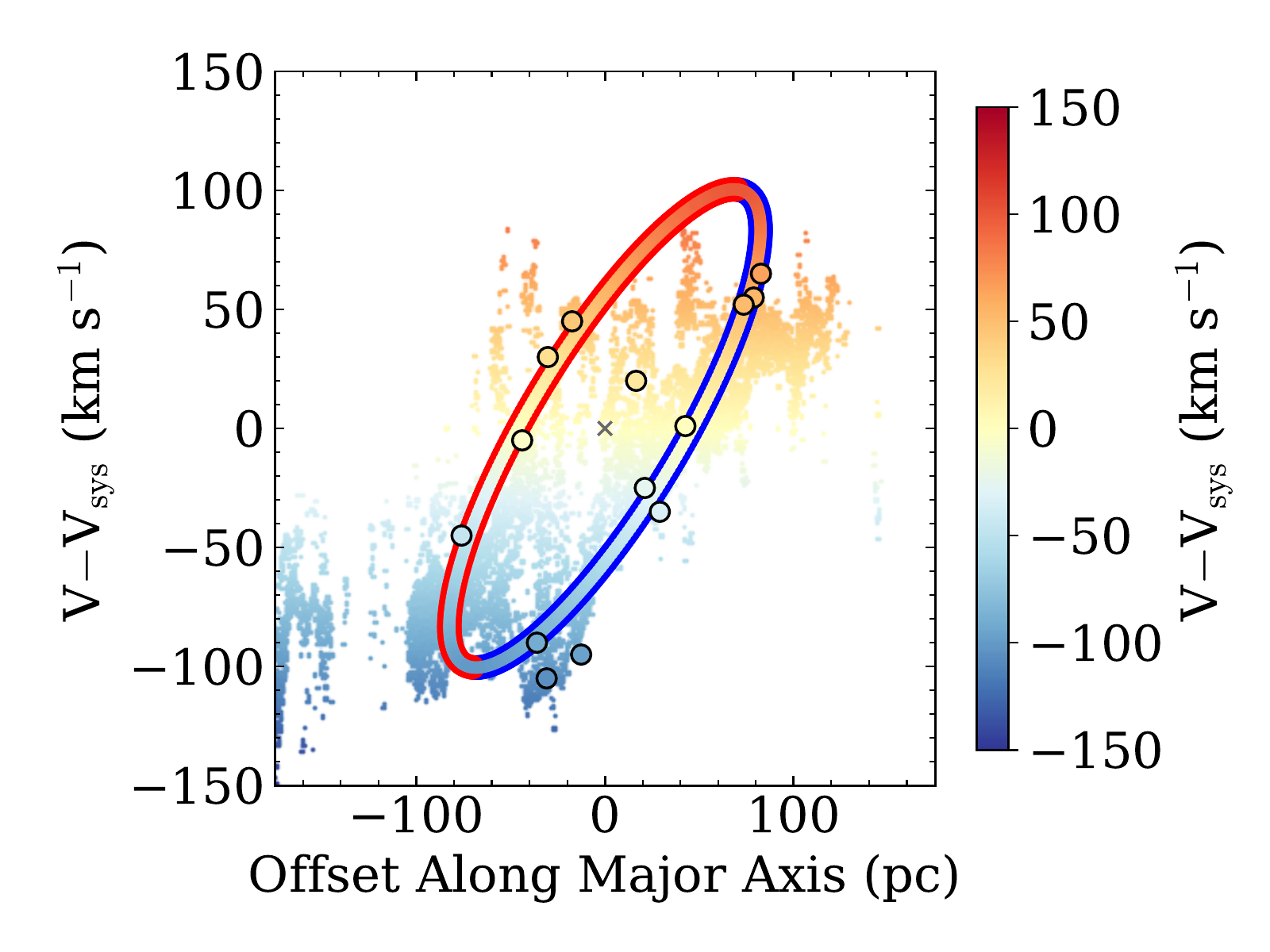}

    \vspace{-6pt}
    \caption{A model of an elliptical, angular momentum-conserving ring. {\em Top:} The ring in galactic coordinates, where $x$ corresponds to galactic longitude, $y$ to the line of sight, and $z$ to galactic latitude. The observer is at $(x,y,z)=(0,-\infty,0)$. The gray $\times$ shows the gas kinematic center. The color coding shows the radial velocity, with the galaxy systemic velocity removed. The blue (red) outlining shows the front (back) portion of the ring along the line of sight with respect to the galaxy center. (An animated version of the top panel, without the observer marked, that rotates in azimuth and elevation is available in the online version.) {\em Middle left:} The R.A.-Decl. projection of the ring with the primary SSCs overplotted. Arrows show the direction of the orbit. {\em Middle right:} A position-velocity projection of the ring and SSCs, where the positional coordinate is along the major axis of the structure (PA~$=235$\D). {\em Bottom row:} The same as the middle row, but with the \cs\ plotted in the background and without the SSC labels for clarity. The arrangement and kinematics of the SSCs and \cs\ are in good agreement with this simple model.}
    \label{fig:angmomconsring}
\end{figure*}

\begin{figure}
    \centering
    \includegraphics[width=\columnwidth]{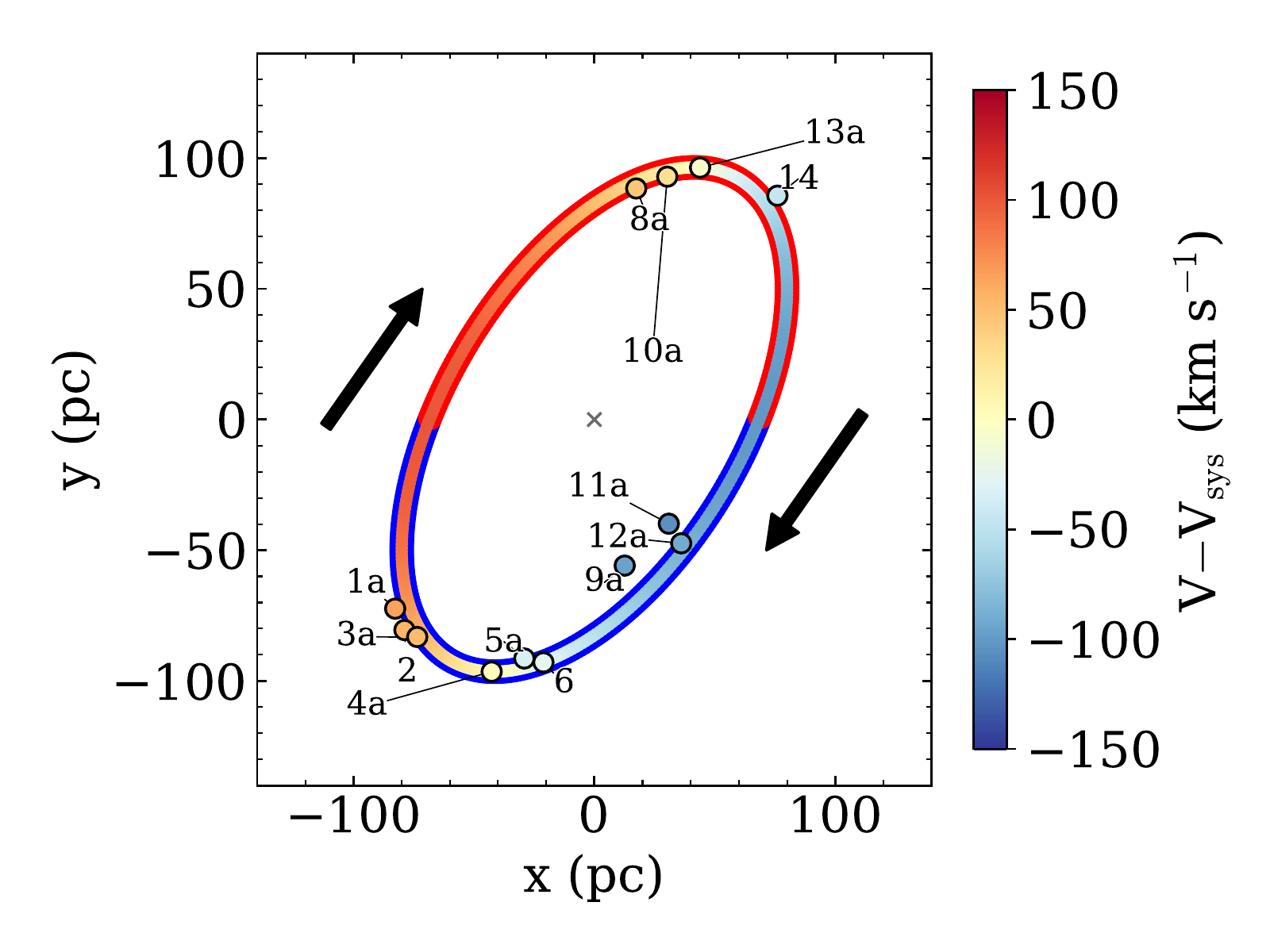}
    \caption{A top-down view of the elliptical ring model shown in the top panel of Figure \ref{fig:angmomconsring}. The observer shown in the top panel of Figure \ref{fig:angmomconsring} is located at $(x,y) = (0,-\infty)$ in this plot. The model-dependent deprojected positions of the SSCs along this ring are shown (except for SSC~7a which is not well fit by this model). Arrows indicate the direction of the orbit. Other annotations and color-coding are as in Figure \ref{fig:angmomconsring}.}
    \label{fig:angmomconsring_topdown}
\end{figure}

Nuclear rings following \xtwo\ orbits are formed spontaneously in a barred potential, as revealed by hydrodynamic simulations \citep[e.g.,][]{tress20}{\change, and are able to power galaxy-scale outflows \citep[e.g.,][]{nguyen22}}. \xtwo\ orbits are mildly elongated perpendicular to the bar and have a nearly-elliptical shape. Their orbital velocity is larger at the pericenter and lower at the apocenter, qualitatively similar to what one would get by assuming that the angular momentum is constant along the orbit. Although the angular momentum is not exactly conserved along \xtwo\ orbits (because a bar potential is non-axisymmetric --- it oscillates around a mean value), a reasonable approximation is to model \xtwo\ orbits as elliptical orbits on which the angular momentum is conserved \citep[see also][]{peters75}. This type of model has been applied to the MW CMZ and can explain the arrangement of dense molecular gas features and star forming regions \citep{tress20}. In the MW, several other types of models have been developed to explain the orbits of gas, stars, and massive star-forming regions in the CMZ, including twisted rings, spirals, and crossing streams \citep[see e.g.,][and references therein]{henshaw22}. We apply some of these models to the CMZ of NGC\,253 in Appendix \ref{app:cmzmodels}.

\subsubsection{A Plausible Model}
Here, we construct a simple kinematic model of the nuclear ring in NGC\,253 by assuming that the gas follows elliptical orbits on which the angular momentum is conserved. This model is built in Cartesian coordinates where the $x$-axis corresponds to galactic longitude, the $y$-axis to the line of sight from the observer, and the $z$-axis to galactic latitude (Figure \ref{fig:angmomconsring}, top). The elliptical, angular momentum-conserving ring is described by four parameters: the semi-major axis length ($a$), the semi-minor axis length ($b$), the orbital velocity at the pericenter of the ring (V$_{\rm orb,0}$), and the position angle of the major axis of the ellipse with respect to the $x$-axis ($\theta_p$). The orbital velocity at every other point along the ellipse is determined by conservation of angular momentum starting from V$_{\rm orb,0}$. The elliptical orbit in this frame is shown in the top panel of Figure \ref{fig:angmomconsring}.

To compare this model with the SSCs, we project it into the sky plane assuming some position angle (PA; defined counterclockwise of north to the receding side the the ring) and an inclination ($i$). The projection of this ring into R.A.-Decl. coordinates is shown in the left column of Figure \ref{fig:angmomconsring}. We also construct a position-velocity diagram, where the position is taken along the major axis of the ring given by the PA (Figure \ref{fig:angmomconsring}, right column).

We adjust the ring parameters by-eye to best fit the arrangement and kinematics of the SSCs. A ring with $a\sim110$~pc, $b\sim60$~pc, V$_{\rm orb,0}\sim115$~\kms, and $\theta_p\sim55$\D\footnote{Ring-like models for the MW CMZ also find $\theta_p\neq0$\D\ \citep{molinari11}, which may be related to the angle at which material from the \xtwo\ orbits flows into this ring. A $\theta_p$ of this magnitude is required to reproduce the kinematics of SSCs 9a, 11a, and 12a, which have the most blue-shifted velocities but are not located at the end of the structure (Figure \ref{fig:angmomconsring}).}, PA~$\approx235$\D, and $i\approx85$\D\ is a good representation of the SSCs. We reiterate that these parameters were fit by-eye and that there are degeneracies among them. This is not meant to be an exact measurement of the size of the SSC structure, but rather to show that this is a possible configuration with reasonable parameters.

As shown in Figure \ref{fig:angmomconsring}, the SSCs agree very well with this simple model. In Figure \ref{fig:angmomconsring_topdown}, we show a top-down ("face-on") view of this ring. Using this model, we place the SSCs along the ring; the positions along the line of sight (y-axis) are entirely model dependent. Under this model, SSCs 1a, 2, 3a, 4a, 5a, 9a, 11a, and 12 would be on the front (near) side of the ring, shown as the blue outlines in Figures \ref{fig:angmomconsring} and \ref{fig:angmomconsring_topdown}. SSCs 8a, 10a, 13a, and 14 would be on the back (far) side of the ring, shown as the red outlines.

The orbital period along the elliptical ring shown in Figures \ref{fig:angmomconsring} and \ref{fig:angmomconsring_topdown} is $\approx6$~Myr. Given the limits on the SSC ages (see Section \ref{sssec:ages}), this means that the SSCs have not completed a full orbit. On average, the SSCs will complete half an orbit before the massive stars within them explode as supernovae.

\subsubsection{Caveats of this Simple Model}
SSC 7a is not well fit by this model. Although this cluster is relatively weak compared to others, its systemic velocity was well constrained by \citet{levy21} and its velocity agrees well with the \cs\ (Figure \ref{fig:sscs_PPV}). Unlike \citet{leroy18} and \citet{mills21}, we find evidence for a single velocity component towards this SSC, which may be due to the increased spatial resolution. There are multiple spatial components near SSC 7a (see e.g., Figure \ref{fig:ngc253_cont} bottom right and Figure 1 of \citealt{levy21}), which can be blended together at lower resolution leading to a second velocity component. Moreover, a shift in velocity cannot fully bring SSC 7a into agreement with the model; a change in the cluster's velocity would shift its position vertically in position-velocity diagram (Figure \ref{fig:angmomconsring}; right column) which does not bring it into agreement with the model for reasonable adjustments to the cluster's velocity. Because SSC~7a is not well described by this model, we do not show it in Figure \ref{fig:angmomconsring_topdown}. Hydrodynamical simulations of the MW CMZ find gas and star formation inside the CMZ ring, and it is possible that this star formation may be happening in the vicinity of the supermassive black hole Sagittarius A$^*$ or associated with the nuclear star cluster \citep[][and references therein]{sormani20}. In NGC253, we also detect dense molecular gas inside the elliptical ring model (e.g., Figure \ref{fig:angmomconsring}). SSC 7a could be evidence of massive star formation close to the galactic center (whose precise position is unknown) though, with only a single cluster, this is only speculative.

Along the ring, there are more extreme radial velocities than represented by the SSCs or the dense molecular gas traced by \cs, particularly on the redshifted side of the structure (Figure \ref{fig:angmomconsring}; bottom row). In the case of the \cs\ data, this is not an effect of the interferometer filtering out emission on large scales as these data include the zero-spacing (total power) data \citep{krieger19,krieger20a}. The \cs\ data cube covers LSRK velocities from 88 to 373~\kms, or -162 to +123~\kms\ about the systemic velocity of the galaxy. This could mean that we are missing some of the most redshifted \cs\ emission if it falls outside the bandpass. As a check, we examine the spectra in the \cs\ cube, but we do not find evidence that the line profiles are cut off by the edge of the bandpass.

As noted above, simulations show that the angular momentum in an \xtwo\ orbit is approximately constant in a time-averaged sense \citep[e.g.,][]{sormani18,tress20}. As the major-to-minor axis ratio of the orbit increases, the amplitude of the oscillations in angular momentum increase and the assumption of angular momentum conservation breaks down. For the CMZ of the MW, an \xtwo\ orbit with an axis ratio of $110/60\approx1.8$ would yield variations of 25\% in the angular momentum over time \citep{sormani18}. This means that for an axis ratio like we estimate for NGC\,253, the variation in angular momentum are not negligible; this is a limitation of this simple model. 

{\change As described in Section \ref{sec:intro}, we assume that the dynamical center of NGC\,253 is the center determined from ionized gas kinematics by \citet{anantharamaiah96}. This center position has a 1-$\sigma$ uncertainty of $\sim0.3$\arcsec, which corresponds to 5~pc. This is a negligible uncertainty in terms of the model fit shown in Figure \ref{fig:angmomconsring}, so we conclude that uncertainties in the center position determined by \citet{anantharamaiah96} will not affect the results from the model. Other determinations of the dynamical center of the galaxy, however, differ by more than this measurement uncertainty. For example, the center determined from stellar kinematics by \citet{muller-sanchez10} is 1.3\arcsec\ (22~pc) away to the northeast in projection (between SSCs 8 and 10/11). From the standpoint of the modeling, varying the location of the dynamical center will change how the SSCs are distributed along the ring. If, for example, we instead choose the center determined by \citet{muller-sanchez10}, a larger ring with a higher orbital velocity is required. More challenging, however, is that a single (symmetric) ring cannot fit clusters on both the near- and far-sides of the ring simultaneously. In other words, compared to Figure \ref{fig:angmomconsring} (right panels), a model using the center from \citet{muller-sanchez10} can either fit clusters on the near (blue outlined) side of the ring or on the far (red outlined) side, but not both simultaneously. Therefore, if the true dynamical center of NGC\,253 is that from \citet{muller-sanchez10}, then this model of a symmetric angular momentum-conserving ring is not a good fit to the data.}

\subsection{A Cautionary Note about SSC Ages and Age Gradients}
\label{ssec:agegrad}

\subsubsection{Previous Estimates of the SSC Ages in NGC\,253}
\label{sssec:ages}

In terms of the relative ages of the SSCs, both \citet{rico-villas20} and \citet{mills21} found evidence for an inside-out formation scenario, where clusters towards the ends of the structure are younger and those towards the middle are older (i.e., SSCs 1, 2, 3, 13, and 14 are among the youngest and SSCs 4, 5, 7, 8, 10, 11, and 12 are among the oldest). On the other hand, \citet{krieger20a} used line ratios of  HCN/HC$_3$N to construct a relative chemical age gradient, which does not follow a pattern with distance from the center and where the progression from youngest to oldest is SSC 13, 14, 1, 8, 3, 2. 

The absolute ages of the SSCs in NGC\,253 are highly uncertain, but limits can be placed on them. \citet{rico-villas20} estimated that the SSCs in NGC\,253 have ages $\approx0.01-1$~Myr based on the ratio of luminosity of in stars (from free-free emission) and protostars. These ages are likely a lower limit, however, since the ionizing photons that produce the free-free emission may be absorbed by dust within the SSCs \citep[see][for a further discussion]{levy21}. \citet{rico-villas20} also found a weak trend between their estimated ages and the HNCO/CS line ratio, where younger clusters tend to have higher line ratios. Hydrodynamical simulations of the Milky Way show that the stars and dense molecular gas are well coupled until the stars are $\sim$5~Myr old \citep{sormani20}. In NGC\,253, the SSCs and \cs\ are well matched in terms of their locations and kinematics (Figure \ref{fig:sscs_PPV}), indicating that the SSCs are younger than $\sim$5~Myr. In a separate study, \citet{mills21} found that the clusters in NGC\,253 cannot be older than $\sim3$~Myr due to the presence of He recombination lines and the lack of appreciable synchrotron emission towards most of the clusters (indicating a lack supernovae in the SSCs). Therefore, the ages of the SSCs are likely $\sim0.01-3$~Myr.

\subsubsection{Models of Star Formation and Predictions for Age Gradients in a Circumnuclear Ring}
There are three main models that describe where and when star formation occurs in a circumnuclear ring. These models make predictions for age gradients (or lack thereof) along the orbit. First, the so-called "pearls-on-a-string" scenario predicts that star formation occurs just downstream of the contact points between the gas inflow from the bar and the circumnuclear ring (i.e., downstream of the apocenters of the orbit; e.g., \citealt{boker08,mazzuca08,sormani20}). Moreover, gas in the ring will have the slowest orbital velocities at the apocenters, more easily allowing it to pile up and become dense. The star clusters age as they orbit along the ring, so the youngest clusters should be found near the apocenters with an increase in age downstream. Alternately, it has been suggested that star formation could be triggered when clouds are compressed due to close pericenter passages \citep{longmore13,kruijssen15}. Under this scenario, the youngest stars should be found closest to pericenter and stellar (or cluster) ages should increase downstream. Finally, the "popcorn" model describes a scenario in which the ring forms from gravitational collapse or turbulence and star formation is distributed uniformly along the ring \citep{boker08}. In this scenario, no age gradients are expected as either the clusters have approximately the same age or the ages are randomly distributed along the ring.

It is tempting to use the elliptical angular momentum conserving ring model to infer the relative ages of the SSC. However, the observations of the SSCs in NGC\,253 are a single snapshot in time. Hydrodynamical simulations show that the instantaneous star formation distribution and ring morphology vary substantially about the time-averaged behavior \citep[e.g.,][]{tress20,sormani20}. While the time-averaged simulation results favor the "pearls-on-a-string" model, it is much less clear if signatures of the latter can be seen in the instantaneous simulation results due to large time-fluctuations (see especially Figure 9 of \citealt{sormani20}). Therefore, with only this single snapshot in time and a small number of SSCs, we cannot simply use the "pearls-on-a-string" (or other) model to infer the expected cluster age gradients from their locations along the elliptical ring.

\section{Summary}
\label{sec:summary4}

The SSCs in the center of NGC\,253 are bright continuum sources at 350~GHz, which primarily traces thermal emission from warm dust \citep[e.g.,][]{leroy18,mills21}. Here, we combine ALMA data from three 12-m configurations and the 7-m array to construct maps of the dust emission covering scales from 0.028\arcsec$-$24.8\arcsec\ (0.48~pc$-$421~pc). This enables us to measure the compact dust emission associated with the clusters themselves as well as the more extended emission in which they are embedded (Figure \ref{fig:ngc253_cont}). We summarize our main results below, indicating the relevant figures and/or tables.

\begin{enumerate}
    \itemsep0em
    \item For the first time, we detect the galaxy-scale outflow in dust continuum emission. As shown in Figures \ref{fig:ngc253_cont}, \ref{fig:SWstreamer_chanmaps}, and \ref{fig:SWstreamer_CO}, we find dust emission along the SW streamer that is located at the edge of the CO emission and hence on the exterior edge of the outflow cone. The dust streamer has a FWHM of 8~pc. From the dust, we estimate that the SW streamer has a molecular gas mass of $\sim(8-17)\times10^6$~\msun, consistent with other measurements \citep[e.g.,][]{walter17}.
    \item We measure the sizes of the SSCs using Gaussian fits to the cluster radial profiles. We deconvolve the Gaussian beam from the size measurements to provide beam-deconvolved cluster sizes, finding radii of $0.4-0.7$~pc (Table \ref{tab:sscparams_radialprofile_deconv}; Figures \ref{fig:radialprofiles} and \ref{fig:intrinsicradii}). Compared to star clusters in the LEGUS survey, the SSCs in NGC\,253 tend to be smaller, likely because they are younger (Figure \ref{fig:intrinsicradii}). 
    \item We investigate the morpho-kinematic arrangement of the SSCs and their possible connection to the bar. The SSC structure is on the same scale as the \xtwo\ orbits, suggesting that gas is transported down to these scales by the bar where it can then become dense enough to form massive star-forming regions (Figures \ref{fig:barresonances} and \ref{fig:sscs_PPV}). We find that the SSCs have a similar distribution and kinematics as the dense molecular gas (Figure \ref{fig:sscs_PPV}). We are able to describe the SSC morphology and kinematics with a simple elliptical, angular momentum-conserving model (Figures \ref{fig:angmomconsring} and \ref{fig:angmomconsring_topdown}), which has a semi-major axis of $\sim110$~pc, a semi-minor axis of $\sim60$~pc, and an orbital period of $\approx6$~Myr. From our perspective, this ring would appear nearly edge-on, leading to the observed nearly linear SSC distribution. 
\end{enumerate}

As described in Section \ref{sssec:ages}, estimates of the (relative) SSC ages in NGC\,253 do not all agree \citep{rico-villas20,krieger20a,mills21}. Constraints on the absolute ages of the clusters are relatively weak, though the SSCs must be young ($\sim0.01-3$~Myr; \citealt{rico-villas20,mills21}). In the future, approved observations with the MIRI integral field unit onboard {\em JWST} will measure the ionizing radiation field of these clusters and hence provide independent constraints on the cluster ages. 

\begin{acknowledgments}
{\change The authors thank Keaton Donaghue for carefully checking the values in Table \ref{tab:sscparams_radialprofile_deconv}.} R.C.L. acknowledges partial support for this work provided by the National Science Foundation (NSF) Astronomy and Astrophysics Postdoctoral Fellowship under award AST-2102625 and through Student Observing Support Program (SOSP) award 7-011 from the NRAO. A.D.B. acknowledges partial support from NSF-AST2108140. {\change M.C.S. acknowledges support by the European Research Council via the ERC Synergy Grant ``ECOGAL – Understanding our Galactic ecosystem: from the disk of the Milky Way to the formation sites of stars and planets'' (grant 855130) and by the Deutsche Forschungsgemeinschaft (DFG, German Research Foundation) – Project-ID 138713538 – SFB 881 (“The Milky Way System”, subprojects B1, B2, B8).} This paper makes use of the following ALMA data: ADS/JAO.ALMA\#2017.1.00433.S and 2015.1.00274.S. ALMA is a partnership of ESO (representing its member states), NSF (USA) and NINS (Japan), together with NRC (Canada), MOST and ASIAA (Taiwan), and KASI (Republic of Korea), in cooperation with the Republic of Chile. The Joint ALMA Observatory is operated by ESO, AUI/NRAO and NAOJ. The National Radio Astronomy Observatory is a facility of the National Science Foundation operated under cooperative agreement by Associated Universities, Inc. This research has made use of NASA’s Astrophysics Data System Bibliographic Services. 
\end{acknowledgments}

\vspace{5mm}
\facilities{ALMA}

\software{adjustText\footnote{\url{https://github.com/Phlya/adjustText}} \citep{adjusttext}, Astropy \citep{astropy}, \textsc{casa} \citep{casa}, \emcee\ \citep{emcee}, imageio \citep{imageio}, MatPlotLib \citep{matplotlib}, NumPy \citep{numpy}, pandas \citep{pandas}, photutils \citep{photutils}, re \citep{re}, SciPy \citep{scipy}, seaborn \citep{seaborn}, WebPlotDigitizer \citep{webplotdigitizer}}

\appendix

\section{Milky Way CMZ Models Applied to NGC\,253}
\label{app:cmzmodels}

\begin{figure*}
    \centering
    \includegraphics[width=0.38\textwidth]{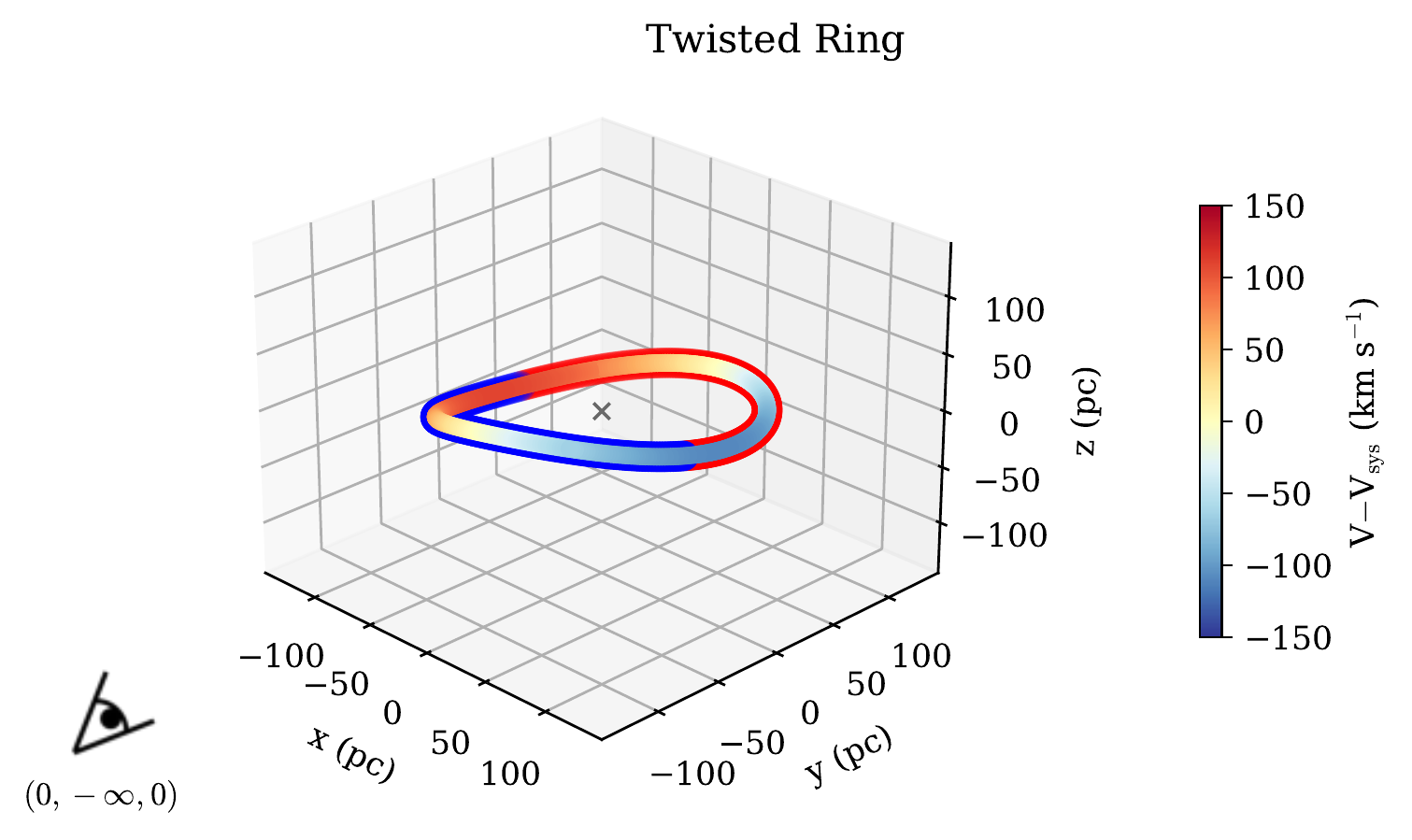}
    \includegraphics[width=0.3\textwidth]{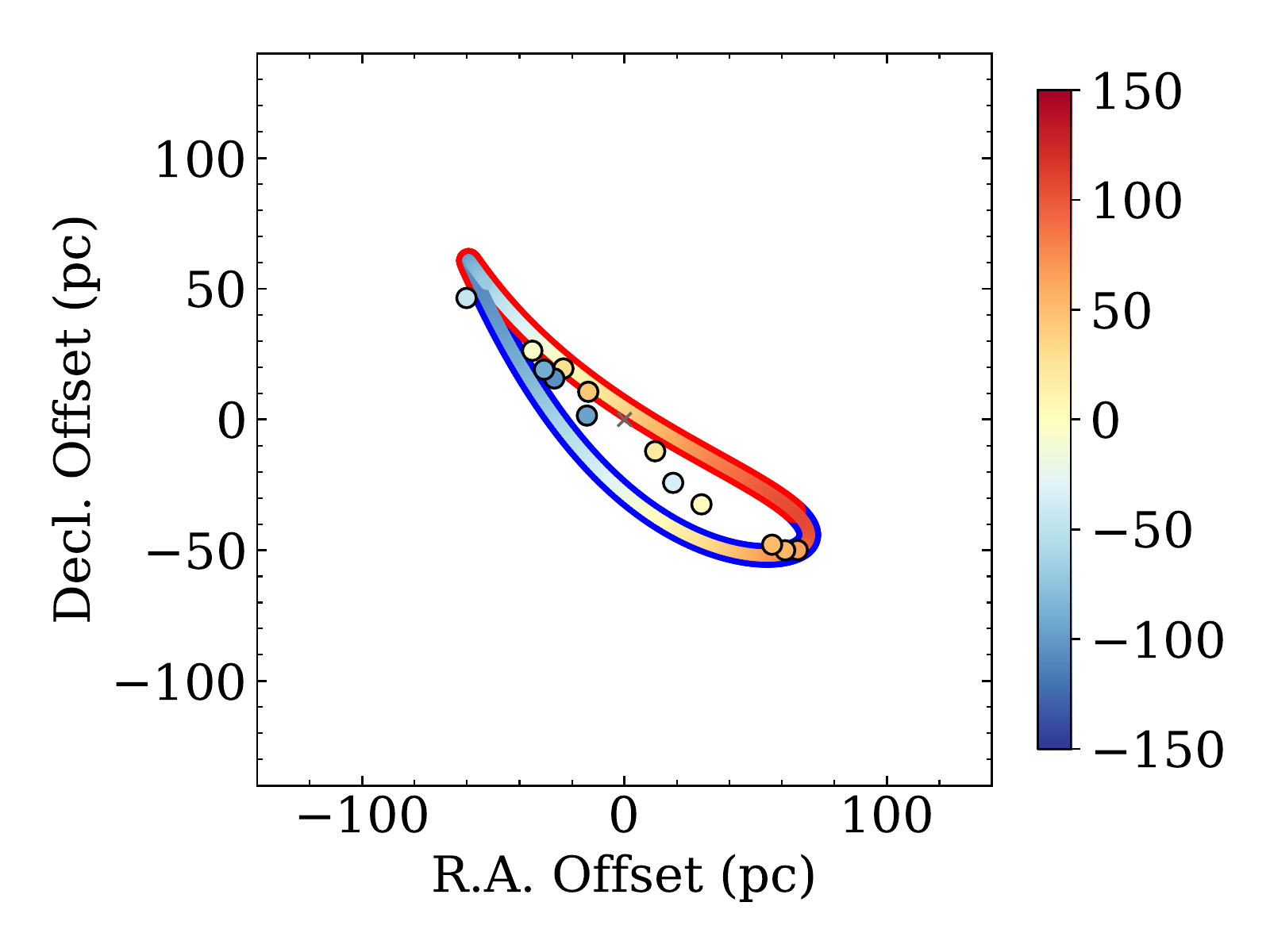}
    \includegraphics[width=0.3\textwidth]{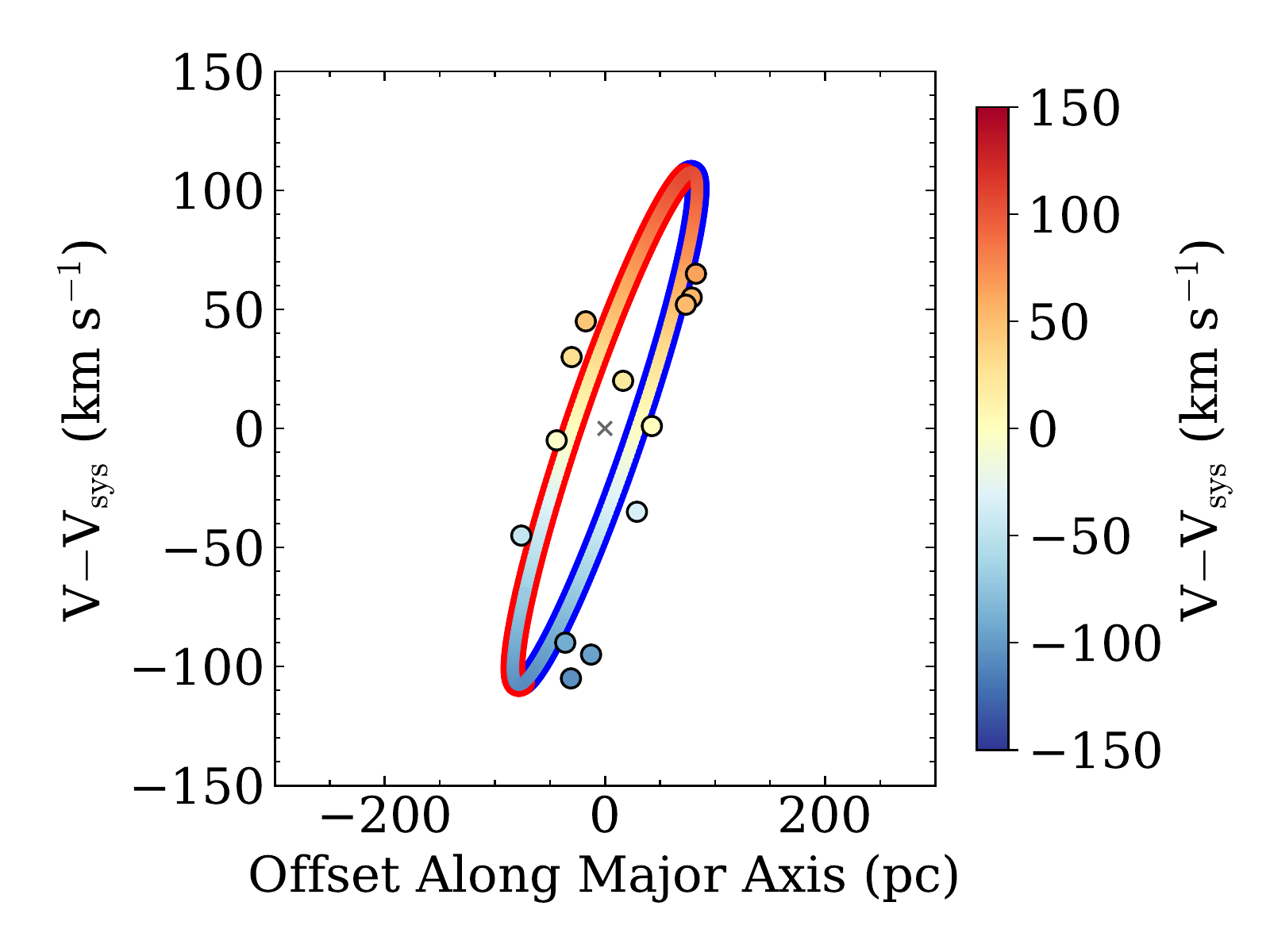}
    
    \includegraphics[width=0.38\textwidth]{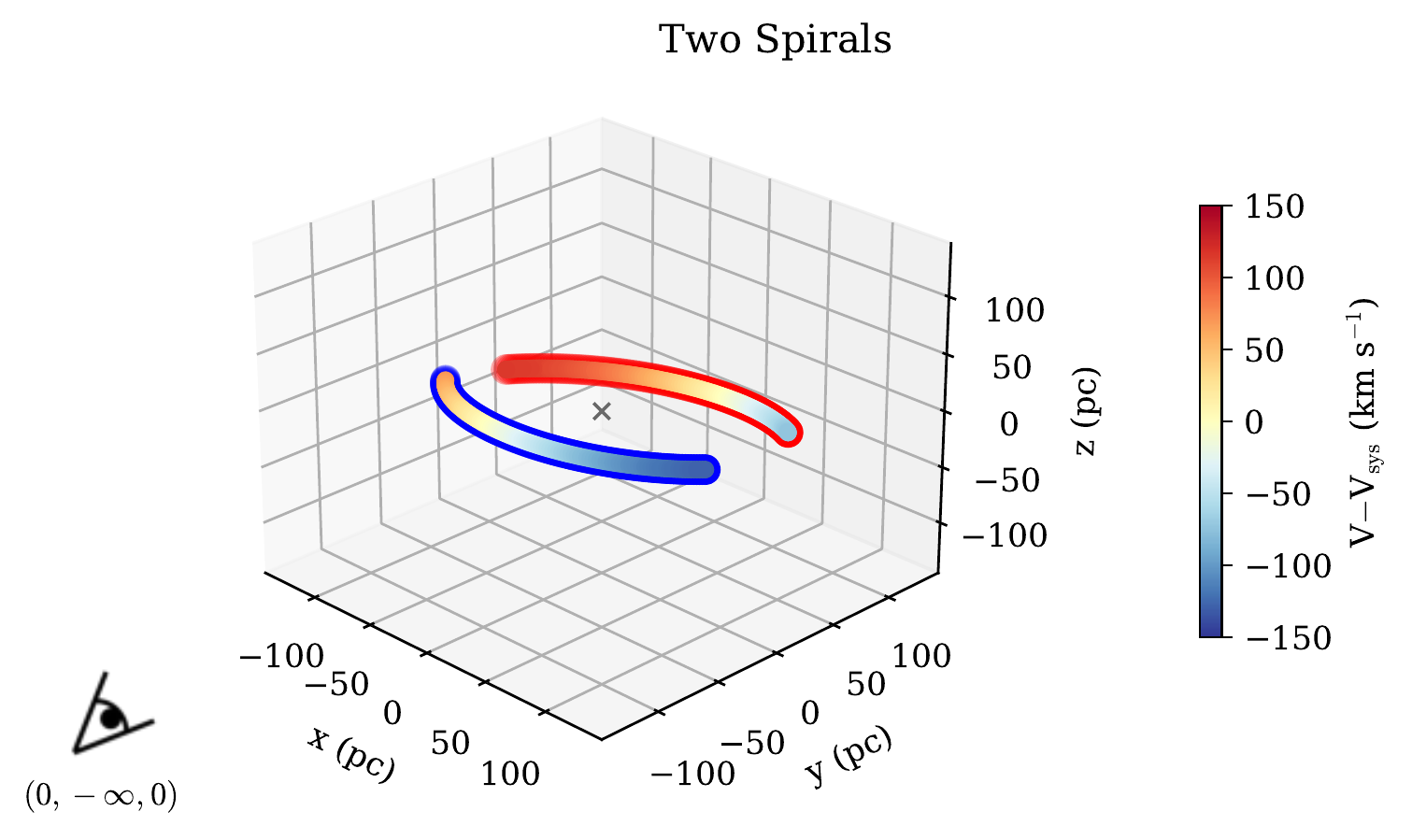}
    \includegraphics[width=0.3\textwidth]{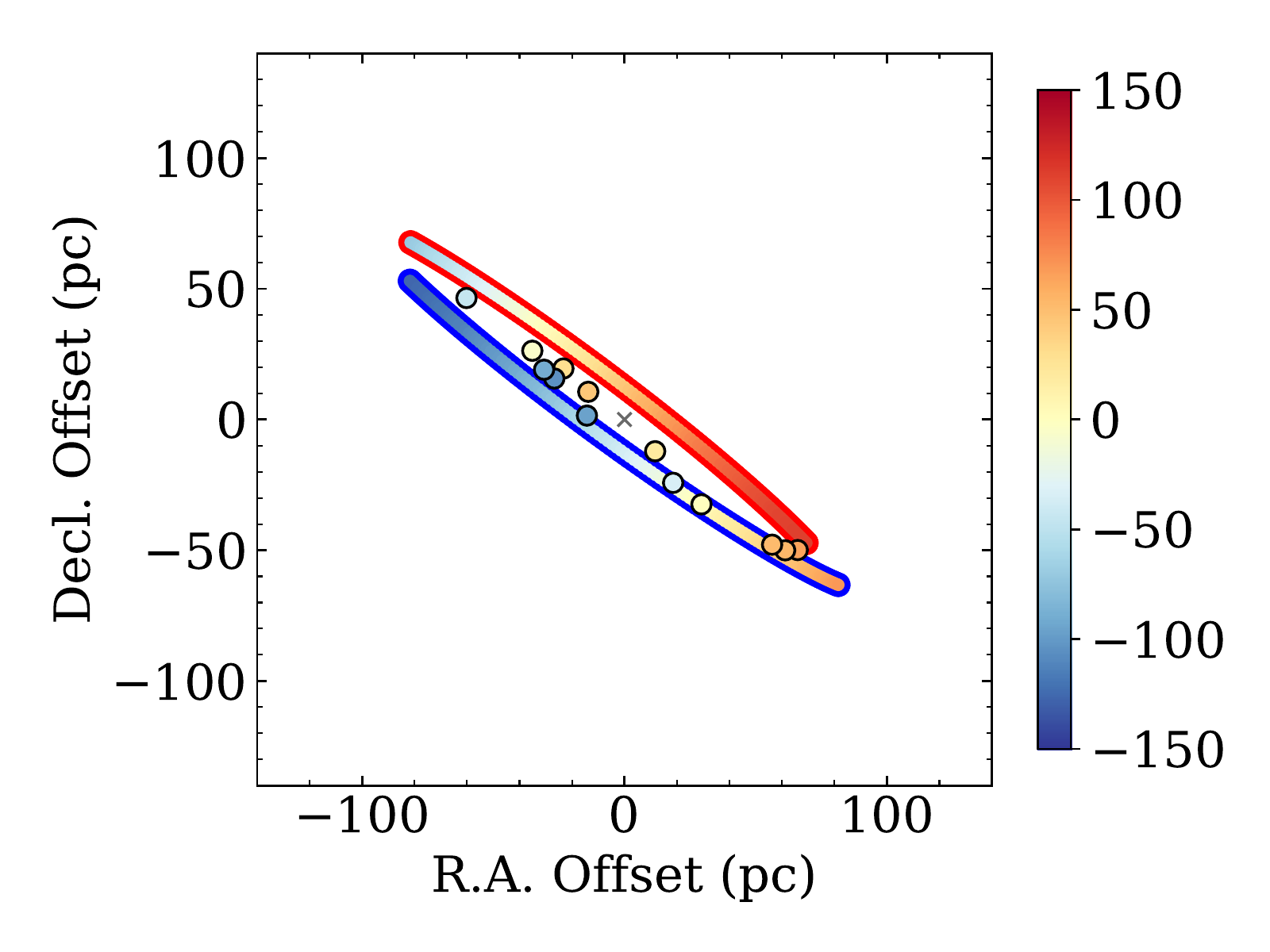}
    \includegraphics[width=0.3\textwidth]{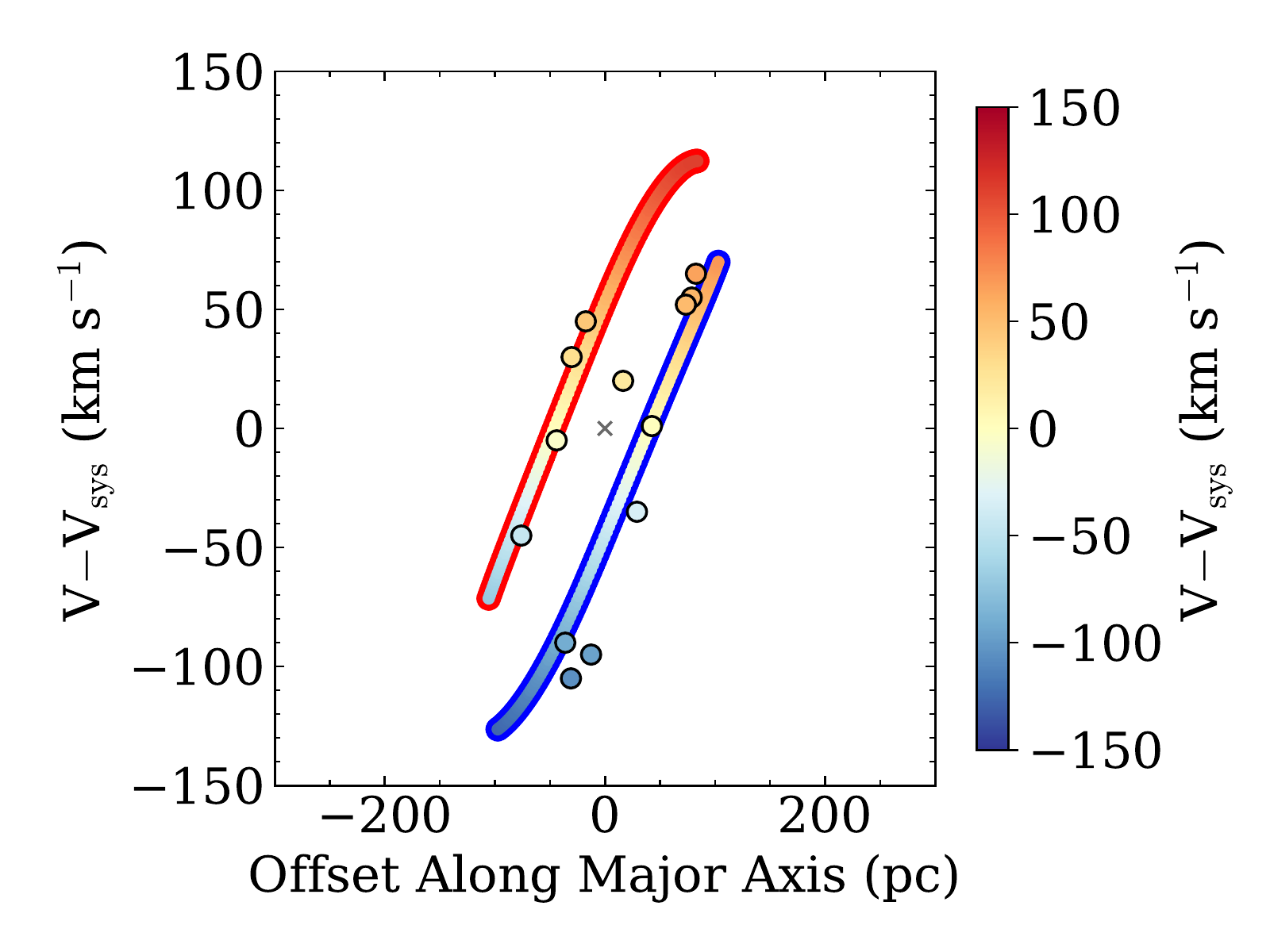}
    
    \includegraphics[width=0.38\textwidth]{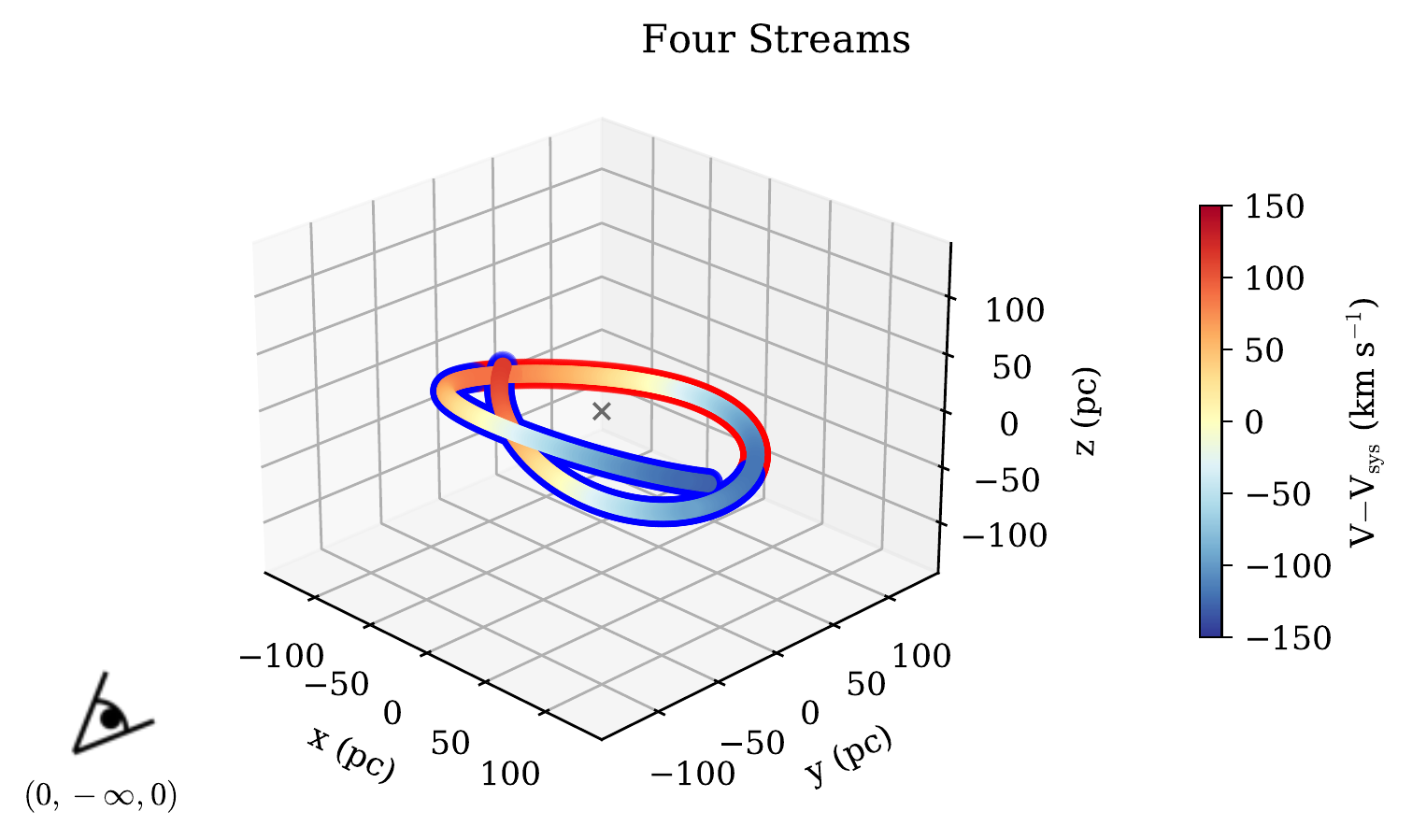}
    \includegraphics[width=0.3\textwidth]{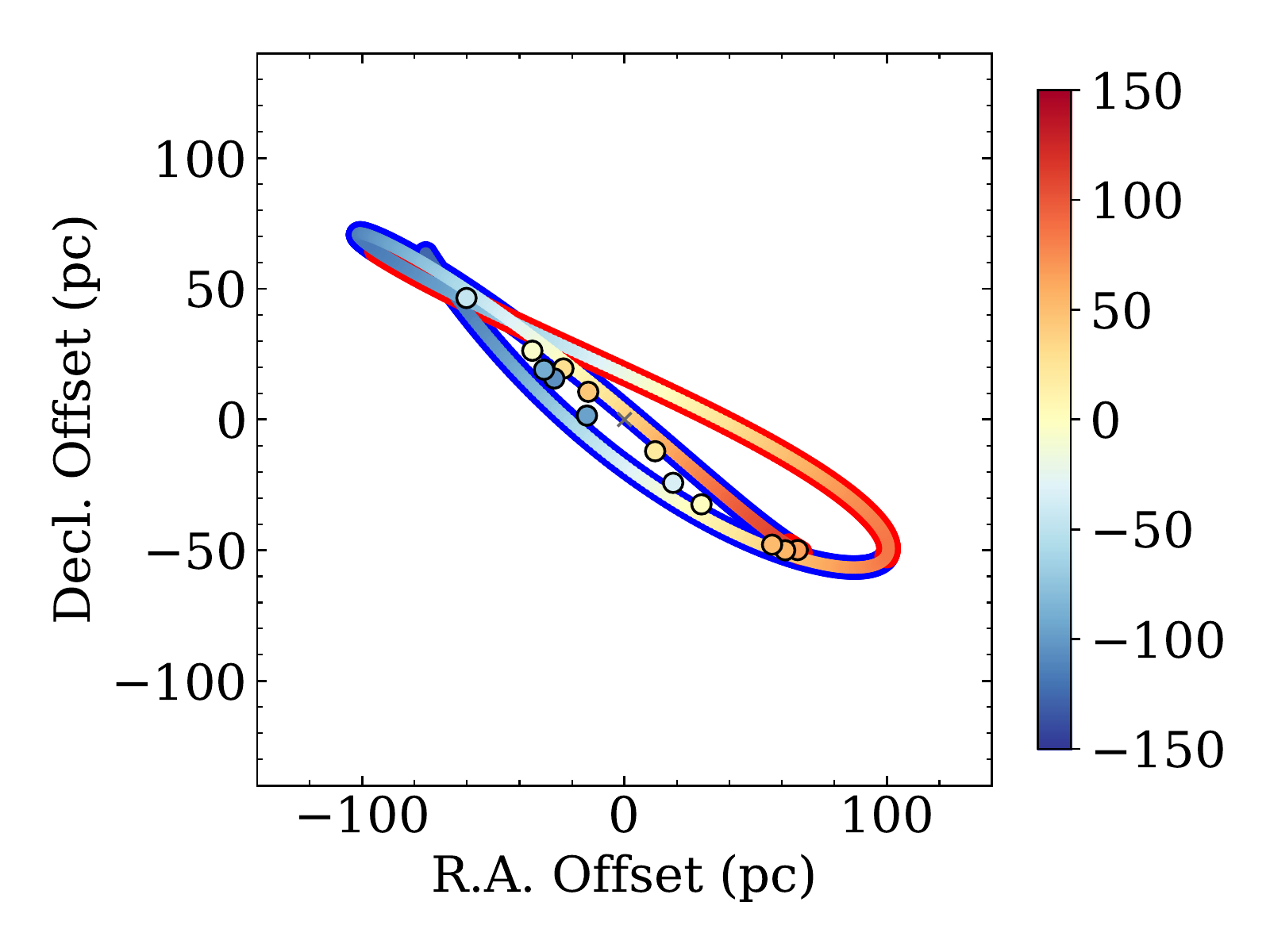}
    \includegraphics[width=0.3\textwidth]{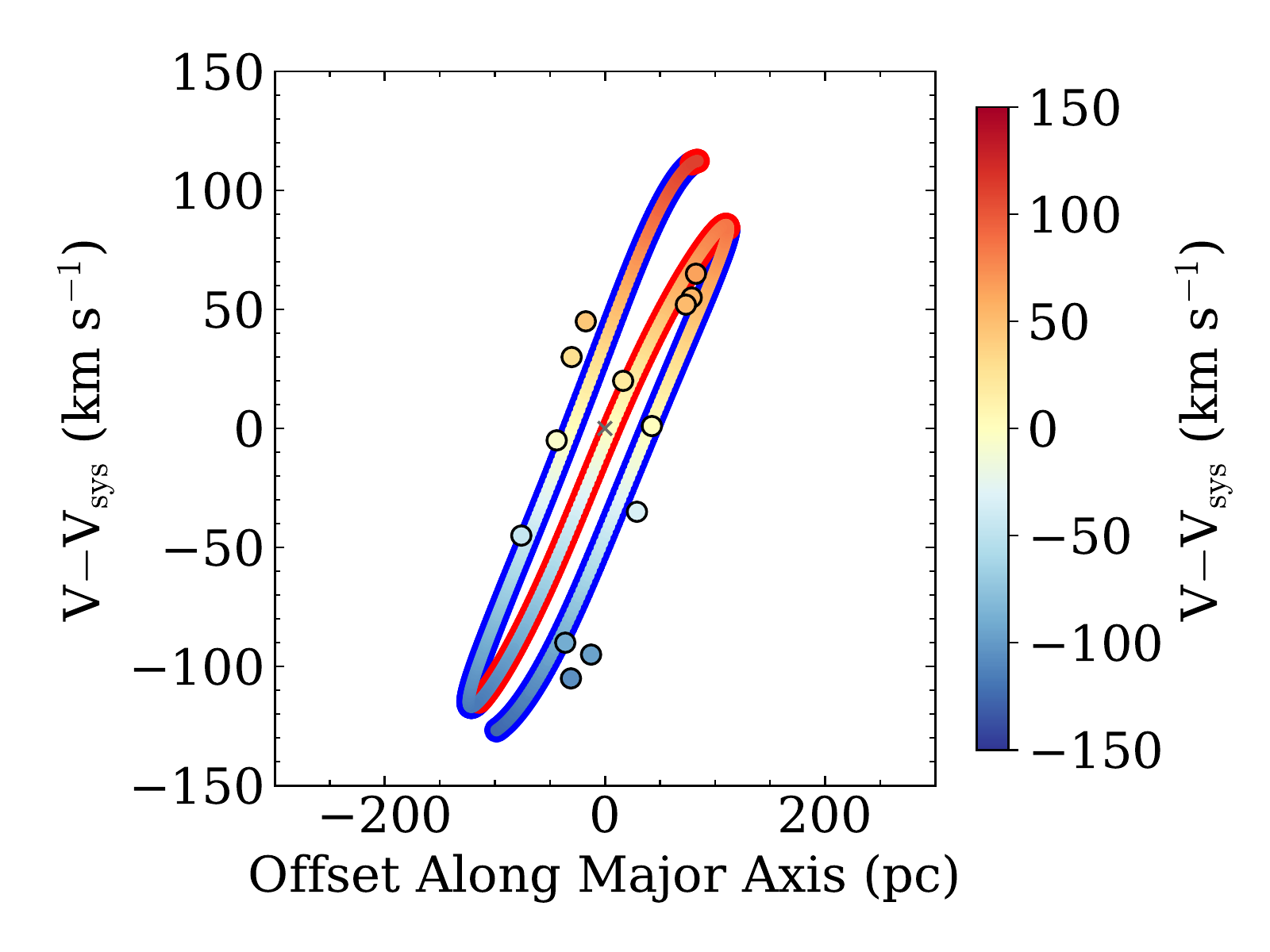}
    \caption{Similar to Figure \ref{fig:angmomconsring}, but for the CMZ models; see the caption of Figure \ref{fig:angmomconsring} for more details. The top row shows a twisted ring model \citep{molinari11}, the middle row shows two spirals \citep[e.g.,][]{sofue95,sawada04}, and the bottom row shows a crossing streams model \citep{kruijssen15}.}
    \label{fig:cmz_models}
\end{figure*}

Several models have been developed to explain the arrangement of gas and massive star-forming regions in the CMZ of the MW. Here we explore how well a na\"ive application of these models to NGC\,253 works to represent our data. To project all of the models below into the sky plane, we assume a PA~$=235$\D\ and $i=78$\D.

\subsection{Closed Orbit Models}
\label{ssec:ring}

In addition to the flat elliptical ring we explore in Section \ref{sec:morphokin}, another possible ring-like arrangement is a closed elliptical orbit with a vertical twist \citep[i.e., a twisted ring][]{molinari11}. In the CMZ of the MW, this configuration is thought to follow the \xtwo\ orbits and the twisted shape is induced by a vertical oscillation. The twisted ring model is described by a semi-major axis ($a$), a semi-minor axis ($b$), a constant orbital velocity (V$_{\rm orb}$), and a  vertical oscillation with an amplitude ($z_0$), frequency ($\nu_z$), and phase ($\theta_z$). $z_0$ is the amplitude of the oscillation, measured from the midplane to the peak (so the ring has a thickness of 2$\times z_0$. We follow \citet{molinari11} and assume that $\nu_z$ is twice the orbital frequency and $\theta_z=0$. Because this twisted ring model has a constant orbital velocity, it does not conserve angular momentum. This twisted ring model and the projections are shown in Figure \ref{fig:cmz_models} (top). The best fitting twisted ring model for the CMZ of the MW has $a=100$~pc, $b=60$~pc, $z_0=15$~pc, V$_{\rm orb}=80$~\kms, and $\theta_p=-40$\D. While such a ring in NGC\,253 would have the same major- and minor-axis lengths, the amplitude of the vertical warp is $<10$~pc. Moreover, we require a higher V$_{\rm orb}=110$~\kms\ in NGC\,253. 

\subsection{Open Orbit Models}
\label{ssec:streams}

As pointed out by \citet{kruijssen15} and others for the MW CMZ, non-circular stable closed orbits are only possible if the potential is not axisymmetric at these scales. In NGC\,253, structural modeling by \citet{leroy15a} suggests that there are asymmetries on larger scales (as is also expected for the CMZ), but their best fitting structural model is axisymmetric on the $100-150$~pc scales we study here. We note, however, that their data are not particularly constraining on these scales due to the 35~pc resolution. Therefore, since we know the SSCs are embedded in larger-scale gas structures \citep[e.g.,][]{sakamoto11,leroy15a,krieger19,krieger20a,krieger20b}, it is perhaps more likely that the SSC orbits will be open. 

One possible open-orbit model describes streams stemming from the bar-ends. Several of these kinds of models have been proposed for the CMZ in the MW \citep[e.g.,][]{sofue95,sawada04,kruijssen15,henshaw16,ridley17,henshaw22}. \citet{kruijssen15} develop a model of the MW CMZ that consists of crossing streams that form a ring-like structure. Unlike the closed ring model, this stream model is open and Sagittarius A* is located at one of the foci of the eccentric orbits. This twisted pattern precesses with time \citep[see also][]{tress20}.

First we compare against a model with two spiral arms \citep{sofue95,sawada04}. Following \citet{henshaw16}, we select two of the streams from the \citet{kruijssen15} model, and we fix their vertical positions to 0 (i.e., the streams are in the midplane). This model and the projections are shown in Figure \ref{fig:cmz_models} (middle).

Next, we compare against the full four-stream model developed by \citet{kruijssen15}. To better fit the SSCs in NGC\,253, we stretch this model in the $x$-direction by a factor of 1.5 and in velocity by a factor of 1.25, and we show this model in Figure \ref{fig:cmz_models} (bottom). Unlike any of the other models described here or in Section \ref{ssec:3dstructure}, this model does reproduce the position and velocity of SSC 7a; however, we cannot claim that this model is superior to others based on one SSC.

\bibliographystyle{aasjournal}

\begin{thebibliography}{}
\expandafter\ifx\csname natexlab\endcsname\relax\def\natexlab#1{#1}\fi
\providecommand{\url}[1]{\href{#1}{#1}}
\providecommand{\dodoi}[1]{doi:~\href{http://doi.org/#1}{\nolinkurl{#1}}}
\providecommand{\doeprint}[1]{\href{http://ascl.net/#1}{\nolinkurl{http://ascl.net/#1}}}
\providecommand{\doarXiv}[1]{\href{https://arxiv.org/abs/#1}{\nolinkurl{https://arxiv.org/abs/#1}}}

\bibitem[{{Adamo} {et~al.}(2015){Adamo}, {Kruijssen}, {Bastian}, {Silva-Villa},
  \& {Ryon}}]{adamo15}
{Adamo}, A., {Kruijssen}, J.~M.~D., {Bastian}, N., {Silva-Villa}, E., \&
  {Ryon}, J. 2015, \mnras, 452, 246, \dodoi{10.1093/mnras/stv1203}

\bibitem[{{Adamo} {et~al.}(2017){Adamo}, {Ryon}, {Messa}, {Kim}, {Grasha},
  {Cook}, {Calzetti}, {Lee}, {Whitmore}, {Elmegreen}, {Ubeda}, {Smith},
  {Bright}, {Runnholm}, {Andrews}, {Fumagalli}, {Gouliermis}, {Kahre}, {Nair},
  {Thilker}, {Walterbos}, {Wofford}, {Aloisi}, {Ashworth}, {Brown}, {Chandar},
  {Christian}, {Cignoni}, {Clayton}, {Dale}, {de Mink}, {Dobbs}, {Elmegreen},
  {Evans}, {Gallagher}, {Grebel}, {Herrero}, {Hunter}, {Johnson}, {Kennicutt},
  {Krumholz}, {Lennon}, {Levay}, {Martin}, {Nota}, {{\"O}stlin}, {Pellerin},
  {Prieto}, {Regan}, {Sabbi}, {Sacchi}, {Schaerer}, {Schiminovich}, {Shabani},
  {Tosi}, {Van Dyk}, \& {Zackrisson}}]{adamo17}
{Adamo}, A., {Ryon}, J.~E., {Messa}, M., {et~al.} 2017, \apj, 841, 131,
  \dodoi{10.3847/1538-4357/aa7132}

\bibitem[{{Anantharamaiah} \& {Goss}(1996)}]{anantharamaiah96}
{Anantharamaiah}, K.~R., \& {Goss}, W.~M. 1996, \apjl, 466, L13,
  \dodoi{10.1086/310157}

\bibitem[{{Ando} {et~al.}(2017){Ando}, {Nakanishi}, {Kohno}, {Izumi},
  {Mart{\'\i}n}, {Harada}, {Takano}, {Kuno}, {Nakai}, {Sugai}, {Sorai},
  {Tosaki}, {Matsubayashi}, {Nakajima}, {Nishimura}, \& {Tamura}}]{ando17}
{Ando}, R., {Nakanishi}, K., {Kohno}, K., {et~al.} 2017, \apj, 849, 81,
  \dodoi{10.3847/1538-4357/aa8fd4}

\bibitem[{{Astropy Collaboration} {et~al.}(2018){Astropy Collaboration},
  {Price-Whelan}, {Sip{\H{o}}cz}, {G{\"u}nther}, {Lim}, {Crawford}, {Conseil},
  {Shupe}, {Craig}, {Dencheva}, {Ginsburg}, {VanderPlas}, {Bradley},
  {P{\'e}rez-Su{\'a}rez}, {de Val-Borro}, {Aldcroft}, {Cruz}, {Robitaille},
  {Tollerud}, {Ardelean}, {Babej}, {Bach}, {Bachetti}, {Bakanov}, {Bamford},
  {Barentsen}, {Barmby}, {Baumbach}, {Berry}, {Biscani}, {Boquien}, {Bostroem},
  {Bouma}, {Brammer}, {Bray}, {Breytenbach}, {Buddelmeijer}, {Burke},
  {Calderone}, {Cano Rodr{\'\i}guez}, {Cara}, {Cardoso}, {Cheedella}, {Copin},
  {Corrales}, {Crichton}, {D'Avella}, {Deil}, {Depagne}, {Dietrich}, {Donath},
  {Droettboom}, {Earl}, {Erben}, {Fabbro}, {Ferreira}, {Finethy}, {Fox},
  {Garrison}, {Gibbons}, {Goldstein}, {Gommers}, {Greco}, {Greenfield},
  {Groener}, {Grollier}, {Hagen}, {Hirst}, {Homeier}, {Horton}, {Hosseinzadeh},
  {Hu}, {Hunkeler}, {Ivezi{\'c}}, {Jain}, {Jenness}, {Kanarek}, {Kendrew},
  {Kern}, {Kerzendorf}, {Khvalko}, {King}, {Kirkby}, {Kulkarni}, {Kumar},
  {Lee}, {Lenz}, {Littlefair}, {Ma}, {Macleod}, {Mastropietro}, {McCully},
  {Montagnac}, {Morris}, {Mueller}, {Mumford}, {Muna}, {Murphy}, {Nelson},
  {Nguyen}, {Ninan}, {N{\"o}the}, {Ogaz}, {Oh}, {Parejko}, {Parley}, {Pascual},
  {Patil}, {Patil}, {Plunkett}, {Prochaska}, {Rastogi}, {Reddy Janga},
  {Sabater}, {Sakurikar}, {Seifert}, {Sherbert}, {Sherwood-Taylor}, {Shih},
  {Sick}, {Silbiger}, {Singanamalla}, {Singer}, {Sladen}, {Sooley},
  {Sornarajah}, {Streicher}, {Teuben}, {Thomas}, {Tremblay}, {Turner},
  {Terr{\'o}n}, {van Kerkwijk}, {de la Vega}, {Watkins}, {Weaver}, {Whitmore},
  {Woillez}, {Zabalza}, \& {Astropy Contributors}}]{astropy}
{Astropy Collaboration}, {Price-Whelan}, A.~M., {Sip{\H{o}}cz}, B.~M., {et~al.}
  2018, \aj, 156, 123, \dodoi{10.3847/1538-3881/aabc4f}

\bibitem[{{Athanassoula}(1992{\natexlab{a}})}]{athanassoula92a}
{Athanassoula}, E. 1992{\natexlab{a}}, \mnras, 259, 328,
  \dodoi{10.1093/mnras/259.2.328}

\bibitem[{{Athanassoula}(1992{\natexlab{b}})}]{athanassoula92b}
---. 1992{\natexlab{b}}, \mnras, 259, 345, \dodoi{10.1093/mnras/259.2.345}

\bibitem[{{Binney} {et~al.}(1991){Binney}, {Gerhard}, {Stark}, {Bally}, \&
  {Uchida}}]{binney91}
{Binney}, J., {Gerhard}, O.~E., {Stark}, A.~A., {Bally}, J., \& {Uchida}, K.~I.
  1991, \mnras, 252, 210, \dodoi{10.1093/mnras/252.2.210}

\bibitem[{{Binney} \& {Tremaine}(2008)}]{binney08}
{Binney}, J., \& {Tremaine}, S. 2008, {Galactic Dynamics: Second Edition}
  (Princeton University Press)

\bibitem[{{B{\"o}ker} {et~al.}(2008){B{\"o}ker}, {Falc{\'o}n-Barroso},
  {Schinnerer}, {Knapen}, \& {Ryder}}]{boker08}
{B{\"o}ker}, T., {Falc{\'o}n-Barroso}, J., {Schinnerer}, E., {Knapen}, J.~H.,
  \& {Ryder}, S. 2008, \aj, 135, 479, \dodoi{10.1088/0004-6256/135/2/479}

\bibitem[{{Bolatto} {et~al.}(2013){Bolatto}, {Warren}, {Leroy}, {Walter},
  {Veilleux}, {Ostriker}, {Ott}, {Zwaan}, {Fisher}, {Weiss}, {Rosolowsky}, \&
  {Hodge}}]{bolatto13a}
{Bolatto}, A.~D., {Warren}, S.~R., {Leroy}, A.~K., {et~al.} 2013, \nat, 499,
  450, \dodoi{10.1038/nature12351}

\bibitem[{{Bolatto} {et~al.}(2021){Bolatto}, {Leroy}, {Levy}, {Meier}, {Mills},
  {Thompson}, {Emig}, {Veilleux}, {Ott}, {Gorski}, {Walter}, {Lopez}, \&
  {Lenki{\'c}}}]{bolatto21}
{Bolatto}, A.~D., {Leroy}, A.~K., {Levy}, R.~C., {et~al.} 2021, \apj, 923, 83,
  \dodoi{10.3847/1538-4357/ac2c08}

\bibitem[{Bradley {et~al.}(2021)Bradley, Sipocz, Robitaille, Tollerud,
  Vin\'icius, Deil, Barbary, Wilson, Busko, GÃŒnther, Cara, Conseil,
  Bostroem, Droettboom, Bray, Bratholm, Lim, Barentsen, Craig, Rathi, Pascual,
  Perren, Donath, Georgiev, de~Val-Borro, Kerzendorf, Bach, Quint, Souchereau,
  \& Weaver}]{photutils}
Bradley, L., Sipocz, B., Robitaille, T., {et~al.} 2021, astropy/photutils:
  1.0.2, 1.0.2,  Zenodo, \dodoi{10.5281/zenodo.4453725}

\bibitem[{{Brown} \& {Gnedin}(2021)}]{brown21}
{Brown}, G., \& {Gnedin}, O.~Y. 2021, \mnras, 508, 5935,
  \dodoi{10.1093/mnras/stab2907}

\bibitem[{{Buta} \& {Combes}(1996)}]{buta96}
{Buta}, R., \& {Combes}, F. 1996, \fcp, 17, 95

\bibitem[{{Calzetti} {et~al.}(2015){Calzetti}, {Lee}, {Sabbi}, {Adamo},
  {Smith}, {Andrews}, {Ubeda}, {Bright}, {Thilker}, {Aloisi}, {Brown},
  {Chandar}, {Christian}, {Cignoni}, {Clayton}, {da Silva}, {de Mink}, {Dobbs},
  {Elmegreen}, {Elmegreen}, {Evans}, {Fumagalli}, {Gallagher}, {Gouliermis},
  {Grebel}, {Herrero}, {Hunter}, {Johnson}, {Kennicutt}, {Kim}, {Krumholz},
  {Lennon}, {Levay}, {Martin}, {Nair}, {Nota}, {{\"O}stlin}, {Pellerin},
  {Prieto}, {Regan}, {Ryon}, {Schaerer}, {Schiminovich}, {Tosi}, {Van Dyk},
  {Walterbos}, {Whitmore}, \& {Wofford}}]{calzetti15}
{Calzetti}, D., {Lee}, J.~C., {Sabbi}, E., {et~al.} 2015, \aj, 149, 51,
  \dodoi{10.1088/0004-6256/149/2/51}

\bibitem[{Caswell {et~al.}(2020)Caswell, Droettboom, Lee, Hunter, de~Andrade,
  Firing, Hoffmann, Klymak, Stansby, Varoquaux, Nielsen, Root, May, Elson,
  SeppÃ€nen, Dale, Lee, McDougall, Straw, Hobson, Gohlke, Yu, Ma, Vincent,
  Silvester, Moad, Kniazev, hannah, Ernest, \& Ivanov}]{matplotlib}
Caswell, T.~A., Droettboom, M., Lee, A., {et~al.} 2020, matplotlib/matplotlib:
  REL: v3.3.2, v3.3.2,  Zenodo, \dodoi{10.5281/zenodo.4030140}

\bibitem[{{Chandar} {et~al.}(2017){Chandar}, {Fall}, {Whitmore}, \&
  {Mulia}}]{chandar17}
{Chandar}, R., {Fall}, S.~M., {Whitmore}, B.~C., \& {Mulia}, A.~J. 2017, \apj,
  849, 128, \dodoi{10.3847/1538-4357/aa92ce}

\bibitem[{{Contopoulos} \& {Grosbol}(1989)}]{contopoulos89}
{Contopoulos}, G., \& {Grosbol}, P. 1989, \aapr, 1, 261,
  \dodoi{10.1007/BF00873080}

\bibitem[{{Contopoulos} \& {Mertzanides}(1977)}]{contopoulos77}
{Contopoulos}, G., \& {Mertzanides}, C. 1977, \aap, 61, 477

\bibitem[{{Das} {et~al.}(2001){Das}, {Anantharamaiah}, \& {Yun}}]{das01}
{Das}, M., {Anantharamaiah}, K.~R., \& {Yun}, M.~S. 2001, \apj, 549, 896,
  \dodoi{10.1086/319430}

\bibitem[{{Davis} {et~al.}(2013){Davis}, {Bayet}, {Crocker}, {Topal}, \&
  {Bureau}}]{davis13}
{Davis}, T.~A., {Bayet}, E., {Crocker}, A., {Topal}, S., \& {Bureau}, M. 2013,
  \mnras, 433, 1659, \dodoi{10.1093/mnras/stt842}

\bibitem[{{Emig} {et~al.}(2020){Emig}, {Bolatto}, {Leroy}, {Mills},
  {Jim{\'e}nez Donaire}, {Tielens}, {Ginsburg}, {Gorski}, {Krieger}, {Levy},
  {Meier}, {Ott}, {Rosolowsky}, {Thompson}, \& {Veilleux}}]{emig20}
{Emig}, K.~L., {Bolatto}, A.~D., {Leroy}, A.~K., {et~al.} 2020, \apj, 903, 50,
  \dodoi{10.3847/1538-4357/abb67d}

\bibitem[{{Fall} \& {Chandar}(2012)}]{fall12}
{Fall}, S.~M., \& {Chandar}, R. 2012, \apj, 752, 96,
  \dodoi{10.1088/0004-637X/752/2/96}

\bibitem[{Flyamer {et~al.}(2020)Flyamer, Weber/GwendalD, Xue, Colin, Li, Neste,
  Espinoza, Morshed, Vazquez, Neff, mski\_iksm, \& scaine1}]{adjusttext}
Flyamer, I., Weber/GwendalD, S., Xue, Z., {et~al.} 2020, Phlya/adjustText: 0.8
  beta, 0.8beta2,  Zenodo, \dodoi{10.5281/zenodo.3924114}

\bibitem[{{Foreman-Mackey} {et~al.}(2013){Foreman-Mackey}, {Hogg}, {Lang}, \&
  {Goodman}}]{emcee}
{Foreman-Mackey}, D., {Hogg}, D.~W., {Lang}, D., \& {Goodman}, J. 2013, \pasp,
  125, 306, \dodoi{10.1086/670067}

\bibitem[{{Galliano} {et~al.}(2008){Galliano}, {Dwek}, \&
  {Chanial}}]{galliano08}
{Galliano}, F., {Dwek}, E., \& {Chanial}, P. 2008, \apj, 672, 214,
  \dodoi{10.1086/523621}

\bibitem[{{Gao} \& {Solomon}(2004)}]{gao04}
{Gao}, Y., \& {Solomon}, P.~M. 2004, \apjs, 152, 63, \dodoi{10.1086/383003}

\bibitem[{{Gieles} {et~al.}(2006){Gieles}, {Larsen}, {Scheepmaker}, {Bastian},
  {Haas}, \& {Lamers}}]{gieles06}
{Gieles}, M., {Larsen}, S.~S., {Scheepmaker}, R.~A., {et~al.} 2006, \aap, 446,
  L9, \dodoi{10.1051/0004-6361:200500224}

\bibitem[{{Goldreich} \& {Tremaine}(1979)}]{goldreich79}
{Goldreich}, P., \& {Tremaine}, S. 1979, \apj, 233, 857, \dodoi{10.1086/157448}

\bibitem[{{Gorski} {et~al.}(2017){Gorski}, {Ott}, {Rand}, {Meier}, {Momjian},
  \& {Schinnerer}}]{gorski17}
{Gorski}, M., {Ott}, J., {Rand}, R., {et~al.} 2017, \apj, 842, 124,
  \dodoi{10.3847/1538-4357/aa74af}

\bibitem[{{Gorski} {et~al.}(2019){Gorski}, {Ott}, {Rand}, {Meier}, {Momjian},
  {Schinnerer}, \& {Ellingsen}}]{gorski19}
{Gorski}, M.~D., {Ott}, J., {Rand}, R., {et~al.} 2019, \mnras, 483, 5434,
  \dodoi{10.1093/mnras/sty3077}

\bibitem[{{Grudi{\'c}} {et~al.}(2021){Grudi{\'c}}, {Guszejnov}, {Hopkins},
  {Offner}, \& {Faucher-Gigu{\`e}re}}]{grudic21}
{Grudi{\'c}}, M.~Y., {Guszejnov}, D., {Hopkins}, P.~F., {Offner}, S. S.~R., \&
  {Faucher-Gigu{\`e}re}, C.-A. 2021, \mnras, 506, 2199,
  \dodoi{10.1093/mnras/stab1347}

\bibitem[{{Haasler} {et~al.}(2022){Haasler}, {Rivilla}, {Mart{\'\i}n},
  {Holdship}, {Viti}, {Harada}, {Mangum}, {Sakamoto}, {Muller}, {Tanaka},
  {Yoshimura}, {Nakanishi}, {Colzi}, {Hunt}, {Emig}, {Aladro}, {Humire},
  {Henkel}, \& {van der Werf}}]{haasler22}
{Haasler}, D., {Rivilla}, V.~M., {Mart{\'\i}n}, S., {et~al.} 2022, \aap, 659,
  A158, \dodoi{10.1051/0004-6361/202142032}

\bibitem[{Harris {et~al.}(2020)Harris, Millman, van~der Walt, Gommers,
  Virtanen, Cournapeau, Wieser, Taylor, Berg, Smith, Kern, Picus, Hoyer, van
  Kerkwijk, Brett, Haldane, Fern\'andez~del R\'io, Wiebe, Peterson,
  GÃ©rard-Marchant, Sheppard, Reddy, Weckesser, Abbasi, Gohlke, \&
  Oliphant}]{numpy}
Harris, C.~R., Millman, K.~J., van~der Walt, S.~J., {et~al.} 2020, Nature, 585,
  357, \dodoi{10.1038/s41586-020-2649-2}

\bibitem[{{Heckman} {et~al.}(2000){Heckman}, {Lehnert}, {Strickland }, \&
  {Armus}}]{heckman00}
{Heckman}, T.~M., {Lehnert}, M.~D., {Strickland }, D.~K., \& {Armus}, L. 2000,
  \apjs, 129, 493, \dodoi{10.1086/313421}

\bibitem[{{Heesen} {et~al.}(2011){Heesen}, {Beck}, {Krause}, \&
  {Dettmar}}]{heesen11}
{Heesen}, V., {Beck}, R., {Krause}, M., \& {Dettmar}, R.~J. 2011, \aap, 535,
  A79, \dodoi{10.1051/0004-6361/201117618}

\bibitem[{Henshaw {et~al.}(2022)Henshaw, Barnes, Battersby, Ginsburg, Sormani,
  \& Walker}]{henshaw22}
Henshaw, J.~D., Barnes, A.~T., Battersby, C., {et~al.} 2022

\bibitem[{{Henshaw} {et~al.}(2016){Henshaw}, {Longmore}, {Kruijssen}, {Davies},
  {Bally}, {Barnes}, {Battersby}, {Burton}, {Cunningham}, {Dale}, {Ginsburg},
  {Immer}, {Jones}, {Kendrew}, {Mills}, {Molinari}, {Moore}, {Ott}, {Pillai},
  {Rathborne}, {Schilke}, {Schmiedeke}, {Testi}, {Walker}, {Walsh}, \&
  {Zhang}}]{henshaw16}
{Henshaw}, J.~D., {Longmore}, S.~N., {Kruijssen}, J.~M.~D., {et~al.} 2016,
  \mnras, 457, 2675, \dodoi{10.1093/mnras/stw121}

\bibitem[{{Hollyhead} {et~al.}(2016){Hollyhead}, {Adamo}, {Bastian}, {Gieles},
  \& {Ryon}}]{hollyhead16}
{Hollyhead}, K., {Adamo}, A., {Bastian}, N., {Gieles}, M., \& {Ryon}, J.~E.
  2016, \mnras, 460, 2087, \dodoi{10.1093/mnras/stw1142}

\bibitem[{{Humire} {et~al.}(2022){Humire}, {Henkel}, {Hern{\'a}ndez-G{\'o}mez},
  {Mart{\'\i}n}, {Mangum}, {Harada}, {Muller}, {Sakamoto}, {Tanaka},
  {Yoshimura}, {Nakanishi}, {M{\"u}hle}, {Herrero-Illana}, {Meier}, {Caux},
  {Aladro}, {Mauersberger}, {Viti}, {Colzi}, {Rivilla}, {Gorski}, {Menten},
  {Huang}, {Aalto}, {van der Werf}, \& {Emig}}]{humire22}
{Humire}, P.~K., {Henkel}, C., {Hern{\'a}ndez-G{\'o}mez}, A., {et~al.} 2022,
  arXiv e-prints, arXiv:2205.03281.
\newblock \doarXiv{2205.03281}

\bibitem[{{Johnson} {et~al.}(2017){Johnson}, {Seth}, {Dalcanton}, {Beerman},
  {Fouesneau}, {Weisz}, {Bell}, {Dolphin}, {Sandstrom}, \&
  {Williams}}]{johnson17}
{Johnson}, L.~C., {Seth}, A.~C., {Dalcanton}, J.~J., {et~al.} 2017, \apj, 839,
  78, \dodoi{10.3847/1538-4357/aa6a1f}

\bibitem[{{King}(1962)}]{king62}
{King}, I. 1962, \aj, 67, 471, \dodoi{10.1086/108756}

\bibitem[{Klein {et~al.}(2018)Klein, Silvester, Tanbakuchi, Müller,
  Nunez-Iglesias, Harfouche, McCormick, Rai, OrganicIrradiation, Smith,
  Konowalczyk, rreilink, Nises, jackwalker64, Vaillant, Zulko,
  NiklasRosenstein, Michael~Hirsch, Schambach, Hugo, Kohlgrüber, Dusold, Lee,
  PeterMinin, Schwabedal, Dennis, Barnes, Lee, Levskaya, \& Elliott}]{imageio}
Klein, A., Silvester, S., Tanbakuchi, A., {et~al.} 2018, imageio/imageio:
  V2.4.1, v2.4.1,  Zenodo, \dodoi{10.5281/zenodo.1488562}

\bibitem[{{Knapen}(1999)}]{knapen99}
{Knapen}, J.~H. 1999, in Astronomical Society of the Pacific Conference Series,
  Vol. 187, The Evolution of Galaxies on Cosmological Timescales, ed. J.~E.
  {Beckman} \& T.~J. {Mahoney}, 72--87.
\newblock \doarXiv{astro-ph/9907290}

\bibitem[{{Koribalski} {et~al.}(2004){Koribalski}, {Staveley-Smith}, {Kilborn},
  {Ryder}, {Kraan-Korteweg}, {Ryan-Weber}, {Ekers}, {Jerjen}, {Henning},
  {Putman}, {Zwaan}, {de Blok}, {Calabretta}, {Disney}, {Minchin}, {Bhathal},
  {Boyce}, {Drinkwater}, {Freeman}, {Gibson}, {Green}, {Haynes}, {Juraszek},
  {Kesteven}, {Knezek}, {Mader}, {Marquarding}, {Meyer}, {Mould}, {Oosterloo},
  {O'Brien}, {Price}, {Sadler}, {Schr{\"o}der}, {Stewart}, {Stootman}, {Waugh},
  {Warren}, {Webster}, \& {Wright}}]{koribalski04}
{Koribalski}, B.~S., {Staveley-Smith}, L., {Kilborn}, V.~A., {et~al.} 2004,
  \aj, 128, 16, \dodoi{10.1086/421744}

\bibitem[{{Kormendy} \& {Kennicutt}(2004)}]{kormendy04}
{Kormendy}, J., \& {Kennicutt}, Robert~C., J. 2004, \araa, 42, 603,
  \dodoi{10.1146/annurev.astro.42.053102.134024}

\bibitem[{{Kornei} \& {McCrady}(2009)}]{kornei09}
{Kornei}, K.~A., \& {McCrady}, N. 2009, \apj, 697, 1180,
  \dodoi{10.1088/0004-637X/697/2/1180}

\bibitem[{{Krieger} {et~al.}(2019){Krieger}, {Bolatto}, {Walter}, {Leroy},
  {Zschaechner}, {Meier}, {Ott}, {Weiss}, {Mills}, {Levy}, {Veilleux}, \&
  {Gorski}}]{krieger19}
{Krieger}, N., {Bolatto}, A.~D., {Walter}, F., {et~al.} 2019, \apj, 881, 43,
  \dodoi{10.3847/1538-4357/ab2d9c}

\bibitem[{{Krieger} {et~al.}(2020{\natexlab{a}}){Krieger}, {Bolatto}, {Leroy},
  {Levy}, {Mills}, {Meier}, {Ott}, {Veilleux}, {Walter}, \&
  {Wei{\ss}}}]{krieger20a}
{Krieger}, N., {Bolatto}, A.~D., {Leroy}, A.~K., {et~al.} 2020{\natexlab{a}},
  \apj, 897, 176, \dodoi{10.3847/1538-4357/ab9c23}

\bibitem[{{Krieger} {et~al.}(2020{\natexlab{b}}){Krieger}, {Bolatto}, {Koch},
  {Leroy}, {Rosolowsky}, {Walter}, {Wei{\ss}}, {Eden}, {Levy}, {Meier},
  {Mills}, {Moore}, {Ott}, {Su}, \& {Veilleux}}]{krieger20b}
{Krieger}, N., {Bolatto}, A.~D., {Koch}, E.~W., {et~al.} 2020{\natexlab{b}},
  \apj, 899, 158, \dodoi{10.3847/1538-4357/aba903}

\bibitem[{{Kruijssen} {et~al.}(2015){Kruijssen}, {Dale}, \&
  {Longmore}}]{kruijssen15}
{Kruijssen}, J.~M.~D., {Dale}, J.~E., \& {Longmore}, S.~N. 2015, \mnras, 447,
  1059, \dodoi{10.1093/mnras/stu2526}

\bibitem[{{Krumholz} \& {Kruijssen}(2015)}]{krumholz15}
{Krumholz}, M.~R., \& {Kruijssen}, J.~M.~D. 2015, \mnras, 453, 739,
  \dodoi{10.1093/mnras/stv1670}

\bibitem[{{Krumholz} {et~al.}(2019){Krumholz}, {McKee}, \& {Bland
  -Hawthorn}}]{krumholz19}
{Krumholz}, M.~R., {McKee}, C.~F., \& {Bland -Hawthorn}, J. 2019, \araa, 57,
  227, \dodoi{10.1146/annurev-astro-091918-104430}

\bibitem[{{Leroy} {et~al.}(2015{\natexlab{a}}){Leroy}, {Bolatto}, {Ostriker},
  {Rosolowsky}, {Walter}, {Warren}, {Donovan Meyer}, {Hodge}, {Meier}, {Ott},
  {Sandstrom}, {Schruba}, {Veilleux}, \& {Zwaan}}]{leroy15a}
{Leroy}, A.~K., {Bolatto}, A.~D., {Ostriker}, E.~C., {et~al.}
  2015{\natexlab{a}}, \apj, 801, 25, \dodoi{10.1088/0004-637X/801/1/25}

\bibitem[{{Leroy} {et~al.}(2015{\natexlab{b}}){Leroy}, {Walter}, {Martini},
  {Roussel}, {Sandstrom}, {Ott}, {Weiss}, {Bolatto}, {Schuster}, \&
  {Dessauges-Zavadsky}}]{leroy15b}
{Leroy}, A.~K., {Walter}, F., {Martini}, P., {et~al.} 2015{\natexlab{b}}, \apj,
  814, 83, \dodoi{10.1088/0004-637X/814/2/83}

\bibitem[{{Leroy} {et~al.}(2018){Leroy}, {Bolatto}, {Ostriker}, {Walter},
  {Gorski}, {Ginsburg}, {Krieger}, {Levy}, {Meier}, {Mills}, {Ott},
  {Rosolowsky}, {Thompson}, {Veilleux}, \& {Zschaechner}}]{leroy18}
{Leroy}, A.~K., {Bolatto}, A.~D., {Ostriker}, E.~C., {et~al.} 2018, \apj, 869,
  126, \dodoi{10.3847/1538-4357/aaecd1}

\bibitem[{{Levy} {et~al.}(2021){Levy}, {Bolatto}, {Leroy}, {Emig}, {Gorski},
  {Krieger}, {Lenki{\'c}}, {Meier}, {Mills}, {Ott}, {Rosolowsky}, {Tarantino},
  {Veilleux}, {Walter}, {Wei{\ss}}, \& {Zwaan}}]{levy21}
{Levy}, R.~C., {Bolatto}, A.~D., {Leroy}, A.~K., {et~al.} 2021, \apj, 912, 4,
  \dodoi{10.3847/1538-4357/abec84}

\bibitem[{{Longmore} {et~al.}(2013){Longmore}, {Kruijssen}, {Bally}, {Ott},
  {Testi}, {Rathborne}, {Bastian}, {Bressert}, {Molinari}, {Battersby}, \&
  {Walsh}}]{longmore13}
{Longmore}, S.~N., {Kruijssen}, J.~M.~D., {Bally}, J., {et~al.} 2013, \mnras,
  433, L15, \dodoi{10.1093/mnrasl/slt048}

\bibitem[{{Mangum} {et~al.}(2013){Mangum}, {Darling}, {Henkel}, {Menten},
  {MacGregor}, {Svoboda}, \& {Schinnerer}}]{mangum13}
{Mangum}, J.~G., {Darling}, J., {Henkel}, C., {et~al.} 2013, \apj, 779, 33,
  \dodoi{10.1088/0004-637X/779/1/33}

\bibitem[{{Mart{\'\i}n} {et~al.}(2021){Mart{\'\i}n}, {Mangum}, {Harada},
  {Costagliola}, {Sakamoto}, {Muller}, {Aladro}, {Tanaka}, {Yoshimura},
  {Nakanishi}, {Herrero-Illana}, {M{\"u}hle}, {Aalto}, {Behrens}, {Colzi},
  {Emig}, {Fuller}, {Garc{\'\i}a-Burillo}, {Greve}, {Henkel}, {Holdship},
  {Humire}, {Hunt}, {Izumi}, {Kohno}, {K{\"o}nig}, {Meier}, {Nakajima},
  {Nishimura}, {Padovani}, {Rivilla}, {Takano}, {van der Werf}, {Viti}, \&
  {Yan}}]{martin21}
{Mart{\'\i}n}, S., {Mangum}, J.~G., {Harada}, N., {et~al.} 2021, \aap, 656,
  A46, \dodoi{10.1051/0004-6361/202141567}

\bibitem[{{Mazzuca} {et~al.}(2008){Mazzuca}, {Knapen}, {Veilleux}, \&
  {Regan}}]{mazzuca08}
{Mazzuca}, L.~M., {Knapen}, J.~H., {Veilleux}, S., \& {Regan}, M.~W. 2008,
  \apjs, 174, 337, \dodoi{10.1086/522338}

\bibitem[{{McCormick} {et~al.}(2013){McCormick}, {Veilleux}, \&
  {Rupke}}]{mccormick13}
{McCormick}, A., {Veilleux}, S., \& {Rupke}, D. S.~N. 2013, \apj, 774, 126,
  \dodoi{10.1088/0004-637X/774/2/126}

\bibitem[{{McMullin} {et~al.}(2007){McMullin}, {Waters}, {Schiebel}, {Young},
  \& {Golap}}]{casa}
{McMullin}, J.~P., {Waters}, B., {Schiebel}, D., {Young}, W., \& {Golap}, K.
  2007, in Astronomical Society of the Pacific Conference Series, Vol. 376,
  Astronomical Data Analysis Software and Systems XVI, ed. R.~A. {Shaw},
  F.~{Hill}, \& D.~J. {Bell}, 127

\bibitem[{{Meier} {et~al.}(2015){Meier}, {Walter}, {Bolatto}, {Leroy}, {Ott},
  {Rosolowsky}, {Veilleux}, {Warren}, {Wei{\ss}}, {Zwaan}, \&
  {Zschaechner}}]{meier15}
{Meier}, D.~S., {Walter}, F., {Bolatto}, A.~D., {et~al.} 2015, \apj, 801, 63,
  \dodoi{10.1088/0004-637X/801/1/63}

\bibitem[{{Messa} {et~al.}(2018{\natexlab{a}}){Messa}, {Adamo}, {{\"O}stlin},
  {Calzetti}, {Grasha}, {Grebel}, {Shabani}, {Chandar}, {Dale}, {Dobbs},
  {Elmegreen}, {Fumagalli}, {Gouliermis}, {Kim}, {Smith}, {Thilker}, {Tosi},
  {Ubeda}, {Walterbos}, {Whitmore}, {Fedorenko}, {Mahadevan}, {Andrews},
  {Bright}, {Cook}, {Kahre}, {Nair}, {Pellerin}, {Ryon}, {Ahmad}, {Beale},
  {Brown}, {Clarkson}, {Guidarelli}, {Parziale}, {Turner}, \&
  {Weber}}]{messa18a}
{Messa}, M., {Adamo}, A., {{\"O}stlin}, G., {et~al.} 2018{\natexlab{a}},
  \mnras, 473, 996, \dodoi{10.1093/mnras/stx2403}

\bibitem[{{Messa} {et~al.}(2018{\natexlab{b}}){Messa}, {Adamo}, {Calzetti},
  {Reina-Campos}, {Colombo}, {Schinnerer}, {Chandar}, {Dale}, {Gouliermis},
  {Grasha}, {Grebel}, {Elmegreen}, {Fumagalli}, {Johnson}, {Kruijssen},
  {{\"O}stlin}, {Shabani}, {Smith}, \& {Whitmore}}]{messa18b}
{Messa}, M., {Adamo}, A., {Calzetti}, D., {et~al.} 2018{\natexlab{b}}, \mnras,
  477, 1683, \dodoi{10.1093/mnras/sty577}

\bibitem[{{Mills} {et~al.}(2021){Mills}, {Gorski}, {Emig}, {Bolatto}, {Levy},
  {Leroy}, {Ginsburg}, {Henshaw}, {Zschaechner}, {Veilleux}, {Tanaka}, {Meier},
  {Walter}, {Krieger}, \& {Ott}}]{mills21}
{Mills}, E.~A.~C., {Gorski}, M., {Emig}, K.~L., {et~al.} 2021, \apj, 919, 105,
  \dodoi{10.3847/1538-4357/ac0fe8}

\bibitem[{{Mok} {et~al.}(2019){Mok}, {Chandar}, \& {Fall}}]{mok19}
{Mok}, A., {Chandar}, R., \& {Fall}, S.~M. 2019, \apj, 872, 93,
  \dodoi{10.3847/1538-4357/aaf6ea}

\bibitem[{{Mok} {et~al.}(2020){Mok}, {Chandar}, \& {Fall}}]{mok20}
---. 2020, \apj, 893, 135, \dodoi{10.3847/1538-4357/ab7a14}

\bibitem[{{Molinari} {et~al.}(2011){Molinari}, {Bally}, {Noriega-Crespo},
  {Compi{\`e}gne}, {Bernard}, {Paradis}, {Martin}, {Testi}, {Barlow}, {Moore},
  {Plume}, {Swinyard}, {Zavagno}, {Calzoletti}, {Di Giorgio}, {Elia},
  {Faustini}, {Natoli}, {Pestalozzi}, {Pezzuto}, {Piacentini}, {Polenta},
  {Polychroni}, {Schisano}, {Traficante}, {Veneziani}, {Battersby}, {Burton},
  {Carey}, {Fukui}, {Li}, {Lord}, {Morgan}, {Motte}, {Schuller},
  {Stringfellow}, {Tan}, {Thompson}, {Ward-Thompson}, {White}, \&
  {Umana}}]{molinari11}
{Molinari}, S., {Bally}, J., {Noriega-Crespo}, A., {et~al.} 2011, \apjl, 735,
  L33, \dodoi{10.1088/2041-8205/735/2/L33}

\bibitem[{{M{\"u}ller-S{\'a}nchez} {et~al.}(2010){M{\"u}ller-S{\'a}nchez},
  {Gonz{\'a}lez-Mart{\'\i}n}, {Fern{\'a}ndez-Ontiveros}, {Acosta-Pulido}, \&
  {Prieto}}]{muller-sanchez10}
{M{\"u}ller-S{\'a}nchez}, F., {Gonz{\'a}lez-Mart{\'\i}n}, O.,
  {Fern{\'a}ndez-Ontiveros}, J.~A., {Acosta-Pulido}, J.~A., \& {Prieto}, M.~A.
  2010, \apj, 716, 1166, \dodoi{10.1088/0004-637X/716/2/1166}

\bibitem[{{Nguyen} \& {Thompson}(2022)}]{nguyen22}
{Nguyen}, D.~D., \& {Thompson}, T.~A. 2022, arXiv e-prints, arXiv:2205.13465.
\newblock \doarXiv{2205.13465}

\bibitem[{{Ossenkopf} \& {Henning}(1994)}]{ossenkopf94}
{Ossenkopf}, V., \& {Henning}, T. 1994, \aap, 291, 943

\bibitem[{{Paglione} {et~al.}(2004){Paglione}, {Yam}, {Tosaki}, \&
  {Jackson}}]{paglione04}
{Paglione}, T. A.~D., {Yam}, O., {Tosaki}, T., \& {Jackson}, J.~M. 2004, \apj,
  611, 835, \dodoi{10.1086/422354}

\bibitem[{{Pence}(1980)}]{pence80}
{Pence}, W.~D. 1980, \apj, 239, 54, \dodoi{10.1086/158088}

\bibitem[{{P{\'e}rez-Ram{\'\i}rez} {et~al.}(2000){P{\'e}rez-Ram{\'\i}rez},
  {Knapen}, {Peletier}, {Laine}, {Doyon}, \& {Nadeau}}]{perez-ramirez00}
{P{\'e}rez-Ram{\'\i}rez}, D., {Knapen}, J.~H., {Peletier}, R.~F., {et~al.}
  2000, \mnras, 317, 234, \dodoi{10.1046/j.1365-8711.2000.03521.x}

\bibitem[{{Peters}(1975)}]{peters75}
{Peters}, W.~L., I. 1975, \apj, 195, 617, \dodoi{10.1086/153363}

\bibitem[{{Plummer}(1911)}]{plummer1911}
{Plummer}, H.~C. 1911, \mnras, 71, 460, \dodoi{10.1093/mnras/71.5.460}

\bibitem[{{Portegies Zwart} {et~al.}(2010){Portegies Zwart}, {McMillan}, \&
  {Gieles}}]{portegieszwart10}
{Portegies Zwart}, S.~F., {McMillan}, S. L.~W., \& {Gieles}, M. 2010, \araa,
  48, 431, \dodoi{10.1146/annurev-astro-081309-130834}

\bibitem[{Reback {et~al.}(2020)Reback, McKinney, jbrockmendel, den Bossche,
  Augspurger, Cloud, gfyoung, Sinhrks, Hawkins, Klein, Roeschke, Tratner,
  Petersen, She, Ayd, MomIsBestFriend, Garcia, Schendel, Hayden, Saxton,
  Jancauskas, McMaster, Battiston, Seabold, chris b1, h~vetinari, Dong, Hoyer,
  Overmeire, \& Winkel}]{pandas}
Reback, J., McKinney, W., jbrockmendel, {et~al.} 2020, pandas-dev/pandas:
  Pandas 1.1.3, v1.1.3,  Zenodo, \dodoi{10.5281/zenodo.4067057}

\bibitem[{{Regan} \& {Teuben}(2003)}]{regan03}
{Regan}, M.~W., \& {Teuben}, P. 2003, \apj, 582, 723, \dodoi{10.1086/344721}

\bibitem[{{Rekola} {et~al.}(2005){Rekola}, {Richer}, {McCall}, {Valtonen},
  {Kotilainen}, \& {Flynn}}]{rekola05}
{Rekola}, R., {Richer}, M.~G., {McCall}, M.~L., {et~al.} 2005, \mnras, 361,
  330, \dodoi{10.1111/j.1365-2966.2005.09166.x}

\bibitem[{{Rico-Villas} {et~al.}(2020){Rico-Villas}, {Mart{\'\i}n-Pintado},
  {Gonz{\'a}lez-Alfonso}, {Mart{\'\i}n}, \& {Rivilla}}]{rico-villas20}
{Rico-Villas}, F., {Mart{\'\i}n-Pintado}, J., {Gonz{\'a}lez-Alfonso}, E.,
  {Mart{\'\i}n}, S., \& {Rivilla}, V.~M. 2020, \mnras, 491, 4573,
  \dodoi{10.1093/mnras/stz3347}

\bibitem[{{Ridley} {et~al.}(2017){Ridley}, {Sormani}, {Tre{\ss}}, {Magorrian},
  \& {Klessen}}]{ridley17}
{Ridley}, M. G.~L., {Sormani}, M.~C., {Tre{\ss}}, R.~G., {Magorrian}, J., \&
  {Klessen}, R.~S. 2017, \mnras, 469, 2251, \dodoi{10.1093/mnras/stx944}

\bibitem[{Rohatgi(2021)}]{webplotdigitizer}
Rohatgi, A. 2021, Webplotdigitizer: Version 4.5.
\newblock \url{https://automeris.io/WebPlotDigitizer}

\bibitem[{{Sakamoto} {et~al.}(2011){Sakamoto}, {Mao}, {Matsushita}, {Peck},
  {Sawada}, \& {Wiedner}}]{sakamoto11}
{Sakamoto}, K., {Mao}, R.-Q., {Matsushita}, S., {et~al.} 2011, \apj, 735, 19,
  \dodoi{10.1088/0004-637X/735/1/19}

\bibitem[{{Sawada} {et~al.}(2004){Sawada}, {Hasegawa}, {Handa}, \&
  {Cohen}}]{sawada04}
{Sawada}, T., {Hasegawa}, T., {Handa}, T., \& {Cohen}, R.~J. 2004, \mnras, 349,
  1167, \dodoi{10.1111/j.1365-2966.2004.07603.x}

\bibitem[{{Scoville} {et~al.}(1985){Scoville}, {Soifer}, {Neugebauer}, {Young},
  {Matthews}, \& {Yerka}}]{scoville85}
{Scoville}, N.~Z., {Soifer}, B.~T., {Neugebauer}, G., {et~al.} 1985, \apj, 289,
  129, \dodoi{10.1086/162871}

\bibitem[{{Sharp} \& {Bland-Hawthorn}(2010)}]{sharp10}
{Sharp}, R.~G., \& {Bland-Hawthorn}, J. 2010, \apj, 711, 818,
  \dodoi{10.1088/0004-637X/711/2/818}

\bibitem[{{Shin} {et~al.}(2017){Shin}, {Kim}, {Baba}, {Saitoh}, {Hwang},
  {Chun}, \& {Hozumi}}]{shin17}
{Shin}, J., {Kim}, S.~S., {Baba}, J., {et~al.} 2017, \apj, 841, 74,
  \dodoi{10.3847/1538-4357/aa7061}

\bibitem[{{Sofue}(1995)}]{sofue95}
{Sofue}, Y. 1995, \pasj, 47, 527.
\newblock \doarXiv{astro-ph/9508110}

\bibitem[{{Sofue} \& {Rubin}(2001)}]{sofue01}
{Sofue}, Y., \& {Rubin}, V. 2001, \araa, 39, 137,
  \dodoi{10.1146/annurev.astro.39.1.137}

\bibitem[{{Sorai} {et~al.}(2000){Sorai}, {Nakai}, {Kuno}, {Nishiyama}, \&
  {Hasegawa}}]{sorai00}
{Sorai}, K., {Nakai}, N., {Kuno}, N., {Nishiyama}, K., \& {Hasegawa}, T. 2000,
  \pasj, 52, 785, \dodoi{10.1093/pasj/52.5.785}

\bibitem[{{Sormani} {et~al.}(2018){Sormani}, {Sobacchi}, {Fragkoudi}, {Ridley},
  {Tre{\ss}}, {Glover}, \& {Klessen}}]{sormani18}
{Sormani}, M.~C., {Sobacchi}, E., {Fragkoudi}, F., {et~al.} 2018, \mnras, 481,
  2, \dodoi{10.1093/mnras/sty2246}

\bibitem[{{Sormani} {et~al.}(2020){Sormani}, {Tress}, {Glover}, {Klessen},
  {Battersby}, {Clark}, {Hatchfield}, \& {Smith}}]{sormani20}
{Sormani}, M.~C., {Tress}, R.~G., {Glover}, S. C.~O., {et~al.} 2020, \mnras,
  497, 5024, \dodoi{10.1093/mnras/staa1999}

\bibitem[{{Sormani} {et~al.}(2022){Sormani}, {Sanders}, {Fritz}, {Smith},
  {Gerhard}, {Sch{\"o}del}, {Magorrian}, {Neumayer}, {Nogueras-Lara},
  {Feldmeier-Krause}, {Mastrobuono-Battisti}, {Schultheis}, {Shahzamanian},
  {Vasiliev}, {Klessen}, {Lucas}, \& {Minniti}}]{sormani22}
{Sormani}, M.~C., {Sanders}, J.~L., {Fritz}, T.~K., {et~al.} 2022, \mnras, 512,
  1857, \dodoi{10.1093/mnras/stac639}

\bibitem[{{Strickland} {et~al.}(2000){Strickland}, {Heckman}, {Weaver}, \&
  {Dahlem}}]{strickland00}
{Strickland}, D.~K., {Heckman}, T.~M., {Weaver}, K.~A., \& {Dahlem}, M. 2000,
  \aj, 120, 2965, \dodoi{10.1086/316846}

\bibitem[{{Strickland} {et~al.}(2002){Strickland}, {Heckman}, {Weaver},
  {Hoopes}, \& {Dahlem}}]{strickland02}
{Strickland}, D.~K., {Heckman}, T.~M., {Weaver}, K.~A., {Hoopes}, C.~G., \&
  {Dahlem}, M. 2002, \apj, 568, 689, \dodoi{10.1086/338889}

\bibitem[{{Sturm} {et~al.}(2011){Sturm}, {Gonz{\'a}lez-Alfonso}, {Veilleux},
  {Fischer}, {Graci{\'a}-Carpio}, {Hailey-Dunsheath}, {Contursi}, {Poglitsch},
  {Sternberg}, {Davies}, {Genzel}, {Lutz}, {Tacconi}, {Verma}, {Maiolino}, \&
  {de Jong}}]{sturm11}
{Sturm}, E., {Gonz{\'a}lez-Alfonso}, E., {Veilleux}, S., {et~al.} 2011, \apjl,
  733, L16, \dodoi{10.1088/2041-8205/733/1/L16}

\bibitem[{{Sugai} {et~al.}(2003){Sugai}, {Davies}, \& {Ward}}]{sugai03}
{Sugai}, H., {Davies}, R.~I., \& {Ward}, M.~J. 2003, \apjl, 584, L9,
  \dodoi{10.1086/368271}

\bibitem[{{Tress} {et~al.}(2020){Tress}, {Sormani}, {Glover}, {Klessen},
  {Battersby}, {Clark}, {Hatchfield}, \& {Smith}}]{tress20}
{Tress}, R.~G., {Sormani}, M.~C., {Glover}, S. C.~O., {et~al.} 2020, \mnras,
  499, 4455, \dodoi{10.1093/mnras/staa3120}

\bibitem[{{Turner} \& {Ho}(1985)}]{turner85}
{Turner}, J.~L., \& {Ho}, P.~T.~P. 1985, \apjl, 299, L77,
  \dodoi{10.1086/184584}

\bibitem[{{Ulvestad} \& {Antonucci}(1997)}]{ulvestad97}
{Ulvestad}, J.~S., \& {Antonucci}, R. R.~J. 1997, \apj, 488, 621,
  \dodoi{10.1086/304739}

\bibitem[{{van Albada} \& {Sanders}(1982)}]{vanalbada82}
{van Albada}, T.~S., \& {Sanders}, R.~H. 1982, \mnras, 201, 303,
  \dodoi{10.1093/mnras/201.2.303}

\bibitem[{Van~Rossum(2020)}]{re}
Van~Rossum, G. 2020, The Python Library Reference, release 3.8.2 (Python
  Software Foundation)

\bibitem[{Virtanen {et~al.}(2020)Virtanen, Gommers, Oliphant, Haberland, Reddy,
  Cournapeau, Burovski, Peterson, Weckesser, Bright, {van der Walt}, Brett,
  Wilson, Millman, Mayorov, Nelson, Jones, Kern, Larson, Carey, Polat, Feng,
  Moore, {VanderPlas}, Laxalde, Perktold, Cimrman, Henriksen, Quintero, Harris,
  Archibald, Ribeiro, Pedregosa, {van Mulbregt}, \& {SciPy 1.0
  Contributors}}]{scipy}
Virtanen, P., Gommers, R., Oliphant, T.~E., {et~al.} 2020, Nature Methods, 17,
  261, \dodoi{10.1038/s41592-019-0686-2}

\bibitem[{{Wada} \& {Habe}(1992)}]{wada92}
{Wada}, K., \& {Habe}, A. 1992, \mnras, 258, 82, \dodoi{10.1093/mnras/258.1.82}

\bibitem[{{Walter} {et~al.}(2017){Walter}, {Bolatto}, {Leroy}, {Veilleux},
  {Warren}, {Hodge}, {Levy}, {Meier}, {Ostriker}, \& {Ott}}]{walter17}
{Walter}, F., {Bolatto}, A.~D., {Leroy}, A.~K., {et~al.} 2017, \apj, 835, 265,
  \dodoi{10.3847/1538-4357/835/2/265}

\bibitem[{{Waskom} {et~al.}(2014){Waskom}, {Botvinnik}, {Hobson}, {Cole},
  {Halchenko}, {Hoyer}, {Miles}, {Augspurger}, {Yarkoni}, {Megies}, {Coelho},
  {Wehner}, {cynddl}, {Ziegler}, {diego0020}, {Zaytsev}, {Hoppe}, {Seabold},
  {Cloud}, {Koskinen}, {Meyer}, {Qalieh}, \& {Allan}}]{seaborn}
{Waskom}, M., {Botvinnik}, O., {Hobson}, P., {et~al.} 2014, {Seaborn: V0.5.0
  (November 2014)}, v0.5.0,  Zenodo, \dodoi{10.5281/zenodo.12710}

\bibitem[{{Watson} {et~al.}(1996){Watson}, {Gallagher}, {Holtzman}, {Hester},
  {Mould}, {Ballester}, {Burrows}, {Casertano}, {Clarke}, {Crisp}, {Evans},
  {Griffiths}, {Hoessel}, {Scowen}, {Stapelfeldt}, {Trauger}, \&
  {Westphal}}]{watson96}
{Watson}, A.~M., {Gallagher}, J.~S., I., {Holtzman}, J.~A., {et~al.} 1996, \aj,
  112, 534, \dodoi{10.1086/118032}

\bibitem[{{Wei{\ss}} {et~al.}(2008){Wei{\ss}}, {Kov{\'a}cs}, {G{\"u}sten},
  {Menten}, {Schuller}, {Siringo}, \& {Kreysa}}]{weiss08}
{Wei{\ss}}, A., {Kov{\'a}cs}, A., {G{\"u}sten}, R., {et~al.} 2008, \aap, 490,
  77, \dodoi{10.1051/0004-6361:200809909}

\bibitem[{{Westmoquette} {et~al.}(2011){Westmoquette}, {Smith}, \&
  {Gallagher}}]{westmoquette11}
{Westmoquette}, M.~S., {Smith}, L.~J., \& {Gallagher}, J.~S., I. 2011, \mnras,
  414, 3719, \dodoi{10.1111/j.1365-2966.2011.18675.x}

\bibitem[{{Whitmore} {et~al.}(2010){Whitmore}, {Chandar}, {Schweizer},
  {Rothberg}, {Leitherer}, {Rieke}, {Rieke}, {Blair}, {Mengel}, \&
  {Alonso-Herrero}}]{whitmore10}
{Whitmore}, B.~C., {Chandar}, R., {Schweizer}, F., {et~al.} 2010, \aj, 140, 75,
  \dodoi{10.1088/0004-6256/140/1/75}

\bibitem[{{Zhang} \& {Fall}(1999)}]{zhang99}
{Zhang}, Q., \& {Fall}, S.~M. 1999, \apjl, 527, L81, \dodoi{10.1086/312412}

\bibitem[{{Zschaechner} {et~al.}(2018){Zschaechner}, {Bolatto}, {Walter},
  {Leroy}, {Herrera}, {Krieger}, {Kruijssen}, {Meier}, {Mills}, {Ott},
  {Veilleux}, \& {Weiss}}]{zschaechner18}
{Zschaechner}, L.~K., {Bolatto}, A.~D., {Walter}, F., {et~al.} 2018, \apj, 867,
  111, \dodoi{10.3847/1538-4357/aadf32}

\end{thebibliography}

\end{document}